\documentclass[12pt]{article}
\usepackage{epsfig,amsfonts,amsmath, amsthm}

\newcommand{\fref}[1]{Figure~\ref{#1}}
\newcommand{\cref}[1]{Chapter~\ref{#1}}
\newcommand{\beq}{\begin{equation}}
\newcommand{\eeq}{\end{equation}}
\newcommand{\ba}{\begin{array}}
\newcommand{\ea}{\end{array}}
\newcommand{\bcenter}{\begin{center}}
\newcommand{\ecenter}{\end{center}}

\def\IC{\mathbb{C}}

\def\IGa{\relax\hbox{${\rm I}\kern-.18em\Gamma$}}

\def\IZ{\mathbb{Z}}

\def\Tr{{\rm Tr}}

\def\smiley{\hbox{\large$\bigcirc$\hspace{-0.80em}\raise.2ex
\hbox{$\cdot\cdot$}\kern-.61em\lower.2ex\hbox{\scriptsize$\smile$}}\ }
\def\frowny{\hbox{\large$\bigcirc$\hspace{-0.80em}\raise.2ex
\hbox{$\cdot\cdot$}\kern-.635em\lower.2ex\hbox{\scriptsize$\frown$}}\ }

\makeatletter
\let\hangafter\@hangfrom
\makeatother

\newcommand{\be}{\begin{equation}}
\newcommand{\ee}{\end{equation}}
\newcommand{\bea}{\begin{eqnarray}}
\newcommand{\eea}{\end{eqnarray}}
\newcommand{\bean}{\begin{eqnarray*}}
\newcommand{\eean}{\end{eqnarray*}}
\newcommand{\bc}{\begin{center}}
\newcommand{\ec}{\end{center}}

\newcommand{\comment}[1]{ }

\begin{document}

~\vskip -2.5truecm
\rightline{Bicocca-FT-07-07} \rightline{SISSA 25/2007/EP}
\rightline{MIT-CTP-3837} \rightline{LPTENS 07/19}  \vskip 1.0cm

\centerline{\Huge Counting Chiral Operators}

\centerline{\Huge in Quiver Gauge Theories}

~\\

\renewcommand{\thefootnote}{\fnsymbol{footnote}}
\centerline{
 \bf Agostino Butti${}^{1}$
, \bf Davide Forcella${}^{1,2,3}$
}
\centerline{
 \bf Amihay Hanany${}^{4}$
,
 \bf David Vegh${}^{5}$
,
 \bf Alberto Zaffaroni${}^{6}$
\footnote{{\bf e-mail}: agostino.butti@lpt.ens.fr, forcella@sissa.it,
 ahanany@perimeterinstitute.ca, dvegh@mit.edu,
 alberto.zaffaroni@mib.infn.it}
}
~\vskip -1truecm
{\small
\begin{center}
${}^1$ Laboratoire de Physique Th\'eorique de l'\'Ecole Normale Sup\'erieure\\
24, rue Lhomond, 75321 Paris Cedex 05, France \\
~\vskip -0.4truecm
${}^2$ International School for Advanced Studies (SISSA/ISAS) \\
and INFN-Sezione di Trieste, via Beirut 2, I-34014, Trieste, Italy\\
~\vskip -0.4truecm
${}^3$ LPTHE, Universit\'es Paris VI et VII, Jussieu 75252 Paris, France\\
~\vskip -0.4truecm
${}^4$ Perimeter Institute for Theoretical Physics\\
31 Caroline Street North,
Waterloo Ontario N2L 2Y5, Canada\\
~\vskip -0.4truecm
${}^5$ Center for Theoretical Physics,
Massachusetts Institute of Technology\\
77 Massachusetts Avenue, Cambridge, MA 02139, USA\\
~\vskip -0.4truecm
${}^6$ Universit\`{a} di Milano-Bicocca
 and INFN, sezione di  Milano-Bicocca\\
Piazza della Scienza, 3; I-20126 Milano, Italy
\end{center}
}
~\\
We discuss in detail the problem of counting BPS gauge invariant operators in the chiral ring of
quiver gauge theories living on D-branes probing generic toric $CY$ singularities. The
computation of generating functions that include counting of baryonic operators is based on a
relation between the baryonic charges in field theory and the K\"ahler moduli of the $CY$
singularities. A study of the interplay between gauge theory and geometry shows that given
geometrical sectors appear more than once in the field theory, leading to a notion of
``multiplicities".
We explain in detail how to decompose the generating function for one D-brane into different sectors and how to compute their relevant multiplicities by introducing geometric and anomalous baryonic charges. The Plethystic
Exponential remains a major tool for passing from one D-brane to arbitrary number $N$ of D-branes.
Explicit formulae are given for few examples, including $\mathbb{C}^3/\mathbb{Z}_3$,
$\mathbb{F}_0$, and $dP_1$.

\setcounter{footnote}{0}
\renewcommand{\thefootnote}{\arabic{footnote}}
\vskip -1.8in


\newpage

\tableofcontents

\newpage

\section{Introduction}

Recently, there has been growing interest in ``counting'' chiral BPS operators in field theories
which arise on the world-volume of branes probing Calabi-Yau singularities
\cite{Romelsberger:2005eg, Kinney:2005ej, Nakayama:2005mf, Biswas:2006tj, Mandal:2006tk,
Benvenuti:2006qr, Martelli:2006vh, Basu:2006id, Butti:2006au, Hanany:2006uc, Feng:2007ur,
Forcella:2007wk, Grant:2007ze, Nakayama:2007jy, Lee:2007kv, Dolan:2007rq, Dav}. Determining the
matter content and the interactions of these field theories is an interesting and nontrivial
question in itself. The study of this problem began with orbifolds \cite{Douglas:1996sw,
Johnson:1996py, Kachru:1998ys, Lawrence:1998ja} and much progress has been made in understanding
toric singularities as well \cite{Douglas:1997de, Morrison:1998cs, Beasley:1999uz, Feng:2000mi,Cachazo:2001sg, Feng:2001xr, Feng:2002zw}.
In the toric case brane tilings,
a generalization of the brane boxes
\cite{Hanany:1997tb,Hanany:1998ru,Hanany:1998it},
allow for a great simplification of the problem by providing a very geometric viewpoint \cite{Hanany:2005ve,
Franco:2005rj, Butti:2005vn, Hanany:2005ss, Feng:2005gw, Hanany:2006nm,
Garcia-Etxebarria:2006aq, Brini:2006ej, Butti:2006hc}.

The chiral BPS operators in question are dual to D3-branes wrapped on generically nontrivial
three-cycles on the gravity side \cite{Gubser:1998fp}.
Branes on trivial cycles are termed (dual) giant gravitons \cite{McGreevy:2000cw,
Hashimoto:2000zp,Balasubramanian:2001nh} and are dual to mesonic operators while branes on
nontrivial cycles are dual to baryonic operators \cite{Gubser:1998fp, Berenstein:2002ke,
Herzog:2003wt}. There is a relation between giant gravitons or baryons and holomorphic curves in
the Calabi-Yau which was first discussed in \cite{Mikhailov:2000ya,Beasley:2002xv}. As a
consequence, combinatorial data of BPS operators can be packed into generating functions for
holomorphic curves \cite{Butti:2006au,Martelli:2006yb}, which contain ample information about
the geometry of the singularity.


This paper is devoted to the study of the (baryonic and mesonic) generating function for the chiral ring in
quiver gauge theories. Extending the results of \cite{Forcella:2007wk}, we compute the
generating functions including baryonic degrees of freedom for various theories. We first study
in detail the generating function for one D-brane and we decompose it into sectors with definite
baryonic charges. This decomposition is closely related to the geometry and to the generating
functions for holomorphic curves obtained by localization in the Calabi-Yau manifold.
We conjecture that  the generating function for a number $N$ of D-branes is completely determined by
the generating function for a single D-brane and it is obtained by applying the plethystic
exponential to each sector \cite{Benvenuti:2006qr, Butti:2006au, Hanany:2006uc, Feng:2007ur,
Forcella:2007wk}. This conjecture, which can be proved in the case of mesonic operators, is
inspired by the geometrical quantization of the classical D3-brane configurations in the
gravitational dual. We explicitly compute the generating functions for a selected set of
singularities, including $\mathbb{C}^3/\mathbb{Z}_3$, $\mathbb{F}_0$ and $dP_1$, and we make
various checks in the dual field theory.

In \cite{Forcella:2007wk} we studied the simpler and elegant
cases of the conifold and the $\mathbb{C}^2/\mathbb{Z}_2$ orbifold. A new feature, which
arises for more involved singularities, like for example $\mathbb{C}^3/\mathbb{Z}_3$,
$\mathbb{F}_0$ and $dP_1$, is the existence of multiplicities, namely the fact
that geometrical sectors appear more than once in field theory.
As we go over these examples in detail, we find that multiplicities have a geometrical
interpretation and can be resolved, with a construction that ties together
in a fascinating way the algebraic geometry of the CY and the combinatorics
of quiver data.



The paper is organized as follows. In Section 2, we discuss the basics of generating functions
and plethystics. In Section 3, we apply these tools and give a brief review of the conifold
example. Section 4 continues with the detailed discussion of the partition functions from both
the field theory and the geometry perspectives. The GKZ decomposition is introduced and the
auxiliary GKZ partition function is defined. Section 5 contains detailed examples, based on $\mathbb{C}^3/\mathbb{Z}_3$, $\mathbb{F}_0$ and $dP_1$, and explicitly
computes generating functions both for $N=1$ and for small $N>1$. Section 6 deals with $N>1$ D-branes and gives a systematic approach to the field theory computation
by means of the Molien formula. Appendices A and B contain some preliminary
discussion and observations about singular horizons and a discussion of the
anomalous baryonic charges from the viewpoint of the dual {\it shiver} construction.

Finally, a useful list of notations is reported in Appendix C.

\section{General structure of generating functions for BPS operators}

In this section we will give general prescriptions on the computation of generating functions
for BPS operators in the chiral ring of a supersymmetric gauge theory that lives on a D-brane
which probes a generic non-compact Calabi-Yau manifold. The simpler cases of the conifold and
the $\mathbb{C}^2 / \mathbb{Z}_2$  orbifold were discussed in detail in \cite{Forcella:2007wk}.

Given an ${\cal N}=1$ supersymmetric gauge theory with a collection of $U(1)$ global symmetries,
$\prod_{i=1}^r U(1)_i$, we have a set of $r$ {\bf chemical potentials} $\{t_i\}_{i=1}^r$. The
generating function for a gauge theory living on a D-brane probing a generic non-compact CY
manifold depends on the set of parameters, $t_i$. There is always at least one such $U(1)$
global symmetry and one such chemical potential $t$, corresponding to the $U(1)_R$ symmetry.

The global charges are divided into classes: {\bf baryonic charges}, and {\bf flavor charges}
(by abuse of language, we will include the $R$-symmetry in this latter class). The number of
non-anomalous flavor symmetries, related to the isometries of the CY, is less than three while
the number of non-anomalous baryonic symmetries, related to the group of divisors in the CY, can
be quite large. In certain cases, we can also have baryonic discrete symmetries. As is
demonstrated below, in addition to the non-anomalous baryonic charges we need to consider the
{\bf anomalous baryonic charges}. We will only consider the case of toric CY where the number of
flavor symmetries is three. When it will be necessary to make distinctions, we will denote with
$x,y$ or $q_i$ the flavor chemical potentials and with $b_i$ the non-anomalous baryonic chemical potentials. Chemical potentials for anomalous charges are denoted by $a_i$.

For a given CY manifold, we denote the generating function for $N$ D-branes by $g_{N}(\{t_i\}; \
CY)$. The generating function for $N=1$ is simple to compute by using field theory arguments.
Recall that the quiver gauge theory living on the world-volume of the D3-branes, which can be
determined in the toric case using dimer technology \cite{Hanany:2005ve,
Franco:2005rj},\footnote{See \cite{ Hanany:2005ss,Feng:2005gw, Hanany:2006nm, Garcia-Etxebarria:2006aq, Brini:2006ej, Butti:2006hc, Benvenuti:2005ja, Benvenuti:2005cz,Franco:2005sm,Butti:2005sw, Butti:2005ps,Franco:2006gc, Benvenuti:2006xg,Lee:2006ru, Imamura:2006ub,Ueda:2006wy,Butti:2006nk,Imamura:2006ie,Oota:2006eg,Kato:2006vx,Imamura:2007dc} for a rich set of subsequent
developments.} consists of a gauge group $SU(N)^G$, adjoint or bi-fundamental chiral
fields\footnote{Henceforth we denote fields by bold characters to distinguish them from global
quantum numbers.} ${\bf X}^J$, which can be considered as $N\times N$ matrices, and a
superpotential $W({\bf X}^J)$.

For $N=1$ the matrices ${\bf X}^J$ reduce to numbers and the F-term conditions become polynomial
equations in the commuting numbers ${\bf X}^J$. We can consider the polynomial ring
$\mathbb{C}[{\bf X}^J]$ to be graded by the weights $t_i$. Since the gauge group is acting
trivially for $N=1$, the ring of gauge invariants is just the quotient ring
$${\cal R}_{N=1}^{inv}=\mathbb{C}[{\bf X}^J]/{\cal I}$$
where $\cal I$ is the set of F-term constraints $dW({\bf X}^J)/d{\bf X}^J$. The generating function
for polynomial rings is called Hilbert series in the mathematical literature and can be
computed in an algorithmic way. In particular, computer algebra programs, like Macaulay2
\cite{M2}, have built-in commands to compute these generating functions. We can therefore assume
that the generating function $g_1(\{t_i\})$ for ${\cal R}_{N=1}^{inv}$ is
known.

We proceed to the determination of $g_N$ with a general conjecture:

\begin{itemize}
\item{
{\it For the class of theories considered here (D-branes probing non-compact CY which are any of
toric, orbifolds or complete intersections), the knowledge of the generating function for $N=1$
is enough to compute the generating function for any $N$.}}
\end{itemize}

This is a familiar fact for mesonic generating functions \cite{Benvenuti:2006qr}, and it is
essentially due to the fact that the operators for finite $N$ are symmetric functions of the
operators for $N=1$. This is also familiar for baryonic generating functions, where the
knowledge of a single generating function, $g_{1,B}$, for one D-brane, $N=1$ and baryon number
$B$, is enough to compute all generating functions for any number of D-branes and for a fixed
baryonic number \cite{Butti:2006au}.

The general construction is as follows. There exists a decomposition of the $N=1$ ring of
invariants ${\cal R}_{N=1}^{inv}$, and consequently of its generating function, into sectors ${\cal S}$ of
definite baryonic charges

\begin{equation}
g_1(\{t_i\}; CY)=\sum_{{\cal S}} g_{1,\cal S}(\{t_i\}; CY)
\end{equation}
where $g_{1,\cal S}$ is the generating function for the subsector ${\cal S}\subset {\cal
R}_{N=1}^{inv}$. All elements in ${\cal S}$ have the same baryonic charges, and, except for a
multiplicative factor, $g_{1,\cal S}$ only depends on the flavor charges $q_i$. In simple cases,
like the conifold, ${\cal S}$ is just a label running over all the possible values of the
non-anomalous baryonic charge. The understanding of the precise decomposition of ${\cal
R}_{N=1}^{inv}$ into subsectors in the general case is a nontrivial task and is one of the
subjects of this paper.

The generating function for $N$ branes is then obtained by taking $N$-fold symmetric products of
elements in each given sector $\cal S$. This is precisely the role which is played by the Plethystic
Exponential (PE) -- to take a generating function for a set of operators and count all possible
symmetric functions of it. If we introduce a chemical potential $\nu$ for the number of
D-branes, the generating function for any number of D-branes is given by
\begin{eqnarray}
\nonumber
g(\nu; \{t_i\}; CY)
&= &\sum_{\cal S} \hbox{PE$_{\nu}$}[g_{1,\cal S}(\{t_i\}; CY)] \equiv \sum_{\cal S} \exp\biggl(\sum_{k=1}^\infty \frac{\nu^k}{k}g_{1,\cal S}(\{t_i^k\}; CY) \biggr)\nonumber \\
&\equiv& \sum_{N=0}^\infty g_{N}(\{t_i\};CY) \nu^N
\label{g1plet}
\end{eqnarray}

The detailed description of the decomposition into sectors $\cal S$ is given
in the rest of this paper, but it is important to notice from the very
beginning that such a decomposition is not unique. As already mentioned above,
gauge invariants in the same sector have the same baryonic
charges. One can take these baryonic charges to be non-anomalous. This however, turns out to be not enough and we seem to need a finer decomposition of the ring of invariants
which is obtained by enlarging the set of non-anomalous baryonic charges to a
larger set. There are two basic extensions, one related to an expansion
in a full set of baryonic charges, anomalous and non-anomalous, and the other extension is related to a full set of discretized K\"ahler moduli on the CY resolutions. We thus have two
complementary points of view:

\begin{itemize}
\item{{\bf Quantum field theory perspective}:
the most general decomposition of the generating function $g_1(\{t_i\})$ is into the full set of
baryonic charges. Let us extend the set of chemical potentials $t_i$ to all the baryonic
charges, including the anomalous ones, denoted by $a_i$. There are $G-1$ independent baryonic
charges, where $G$ is the number of gauge groups. We can thus decompose ${\cal R}_{N=1}^{inv}$ into sectors with definite charges under
$U(1)^{G-1}$. $g_{1}(\{t_i\})$ will decompose into a formal Laurent series in the baryonic
chemical potentials $b_i$ and $a_i$. The ${\cal R}_{N}^{inv}$ rings of invariants for number of colors
$N$ will similarly decompose into sectors of definite baryonic charge.
We can formally extend the gauge group $SU(N)^G$ to $U(N)^G/U(1)$ by gauging the baryonic
symmetries.\footnote{The theory will of course be anomalous. The overall $U(1)$ is discarded
since it acts trivially.} From this perspective, the decomposition of the ring of $SU(N)^G$
invariants into Abelian representations of the extended group $U(N)^G/U(1)$ is sometimes called
an expansion in {\it covariants} and is extremely natural from the point of view of invariant
theory. All sectors ${\cal S}$ appear with multiplicity one in the decomposition of Equation
(\ref{g1plet}).}
\item{{\bf The dual geometrical perspective}: the full set of BPS states
of the dual gauge theory can be obtained by quantizing the classical configuration space of
supersymmetric D3-branes wrapped on the horizon. This problem can be equivalently rephrased in
terms of holomorphic surfaces in the CY with $g_1$ as generating function \cite{Mikhailov:2000ya,Beasley:2002xv}.
Quite remarkably, $g_1$ has a decomposition

\begin{equation}
g_1(\{t_i\}; CY)=\sum_{\cal S} m ( {\cal S} ) \ g_{1,\cal S}(\{t_i\}; CY),
\end{equation}

where the parameters $\cal S$ can be identified with a complete set of {\it discretized  K\"ahler
moduli} and the integers $m ({\cal S})$ are multiplicities. We will call it the {\bf GKZ decomposition}, from the known description of the K\"ahler
cone in terms of a secondary fan given by the GKZ construction (for a useful reference see \cite{Cox:2000vi}). The functions $g_{1,\cal S}$
can be explicitly determined with the computation of a character using the equivariant index
theorem. This geometrical decomposition has multiplicities $m({\cal S})$ which will be interpreted in the
following sections and discussed in detail in Section \ref{decomposition}. The result for finite $N$ is generated by the following function

\begin{equation}
g(\nu;\{t_i\}; CY)=\sum_{\cal S} m ( {\cal S} ) {\rm PE}_\nu [g_{1,\cal S}(\{t_i\}; CY)] ,
\end{equation}

and can be interpreted as the result of quantizing the classical BPS D3-brane
configuration in each sector $\cal S$.}
\end{itemize}

The two decompositions of the $N=1$ generating function are different and complementary. For a
toric CY manifold that has a toric diagram with $d$ external vertices and $I$ internal integral
points, the number of non-anomalous baryonic symmetries is $d-3$, the number of anomalous baryonic symmetries is $2 I$ and the dimension of the K\"ahler moduli space is $d-3+I$. The field theory expansion is thus based on a lattice $\Gamma_{(b,a)}$ of dimension $d-3+2 I$ consisting of all baryonic
charges, anomalous or not, while the geometrical expansion is based on a lattice $\Gamma_{GKZ}$
of dimension $d-3+I$. The two sets have a nontrivial intersection $\Gamma_b$, consisting of
non-anomalous baryonic charges.

At the end,  we will be interested in the generating function for BPS operators
with chemical potential with respect to the {\it non-anomalous} charges.
To this purpose, we must project any of the two lattices on their
intersection, which is the $d-3$ lattice of non-anomalous baryonic symmetries
$\Gamma_b$
$$g_1(\{t_i\})=\sum_{k \in \Gamma_b} m ( k )\ g_{1,k}(\{t_i\})$$
and multiplicities will generically appear.


On the other hand, we could even imagine to enlarge our lattices. Adding the anomalous baryonic
charges to the GKZ fan we obtain a lattice of dimension $d-3+3 I$. The points give hollow polygons
over the GKZ fan. All these issues will be discussed in detail in the rest of the paper.

\section{Review of the conifold example}\label{conif}

To demonstrate our general discussion above and to prepare for more involved cases we start by
reviewing the generating function for the conifold.

\begin{figure}[ht]
\begin{center}
  \epsfxsize = 10cm
  \centerline{\epsfbox{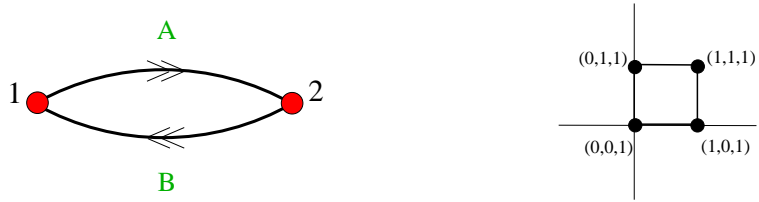}}
  \caption{Quiver and toric diagram for the conifold.}
  \label{coniquiver}
\end{center}
\end{figure}
The gauge theory on the conifold has a global symmetry $SU(2)_1\times SU(2)_2\times U(1)_R\times
U(1)_B$. It has four basic fields ${\bf A}_{1,2}$ and ${\bf B}_{1,2}$ that transform under these
symmetries according to Table \ref{globalconifold}.

\begin{table}[htdp]
\begin{center}
\begin{tabular}{|c||c c|c c|c|c||c|}
\hline
 & \multicolumn{2} {c|} {$SU(2)_1$} & \multicolumn{2} {c|} {$SU(2)_2$} & $U(1)_R$ & $U(1)_B$ & monomial\\ \hline
\cline{2-5} & $j_1$ & $m_1$ & $j_2$ & $m_2$&&&\\ \hline \hline ${\bf A}_1$ & $\frac{1}{2}$ &
$+\frac{1}{2}$ & 0 & 0 & $\frac{1}{2}$ & 1& $t_1 x$\\ \hline ${\bf A}_2$ & $\frac{1}{2}$ &
$-\frac{1}{2}$ & 0 &0 & $\frac{1}{2}$ & 1& $\frac{t_1}{x}$ \\ \hline ${\bf B}_1$ & 0 & 0 &
$\frac{1}{2}$ & $+\frac{1}{2}$ & $\frac{1}{2}$ & $-1$& $t_2 y$ \\ \hline ${\bf B}_2$ & 0 & 0 &
$\frac{1}{2}$ & $-\frac{1}{2}$ & $\frac{1}{2}$ & $-1$& $\frac{t_2}{y}$ \\ \hline
\end{tabular}
\end{center}
\caption{Global charges for the basic fields of the quiver gauge theory
living on the D-brane probing the conifold.}
\label{globalconifold}
\end{table}

The last column represents the corresponding monomial in the generating function for BPS
operators in the chiral ring. $t_1$ is the chemical potential for the number of ${\bf A}$
fields, $t_2$ is the chemical potential for the number of ${\bf B}$ fields, $x$ is the chemical
potential for the Cartan generator of $SU(2)_1$, and $y$ is the chemical potential for the
Cartan generator of $SU(2)_2$.

The theory has a single baryonic charge $U(1)_B$ which is not anomalous.
We can introduce a corresponding chemical potential $b$. With this notation
we have $t_1=t b$ and $t_2 = \frac{t}{b}$. The chemical potentials
$t$ and $b$ keep track of the R-charge and the baryonic charge $B$,
respectively.

Since the F-terms of the theory are
$$ {\bf A}_1 {\bf B}_i {\bf A}_2 -{\bf A}_2 {\bf B}_i {\bf A}_1 = 0\, \qquad\qquad  {\bf B}_1 {\bf A}_i {\bf B}_2 -{\bf B}_2 {\bf A}_i {\bf B}_1  = 0\qquad\qquad i=1,2 $$
they vanish for $N=1$. The $N=1$ generating function is thus freely generated by the four basic
fields of the conifold gauge theory and it takes the form

\begin{equation}
g_1(t_1,t_2,x,y; {\cal C}) = \frac{1}{(1-t_1 x) (1-\frac{t_1}{x})(1-t_2 y) (1-\frac{t_2}{y})}.
\label{g1coni}
\end{equation}

In the following we set $x=y=1$ for simplicity. General formulae including the $SU(2)$ chemical
potentials can be found in \cite{Forcella:2007wk} and in Section \ref{examples}.

$g_1$ decomposes into sectors with fixed baryonic charge $B$, each with multiplicity
one:
\begin{eqnarray}
g_1(t_1,t_2; {\cal C}) &=& \sum_{B=-\infty}^\infty g_{1,B}(t_1,t_2; {\cal C}),
\nonumber\\
g_{1,B>0}(t_1,t_2; {\cal C}) &=&\sum_{n=0}^\infty (n+1+B)(n+1) t_1^{n+B} t_2^{n}\nonumber\\
 g_{1,B<0}(t_1,t_2; {\cal C}) &=&\sum_{n=0}^\infty (n+1)(n+1+|B|) t_1^{n} t_2^{n+|B|}
\label{g1conit}
\end{eqnarray}

\begin{figure}[h!!!]
\begin{center}
\includegraphics[scale=0.65]{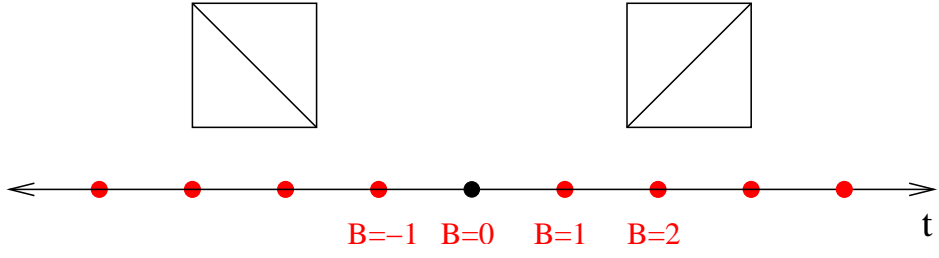}
\caption{The GKZ decomposition for the K\"ahler moduli space of the conifold, consisting of two
one-dimensional cones connected by a flop. The coordinate $t$ on the moduli space is associated
with the volume of the two-cycle in the resolution of the conifold. When $t$ goes to zero, the
cycle vanishes and we can perform a flop on the variety by inflating a different two-cycle. A
natural discretization of the GKZ fan is in correspondence with the decomposition of the $g_1$
generating function. }
\label{GKZconifold}
\end{center}
\end{figure}

It is manifest that each term in $g_{1,B}$ has a monomial $b^B$ corresponding to a baryonic
charge $B$. The decomposition into each baryonic charge can be computed by expanding $g_1(t,b;
{\cal C})$ in a  formal Laurent series in $b$ or by determining the functions $g_{1,B}$ by
resolving the CY, see Figure \ref{GKZconifold}, and using the equivariant index theorem. Both
computations have been discussed in detail in \cite{Forcella:2007wk} and will be reviewed in the
following Sections.

The decomposition has a clear interpretation in terms of supersymmetric D3-branes states:
$g_{1,B}$ counts the supersymmetric D3-branes wrapping cycles of homology $B$ thus corresponding
to states with baryonic number $B$. Quite remarkably, the structure of the integer lattices with
generating function $g_{1,B}$ and the explicit computation with the index theorem, strongly
suggests a relation between $B$ and a discretized K\"ahler modulus of the resolved CY
\cite{Forcella:2007wk}.

The result for generic $N$ is obtained as follows
\begin{eqnarray}
g(\nu; t_1,t_2; {\cal C}) &=& \sum_{B=-\infty}^\infty {\rm PE}_\nu [g_{1,B}(t_1,t_2; {\cal C})], \nonumber\\
g(\nu; t_1,t_2; {\cal C}) &=& \sum_{N=0}^\infty \nu^N g_{N}(t_1,t_2; {\cal C})
\end{eqnarray}

Here we list the generating functions for small values of $N$

\begin{equation}
g_2(t_1,t_2;{\cal C})= \frac{1 + t_1 t_2 + t_1^2 t_2^2 - 3 t_1^4 t_2^2 - 3 t_1^2 t_2^4 + t_1^5 t_2^3 + t_1^3 t_2^5  - 3 t_1^3 t_2^3 + 4 t_1^4 t_2^4}{(1 - t_1^2)^3(1 - t_1 t_2)^3 (1 - t_2^2)^3} .
\label{conifoldN2}
\end{equation}

\begin{equation}\label{conifoldN3}
g_3(t_1,t_2;{\cal C})=\frac{F(t_1,t_2)}{(1-t_1^3)^4(1 - t_1 t_2)^3 (1-t_1^2 t_2^2)^3(1- t_2^3)^4} ,
\end{equation}
\begin{eqnarray}
 F(t_1,t_2)&=& 1 +  t_1^{15}
    t_2^9 + + 3 t_1^{14} t_2^8 (-1 + 2 t_2^3) + t_1 (t_2 + 2 t_2^4)
 + t_1^2(7 t_2^2 - 4 t_2^5) +
        t_1^3(7 t_2^3 - 10 t_2^6) \nonumber\\
& +&  3 t_1^{13} t_2^7(-1 + 2
t_2^6)
+
 t_1^{12} t_2^6(9 - 34 t_2^3 + 22 t_2^6)
  +  t_1^7 t_2^4(-22 + 35 t_2^3 + 8 t_2^6 - 3 t_2^9)
\nonumber\\
& -& t_1^8
    t_2^5(4 - 35 t_2^3 + 10 t_2^6 + 3 t_2^9)
+ t_1^6
          t_2^3(-10 - t_2^3 + 4 t_2^6 + 9 t_2^9) + t_1^{10}(6
        t_2^4 + 8 t_2^7 - 26 t_2^{10})
\nonumber\\
 &+&
        t_1^9 t_2^6(4 + 31 t_2^3 - 34 t_2^6 +
        t_2^9)
+ 2 t_1^{11} t_2^5(3 - 5 t_2^3 -
    7 t_2^6 + 3 t_2^9)
\nonumber\\ &+& 2
        t_1^4(t_2 + t_2^4 - 11 t_2^7 + 3 t_2^{10}) + 2 t_1^5 t_2^2(-2 - 5 t_2^3 - 2
          t_2^6 + 3 t_2^9)
\end{eqnarray}

\section{Expanding the $N=1$ generating function}
\label{decomposition}

Our decomposition of the ring of invariants of the gauge theory is a decomposition into
different types of {\it determinants}. For simplicity, we will use the following notation: given
a pair of gauge groups $(\alpha,\beta),\ \alpha,\beta=1,...,G$, we call {\bf determinant of type $(\alpha,\beta)$} a gauge
invariant of the form
$$\epsilon^{i_1,...,i_N} ({\bf X}_{I_1}^{(\alpha,\beta)})_{i_1}^{j_1} .... ({\bf X}_{I_N}^{(\alpha,\beta)})_{i_N}^{j_N}\epsilon_{j_1,...,j_N}$$
where $({\bf X}_I^{(\alpha,\beta)})_{i}^{j}$ denotes a string of elementary fields with all gauge indices
contracted except two indices, $i$ and $j$, corresponding to the gauge groups $(\alpha,\beta)$. The index $I$ runs over all possible strings of elementary fields
with these properties.
The full
set of invariants is obtained by arbitrary products of these determinants. Using the tensor
relation
$$\epsilon^{i_1,...,i_N}\epsilon_{j_1,...,j_N} = \delta^{i_1}_{[j_1}\cdots \delta^{i_N}_{j_N]} $$
some of these products of determinants are equivalent and some of these are actually equivalent
to mesonic operators made only with traces. In particular, mesons are included in the above
description as determinants of type $(\alpha,\alpha)$.

We can decompose the ring of invariants according to the baryonic charges,
which indeed distinguish between different types of determinants, or baryons.
This decomposition is natural in field theory and it also has a simple interpretation in the dual gravity theory.

In fact, in theories obtained from D3-branes at CY singularities, baryons can be identified with
branes wrapped on nontrivial cycles on the base $H$ of the CY. The non-anomalous symmetries can
be clearly identified in the dual theory. In particular, states with the same non-anomalous
baryonic charges can be continuously deformed into each other: we can thus relate the set of
non-anomalous baryonic charges to the group of three-cycles in $H$. The homology
$H^3(H,\mathbb{Z})=\mathbb{Z}^{d-3}\times \Gamma$ determines $d-3$ continuous baryonic charges
($d$ is the number of vertices of the toric diagram) and possibly a set of discrete baryonic
charges from the torsion part $\Gamma$.


In the case of the conifold there is one non-anomalous baryonic charge (since $d=4$) which is
related to the single three-cycle in the base $T^{1,1}$. There are only two gauge groups and two
types of determinants: $(1,2)$ and $(2,1)$. The invariants decompose according to the baryonic
charge:

\begin{enumerate}
\item $B=0$ corresponds to the mesons (D3-branes wrapping trivial cycles, a.~k.~a. giant
gravitons \cite{McGreevy:2000cw}),
\item $B>0$ corresponds to the sector containing the determinants $(\det {\bf
A})^B$ and all possible mesonic excitations (D3-branes wrapping $B$ times the 3-cycle),
\item finally, $B<0$ corresponds to the sector containing the determinants $(\det {\bf B})^{|B|}$
and all possible mesonic excitations (D3-branes wrapping $|B|$ times the 3-cycle with opposite
orientation).
\end{enumerate}

The conifold picture is nice and in many ways elegant. However, a simple look at any other
quiver gauge theory reveals that this simple picture is too naive. 
Consider, for~example, the case of the orbifold $\mathbb{C}^3/\mathbb{Z}_3$ (see
\fref{dP0quiver}), that already reveals all types of oddities:

\begin{itemize}
\item{Since $d=3$, there is no continuous non-anomalous baryonic symmetry.
However, $H^3(S^5/\mathbb{Z}_3,\mathbb{Z})=\mathbb{Z}_3$ and there is a discrete baryonic
symmetry. We can indeed construct determinants, for example, using the fields ${\bf U},{\bf V}$
and ${\bf W}$ with $\mathbb{Z}_3$ charge $+1$. These do not carry any continuous conserved
charge since the product $\det {\bf U} \det {\bf V} \det {\bf W}$ can be rewritten as a meson in
terms of traces; for example, using $\epsilon^{i_1,...,i_N}\epsilon_{j_1,...,j_N} = \delta^{i_1}_{[j_1}\cdots \delta^{i_N}_{j_N]} $ we can write,
$$ \det {\bf U}_1\det{\bf V}_1\det{\bf W}_1 =\det ( {\bf U}_1{\bf V}_1 {\bf W}_1) = \Tr( {\bf U}_1{\bf V}_1{\bf W}_1)^N +... \pm (\Tr {\bf U}_1{\bf V}_1{\bf W}_1 )^N $$
On the other hand, $(\det {\bf U}_1)^3$ cannot be reduced to traces simply because there are no
gauge invariant traces we can make with ${\bf U}_1$ alone:  we have actually an infinite number
of products of determinants ($(\det {\bf U})^n$ for $n=1,2,...$ for example) that cannot be
rewritten in terms of mesons. All these operators correspond in the ring  of invariants to
sectors that cannot be distinguished by the discrete baryonic charge.}
\item{The BPS D3-brane configurations wrap divisors in the CY: for
$\mathbb{C}^3/\mathbb{Z}_3$ we have just a single divisor $D$ satisfying $3 D=0$ and this agrees
with the homology of the base $S^5/\mathbb{Z}_3$. However, we also have a vanishing compact
four-cycle which is represented in toric geometry by the integer internal point of the toric
diagram. The size of this cycle becomes finite when we blow up the orbifold. It is conceivable
that the inclusion of compact four-cycles such as this one will affect the description of the classical
configuration space of D3-branes. This will add a new parameter, related to the group of
divisors on the CY resolution, which has dimension one.}
\item{We could distinguish among elementary fields and types of determinants
by using all the possible baryonic charges, including the anomalous ones.
For $\mathbb{C}^3/\mathbb{Z}_3$ this would lead to the inclusion of the
two existing anomalous baryonic charges. As we will explain,
this set of charges is different and complementary with respect to that
related to the group of divisors on the resolution. }
\end{itemize}

Having this example in mind, we now discuss the two possible expansions
of the ring of invariants. Explicit examples of the decompositions will be
presented in Section \ref{examples}. We encourage the reader to jump forth
and back with  Section \ref{examples} for a better understanding of the
material.

\subsection{Expanding in a complete set of baryonic charges}\label{full}

The most general decomposition of the $g_1(\{t_i\})$ generating function is according to the
full set of baryonic charges, including the anomalous ones, denoted by $a_i$. The sectors ${\cal
S}$ in this case correspond to sectors with definite anomalous and non-anomalous baryonic
charges.

There are $G-1$ independent baryonic charges, where $G$ is the number of gauge
groups. By gauging the baryonic symmetries, we would obtain a quiver theory
with the same fields and superpotential, and a gauge group
$$\prod_{i=1}^G U(N)\, /\, U(1) ,$$
where we factor out the overall decoupled $U(1)$. Some of the $U(1)$ factors will be anomalous,
of course. The baryonic charges have a very natural description: they  correspond to the $U(1)$
factors in $U(N)^G/U(1)$. In this way, different elementary fields have the same baryonic
charges if and only if they are charged under the same gauge groups. This allows to efficiently
distinguish between invariants belonging to different sectors. Notice that non-anomalous
baryonic symmetries alone would not distinguish all inequivalent possibilities. For example, in
$\mathbb{C}^3/\mathbb{Z}_3$, the mesonic operator $\det {\bf U}_1\det {\bf V}_1\det {\bf W}_1$ and the
determinant $(\det {\bf U}_1)^3$ have the same charge under $\mathbb{Z}_3$, but different charges
under the two anomalous baryonic symmetries.

Let us thus extend the set of chemical potentials $t_i$ to all the baryonic charges, including
the anomalous ones.  We can therefore decompose ${\cal R}_{N=1}^{inv}$ into sectors with definite
charges under $U(1)^{G-1}$.

The $N=1$ generating function $g_{1}(\{t_i\})$ will decompose into a formal Laurent series in
the baryonic chemical potentials $b_i$ and $a_i$. The explicit decomposition of $g_1$ into a
formal Laurent series can be done by repeatedly applying the residue theorem; the computation
however quickly becomes involved, since the order of integration becomes crucial and divides the
result into many different cases. The ring of invariants ${\cal R}_{N}^{inv}$, will similarly decompose
into sectors of definite baryonic charges. The generating function $g_N(\{t_i\})$ can then be
computed according to Equation (\ref{g1plet}).

We can understand this decomposition in terms of representation theory.
From this perspective, we have decomposed the ring of $SU(N)^G$ invariants into Abelian
representations of the extended group $U(N)^G/U(1)$. This is sometimes called an expansion in
{\it covariants} and is extremely natural from the point of view of invariant theory. From our
point of view, covariants are just the possible set of independent {\it determinants}.   Each
sector $\cal S$ in ${\cal R}_N^{inv}$ will be specified by a certain number of gauge group pairs $(\alpha_i,\beta_i)$
and is associated to the subsector of the ring of gauge invariants made with products of the
{\it determinants} of type $(\alpha_i,\beta_i)$.

To make connection with the toric quiver gauge theories we notice that we
have the relation \cite{Feng:2002zw}:
\begin{equation}
G-1=2I+(d-3)
\end{equation}
where $d$ in the number of vertices of the toric diagram and $I$ the number of
integer internal points.
Only $d-3$ of these baryonic symmetries are not anomalous.

We expect all sectors ${\cal S}$
to appear with multiplicity one in the decomposition of
Equation (\ref{g1plet}).

\subsection{Supersymmetric D3-branes and the GKZ decomposition}\label{D3sup}
The full set of BPS states of the dual gauge theory can be obtained by quantizing the classical
configuration space of supersymmetric D3-branes wrapped on the CY horizon
\cite{Mikhailov:2000ya,Beasley:2002xv,Biswas:2006tj}. The supersymmetric
D3-brane configurations are in one-to-one correspondence with holomorphic four-cycles in the CY,
or divisors \cite{Mikhailov:2000ya}.
This is clear for static D3-branes wrapping a three-cycle in the horizon: the
corresponding divisor is the cone over the three-cycle. For more general configurations of
excited and rotating D3-branes we obtain a four-cycle by a Euclidean continuation: we can
replace time with the radial coordinate using the isometries of (Euclidean) $AdS_5$. Our problem
can be equivalently rephrased in terms of holomorphic surfaces in the CY manifold with $g_1$ as a
generating function.

From this perspective, we have an obvious decomposition into sectors ${\cal S}$ corresponding to
Euclidean D3-branes that can be continuously deformed into each other in the CY. Such D3-branes
have the same non-anomalous baryonic numbers; indeed, geometrically, the non-anomalous baryonic
charges are identified with the group of divisors modulo linear equivalence. Let us discuss this
point in detail for toric varieties since it will be crucial in the following.

Recall that our conical CY is specified by a toric diagram in the plane with $d$ vertices having
integer coordinates $n_i$. By embedding the plane in three dimensions we have a toric cone with
edges $V_i=(n_i,1)\in \mathbb{Z}^3$ (the toric fan of our conical CY). Call the set of edges
$\Sigma (1) $. Assign a ``homogeneous coordinate'' $x_i$ to each $V_i \in \Sigma (1) $; the $x_i
$ span $\mathbb{C}^d$. Consider the group
\begin{equation}
K=\left\{(\mu_1,...,\mu_d)\in (\mathbb{C}^*)^d \left | \prod_{i=1}^d \mu_i^{\langle m,V_i\rangle } =1, m\in \mathbb{Z}^3 \right. \right\} ,
\label{Kgroup}\end{equation}
which acts on $x_i$ as
$$(x_1,...,x_d) \rightarrow (\mu_1 x_1,...,\mu_d x_d)\, .$$
 $K$ is isomorphic, in general, to $(\mathbb{C}^*)^{d-3}$ times a discrete group.
 Then, the conical CY is defined by the symplectic quotient:
$$CY = (\mathbb{C}^d \ \backslash \ \Delta)/ K$$
where $\Delta$ is a subset fixed by the action of $K$. Geometrically, the $x_i$ can be
interpreted as homogeneous coordinates on the CY, just like the familiar coordinates for
projective spaces. The residual $(\mathbb{C}^*)^{3}$ complex torus action acting on the CY is
dual to the flavor symmetry group in the gauge theory, while the group $K$ is dual to the
non-anomalous baryonic symmetry group. Notice that the flavor and baryonic symmetries nicely
combine in the full group of $d$ non-anomalous charges which act naturally on the $d$
homogeneous coordinates $x_i$ as $(\mathbb{C}^*)^d$ acts on $\mathbb{C}^d$. In the tiling
construction, the $x_i$ are used to assign the non-anomalous charges to each field in the
quiver \cite{Franco:2005sm,Benvenuti:2005cz,Butti:2005vn}.

On the other hand, in toric geometry each edge $V_i$ determines a (not necessarily compact)
four-cycle $D_i$ in the CY. A generic four-cycle is given by a linear combination of basic
divisors $\sum_{i=1}^d a_i D_i$ where $a_i$ are integer coefficients. These divisors are subject to precisely three linear equivalence conditions given by

$$ \sum_{i=1}^d \langle e_k, V_i\rangle  D_i = 0 \, \qquad\qquad k=1,2,3 $$

where $e_k$ is a basis for $\mathbb{Z}^3$. It is then easy to see that the group of four-cycles
modulo linear equivalence is isomorphic to the baryonic group $K$. It follows that the
non-anomalous baryonic symmetry distinguishes deformation equivalence classes of Euclidean
D3-branes.

However this is not the end of the story. The decomposition into non-anomalous baryonic charges
is not fine enough. A D3-brane state with baryonic charge $B$ can form a sort of bound state which
distinguishes it from a set of $B$ D3-brane states with baryonic charge one. This typically
happens in theories where there are elementary fields with multiple non-anomalous baryonic
charges. By going over examples, it easy to convince
oneself that the classical
D3 brane configurations obtained from divisors on the singular CY do not
exhaust all possible sectors of the dual gauge theory. However,
as already mentioned, we have a plethora of compact vanishing four-cycles that are
expected to enter in the description of the set of supersymmetric D3-branes and solve these
ambiguities. We have exactly $I$ compact vanishing four-cycles, one for each integer internal
point in the toric diagram. These cycles become of finite size in the smooth resolutions of the CY. We will see that with the addition of these divisors
we can give a convenient description of all sectors in the dual gauge theory.
It would be interesting to understand the necessity for the inclusions of
these divisors directly from the point of view of the geometric quantization
of classical supersymmetric branes living on the horizon.

We are led to enlarge the set of basic divisors of size $d-3$ to a larger set of size
$d-3+I$ by adding a divisor $D_i$ for each internal point of the
toric diagram. We now have a larger group of divisors which strictly
contains the baryonic symmetry group. The larger set of divisors
immediately leads us to the description of the K\"ahler moduli space of the CY, of dimension $d-3+I$.
This moduli space is still a toric variety described by the so-called secondary fan,
or GKZ fan, and it is indeed parameterized by the divisors $D_i$ in the larger set.

\subsubsection{The GKZ decomposition}\label{dec}

It is well known that there are many different smooth resolutions of the CY corresponding to the
possible complete triangulations of the toric diagram. Different resolutions are connected by
flops. The number of K\"ahler moduli of the CY is $I+d-3$ where $I$ is the number of internal
points; this is the same as the number of geometrical FI terms that appear in the symplectic
quotient description of the resolved manifold. This number can be greater than the number of
non-anomalous baryonic symmetries in field theory, which is $d-3$.

There is an efficient description of the K\"ahler moduli space in terms of divisors \cite{Cox:2000vi}. Take a
complete resolution of the variety and consider the set of all effective divisors

\begin{equation}
\left\{ \sum_{i=1}^{d+I} a_i D_i, \ \mbox{such that} \ a_i\ge 0 \hbox{ , }\forall i \right\}
\end{equation}

modulo the three linear equivalence conditions given by $\sum_{i=1}^{d+I} \langle e_k,V_i\rangle  D_i=0$ where $e_k$ is the standard basis for $\mathbb{Z}^3$ and $V_i$ are the vertices of the toric
diagram, including the internal integer points. The $a_i\in \mathbb{R}^+$ give a parametrization of the $d+I-3$ dimensional K\"ahler
moduli provided we impose a further condition: to have all cycles of positive volumes we must
consider only {\bf convex} divisors. The convexity conditions can be expressed as follows.
Assign a number $a_i$ to the $i$-th point in the toric diagram. To each triangle $\sigma$ in
the triangulation of the toric diagram we assign a vector $m_\sigma\in \mathbb{Z}^3$ which is the integer solution of the system of three linear equations\footnote{ Actually, these equations can be solved for all simplicial resolutions, corresponding to not necessarily maximal triangulations of the toric diagram. If we allow triangles with area greater than one, we have resolved varieties which still have orbifold singularities. For completely smooth resolutions, the vertices of all triangles $\sigma$ are
primitive vectors in $\mathbb{Z}^3$ and Equation (\ref{weights}) has integer solutions.},

\begin{equation}
\langle  m_\sigma, V_i \rangle  = -a_i,\,\,\,   i\in \sigma
\label{weights}
\end{equation}

 and impose the inequalities

\begin{equation}
\langle  m_\sigma, V_i \rangle \,  \ge \,  -a_i,\,\,\,   i\not\in \sigma
\label{convexity}
\end{equation}

The set of inequalities (\ref{convexity}), as $\sigma$ runs over all the triangles, determines
the convexity condition for the divisor. For a given resolution, the set of convex divisors forms a cone in the $\mathbb{R}^{d+I-3}$ vector space, that parameterizes the K\"ahler moduli of the
resolution. The boundary of this cone corresponds to the vanishing of some cycle. If
we can perform a flop, we enter a new region in the moduli space corresponding to a different
resolution. Indeed, the cones constructed via the convexity condition for the various
possible resolutions of the toric diagram form regions in the $\mathbb{R}^{d+I-3}$ vector space
that are adjacent; altogether these reconstruct a collection of adjacent cones
(a fan in toric language) in $\mathbb{R}^{d+I-3}$. The toric variety
constructed from this fan in $\mathbb{R}^{d+I-3}$ is the
K\"ahler moduli space of the CY.
This is known as the GKZ fan, or secondary fan \cite{OdaTadao:19910900,Gelf}.
We move from a cone in the
GKZ fan to another by performing flops (or in case we also consider orbifold resolutions
by flops or further subdivisions of the toric diagram). It is
 sufficient for us to consider smooth varieties and we thus reserve
the name {\bf GKZ fan} to the collections of cones corresponding to smooth
resolutions.

The GKZ fans 
for the conifold, $\mathbb{F}_0$ and $dP_1$ are given in
Figures \ref{GKZconifold},
\ref{FF} and \ref{Y21a}, respectively.

We form a lattice by considering the integer points in the GKZ fan.
We claim that the $N=1$ generating function has an expansion in sectors
corresponding to the integer points of the GKZ lattice. Denote by $P$ an integer point in the GKZ lattice, then

\begin{equation}
g_1(\{t_i\})= \sum_{P\in GKZ} m(P) g_{1,P}(\{t_i\}) , \label{GKZex}
\end{equation}

where $m(P)$ is the multiplicity of the point $P$. Furthermore, we conjecture that the finite $N$ generating function can be obtained as

\begin{equation}
\sum_N g_N(\{ t_i\}) \nu^N =\sum_{P\in GKZ} m(P) {\rm PE}_\nu [g_{1,P}(\{t_i\})]
\end{equation}

\subsubsection{GKZ and field theory content}
\label{auxiliary}

At the heart of the previous formulae, there is a remarkable connection between the GKZ
decomposition and the quiver gauge theory. To fully appreciate it we suggest to the reader to
read this and the following subsections in close parallel with Section 5 where explicit examples
are given.

The integer points in the GKZ fan correspond to
sectors in the quantum field theory Hilbert space made out of determinants. Recall that mesons in
the quiver gauge theories correspond to closed paths in the quiver. We want to associate
similarly the other independent sectors made out of determinants with equivalence classes of open
paths in the dimers. The open paths fall into equivalence classes $A$
specified by the choice of ending points on the dimer.
The open path in a given class can be reinterpreted in the gauge
theory as strings of elementary fields with all gauge indices contracted
except two corresponding to a choice of a specific pair of gauge groups;
let us call these {\it composite fields}.
Baryonic operators are written as  ``$\det
{ A}$'', which is a schematic expression for two epsilons contracted with $N$ composite fields
freely chosen among the representatives of the class $A$. Generic sectors are made with
arbitrary products $\det A \det B ....$etc. Whenever open paths $A$ and $B$ can be composed to
give the open path $C$, there is at least one choice of representatives for $A$ and $B$ such that
we can write $\det A \det B=\det C$ and we want to consider the two sectors $\det A\det B$ and
$\det C$ equivalent. This can be enforced as follows. Denote with letters $a,b,c...$ the
equivalence classes of arrows in the quiver connecting different gauge groups. By decomposing
open paths in strings of letters, we can associate a sector with a string of letters. We should
however take into account the fact that if, for example, the arrows $a,b,c$ make a closed loop,
the operator $$(\det a)( \det b)( \det c)=\det abc$$ is a meson. We take into account this fact
by imposing the constraint $abc=0$. Moreover, composite fields connecting the same pairs of
gauge groups as an elementary fields do not determine the existence of new independent
determinants; to avoid overcountings, the corresponding string of letters should be set to zero.
Analogously, whenever two different strings of letters correspond to open paths
with the same endpoints, these strings should be identified.
We call ${\cal I}$ the set of
constraints obtained in this way and construct the ring
$${\cal R}_{GKZ}=\mathbb{C}[a,b,c...]/{\cal I}$$

Quite remarkably, the monomials in the ring ${\cal R}_{GKZ}$ are in correspondence with the integer
points in the GKZ fan. More precisely, we can grade the ring with $d-3+I$ charges in such a way
that the generating function of ${\cal R}_{GKZ}$, which we call the {\bf auxiliary GKZ partition
function}, and is denoted by $Z_{\rm aux}$, counts the integer points in the GKZ fan. Moreover, any integer point $P$ comes with
a {\bf multiplicity} $m(P)$ which is the one in Equation (\ref{GKZex}).

We have explicitly verified the above statement in all the examples we studied and we conjecture
that it is a general result for all toric diagrams: the auxiliary partition function counting
open paths in the quiver modulo equivalence coincides with the generating function for the GKZ
lattice dressed with field theory multiplicities. Just another remarkable connection between
apparently different objects: combinatorics on the tiling, geometry of the CY and gauge theory!


As another example of this fascinating correspondence, we will see that it is
possible to eliminate multiplicities
by refining the GKZ lattice. This is
done by enlarging the set of charges. In particular, we have found that,
if we refine the GKZ lattice by adding the anomalous baryonic charges,
we obtain a ``hollow cone'' in $\mathbb{Z}^{d-3+3 I}$ with no multiplicities.

As previously explained in a generic toric quiver gauge theory there are
 $2I$ anomalous baryonic symmetries, where $I$ is the number of internal
integer points. The variables in the auxiliary GKZ ring ${\cal R}_{GKZ}$ corresponds to arrows
in the quiver and therefore can be assigned a definite charge under the baryonic symmetries that
are just the ungauged $U(1)$ gauge group factors. We can therefore grade the auxiliary ring
${\cal R}_{GKZ}$ with a set of $d-3+3 I$ weights: the original $d-3 +I$ discretized K\"ahler
parameters $(\beta_1,...,\beta_{d-3+I})$ of the GKZ lattice plus the $2 I$ anomalous baryonic
weights $a_i$. The power series expansion of the auxiliary partition function
$Z_{aux}(\beta_i,a_i)$ will draw a lattice in $\mathbb{Z}^{d-3+3 I}$ which has the shape of an
hollow cone over the GKZ fan: over each point of the GKZ fan there is a hollow polygon $C(P)$
whose shape is related to the pq-web of the toric geometry. Quite remarkably, all points in
the lattice come with multiplicity one. Examples for $\mathbb{C}^3/Z_3,\mathbb{F}0$ and $dP_1$
are presented in Figure \ref{emptyC3}, \ref{emptyF0}, \ref{emptydP1_1} and \ref{emptydP1_2}.


\subsubsection{Computing $g_{1,P}$ for one D brane in a sector $P$ using localization}\label{localization}
We want to demonstrate now how it is possible  to compute the partition functions $g_{1,P}$
using the equivariant index theorem.

Every integer point $P$ in the GKZ fan is associated with a smooth resolution of the CY and a
particular divisor $\sum a_i D_i$ on it.
Extending our interpretation of the BPS states in terms of
supersymmetric D3-branes wrapping holomorphic cycles in the singular CY to its resolution, we have a natural definition for the function $g_{1,P}$:
it  should count all
the sections of the line bundle ${\cal O} ( \sum a_i D_i)$ corresponding to holomorphic surfaces
in the given linear equivalence class. Therefore $g_{1,P}$ is just the character
$${\rm Tr} \left\{ H^0\left(CY,{\cal O}\left(\sum a_i D_i\right)\right) \bigr| q\right\}$$
under the action of the element $q\in T^3$ of the torus of
flavor symmetries. All elements in
 $ H^0(CY,{\cal O}(\sum a_i D_i))$ have the same baryonic charges.
It is important to notice that the higher cohomology groups
of a convex line bundle vanish \cite{Fulton}:
the character then coincides with the
holomorphic Lefschetz number and can be computed with the equivariant index
theorem.

The way of doing this computation is explained in detail in \cite{Butti:2006au}, generalizing the analogous computation for holomorphic functions given in
\cite{Martelli:2006yb},
 and expresses
the result as a sum over the fixed points $P_I$ of the torus $T^3$ action on
the particular smooth resolution of the CY corresponding to the point $P$
in the GKZ lattice. It is known indeed that the torus action has only isolated
fixed points on the resolved CY. The character receives contributions only
from the fixed points and reads
\begin{equation}
g_{1,P}(\{t_i\}; CY) = t^{n_P}\sum_{P_I} \frac{q^{m^{(I)}_P}}{\prod_{i=1}^3 (1-q^{m^{(I)}_i})} ,
\label{loc}
\end{equation}
where the index $I$ denotes the set of isolated fixed points
and the four vectors
$m^{(I)}_i,\, i=1,2,3$, $m^{(I)}_P$ in $\mathbb{Z}^3$ are the weights of the linearized action of $T^3$ on the resolved CY and the fiber of the line bundle
${\cal O} (\sum a_i D_i)$, respectively.

The fixed points of the
torus action are in correspondence
with the triangles in the subdivision of the toric diagram (or, equivalently,
 with the vertices of the pq-web).
The vectors  $m^{(I)}_i,\, i=1,2,3$ in the denominator of
Equation (\ref{loc}) are computed as the three primitive inward normal vectors
of the cone $\sigma_I$ in $\mathbb{Z}^3$ made with the three vertices $V_{i}$ of
the I-th triangle. The vector $m^{(I)}_0$ in the numerator is instead computed as in Equation (\ref{weights})
\begin{equation}
\langle  m^{(I)}_0, V_i \rangle  = -a_i,\,\,\,   i\in \sigma_I
\label{weights2}
\end{equation}
Finally, the prefactor $t^{n_P}$ in Equation (\ref{loc}) is just the charge
of the divisor $\sum a_i D_i$. The full dependence on baryonic charges
in encoded in this prefactor.

In explicit computations, some care should be paid to the choice of charges. There is a natural
geometric basis for the non-anomalous charges of the gauge theory. In fact, the homogeneous
coordinates $x_i$ that are used to define the CY as a symplectic quotient (see Section
\ref{dec}) are extremely useful to assign a full set of $d$ (flavor+baryonic) charges to the
elementary fields in the quiver; this is done using zig-zag paths and standard dimers techniques
\cite{Butti:2005vn,Butti:2005ps}. All the elementary fields have charge which is given by a
product of the $x_i$. We can also assign charge $x_i$ to the divisor $D_i$ of the singular cone.
In all the examples we have considered there is a natural way to assign charges to the enlarged
set of divisors entering the GKZ decomposition. This allows to compute the prefactor $t^{n_P}$.
The $x_i$ decompose into three flavor charges and $d-3$ baryonic charges. The splitting of the
charges $x_i$ into flavor and baryonic charges is not unique in general; flavor charges can be
always redefined by adding a linear combination of the baryonic charges.
However, a
 toric diagram comes with a specific basis for the flavor $T^3$ action
which is determined by the equation
\begin{equation}
q_k=\prod_{i=1}^d x_i^{\langle e_k,V_i\rangle}\,\qquad\qquad k=1,2,3,
\label{rel}
\end{equation}
where $e_k$ are the basis vectors of $\mathbb{Z}^3$ and $V_i$ the vertices of
the toric diagram. Notice that all dependence on baryonic charges drops from
the right hand side by Equation (\ref{Kgroup}). This is the $T^3$ basis
that should be used in the localization formula (\ref{loc}).

\subsubsection{Checks with all charges: GKZ approach vs. field theory}\label{checks}


Having understood how to compute and resolve the multiplicities in the GKZ cones and to compute the partition functions we can also refine our decomposition
of the $N=1$ generating function by adding the anomalous baryonic charges.

Using the equivariant index theorem we compute all the generating functions $g_{1,P}$ for all
points $P$ in the GKZ lattice. We can use $d-3+I$ coordinates for the GKZ cone
$\beta=(\beta_1,...,\beta_{d-3+I})$. Denote also with $B(\beta)$ the non-anomalous baryonic
charge corresponding to the point $P$ of the GKZ lattice. As discussed in Section
\ref{localization}, the generating functions depend on the baryonic charges only by a
multiplicative factor: $g_{1,P}=b^{B(\beta)} g_{1,\beta}(q)$ and all the other dependence is on
the flavor charges $q_i$. Thanks to the auxiliary generating functions we were able to find
expressions for the multiplicities $m(\beta)$ of the fields over each point of the GKZ
lattice. These fuctions sum up to the complete generating functions with $N=1$ and with all the
non-anomalous charges:
\begin{equation}
g_1(q,b)=\sum_{P \hbox{ }\in \hbox{ }GKZ} m(P) \hbox{ }g_{1,\beta}(q) \hbox{ }b^{B(\beta)}
\end{equation}
where $b$ are the chemical potentials for the non-anomalous baryonic charges.
To resolve the multiplicities $m(\beta)$ we construct the hollow cone by
adding the anomalous baryonic charges. Over each point in the GKZ lattice there
is a hollow polygon $C(\beta)$ which can be parametrized
in terms of the set of $2I$ anomalous charges $a_j$
with $j=1,...,2I$, such that:

\begin{equation}
\sum_{K_j \hbox{ } \in \hbox{ } C(\beta)} a_1^{K_1}...\hbox{  }a_{2I}^{K_{2I}} \Big|_{(a_1=1,\hbox{  }...,\hbox{ }a_{2I}=1)} = m(P)
\end{equation}

Using these resolutions we obtain the resolved generating functions for $N=1$ with all the charges, anomalous and non-anomalous:

\begin{equation}\label{allgkz}
g_1(q,b,a)=\sum_{\beta \hbox{ }\in \hbox{ }GKZ}\sum_{K_j \hbox{ }\in\hbox{ } C(\beta)} a_1^{K_1}...\hbox{ }a_{2I}^{K_{2I}} \hbox{   } b^{B(\beta)}\hbox{  } g_{1,\beta}(q)
\end{equation}

The non flavor charges do not appear in the basic generating functions $g_{1,\beta}(q)$, but
they are multiplicative factors over which one has to sum up, in the same way one does for the
usual non-anomalous baryonic charges.

We would like to stress that Equation (\ref{allgkz}) points to a remarkable connection between the
geometry of the CY, which is used to compute the right hand side, and the
field theory, that can be used to determine the left hand side (the
$N=1$ generating function).
In other words, we have two different ways of computing the $N=1$ generating
$g_1(q,b,a)$ function which nicely match:

\begin{itemize}
\item{In the first case we use the GKZ geometric picture explained in the previous sections. We first compute the generating functions $g_{1,\beta}(q)$ for each point in the GKZ lattice which depend on the flavor charges.
We next sum over all the points of the hollow cone by dressing
$g_{1,\beta}(q)$ with the appropriate weight under the baryonic symmetries.}
\item{In the second case we use the field theory picture. We can take
the fields of the gauge theory as basic variables
and we assign to them all the possible charges anomalous and non-anomalous.
This means that we
construct the ring generated by the fundamental fields and we grade it with all the charges. Then we construct the quotient ring by modding out the  ring of elementary fields by the ideal
generated by $F$-term equations. Using Macaulay2 we compute the Hilbert series of the quotient
ring obtaining the completed resolved generating function with all the charges of the field
theory: $g_1(q,b,a)$.}
\end{itemize}
As we will explicitely demonstrate on the examples in Section \ref{examples},
the two computations completely agree.

As a final remark, we notice that Equation (\ref{allgkz}) is the most general decomposition of
the $N=1$ generating function that we can write. We can close our circular discussion and go
back to the initial point. Equation (\ref{allgkz}) has been obtained by enlarging the GKZ
lattice in order to eliminate multiplicities. The hollow cone is a lattice in $d-3+3I$
dimensions. The corresponding $d-3+3I$ charges contain, as a subset, all the anomalous and
non-anomalous baryonic charges that are in number $d-3+2I$. Notice that the terms in series in
Equation (\ref{allgkz}) depend on the extra $I$ GKZ parameters only through the factor
$b^{B(\beta)}$. By projecting the hollow cone on the  $d-3+2I$ dimensional space of baryonic
charges we obtain the explicit expansion of the $N=1$ generating function $g_1(q,b,a)$ in a
complete set of baryonic charges which was discussed in Section \ref{full}. One can compare this
expansion with the one obtained by performing repeated contour integrations. As one can check
explicitly, the points in the baryonic charge lattice have still multiplicity one.

\section{Examples}\label{examples}

In this section we explicitly compute the $N=1$ generating function for a certain number of
toric CY manifolds and decompose it. We start by revisiting the example of the conifold.

\subsection{The conifold revisited}

The conifold has only one baryonic charge, not anomalous, which can be
used to parametrize the K\"ahler moduli space. The two expansions,
one in baryonic charges, the other according to the GKZ lattice, coincide.

\paragraph{Baryonic charge expansion}
We first expand the $N=1$ generating function, Equation (\ref{g1coni}) for the
conifold according to the baryonic charge

\begin{equation}
g_1(t,b,x,y; {\cal C}) = \sum_{B=-\infty}^\infty b^B g_{1,B}(t,x,y; {\cal C}),
\end{equation}
$g_{1,B}(t,x,y; {\cal C})$ can be computed using the inversion formula

\begin{equation}
g_{1,B}(t,x,y; {\cal C}) = \frac{1}{2\pi i} \oint \frac {db} {b^{B+1}} g_1(t,b,x,y; {\cal C}) ,
\label{res1}
\end{equation}
with a careful evaluation of the contour integral for positive and negative values of the baryonic charge $B$.  For $B\ge0$ the contribution of the contour integral comes from the positive powers of the poles for $b$ ($b=x/t,1/(xt)$)
while for $B\le0$ the contribution of the contour integral comes from the negative powers of the poles for $b$ ($b=t y,t/y$)

\begin{eqnarray}
g_{1,B\ge0}(t,x,y; {\cal C}) &=& \frac{t^B x^{B} } { (1 - \frac{1}{x^2}) (1-t^2 x y)  (1-\frac{t^2 x}{y}) }+ \frac{t^B x^{-B}} { (1 - x^2) (1-\frac{t^2 y} {x})  (1-\frac{t^2}{x y}) } , \nonumber \\ \nonumber
g_{1,B\le0}(t,x,y; {\cal C}) &=& \frac{t^{-B} y^{-B} } { (1-\frac{1}{y^2}) (1-t^2 x y)  (1-\frac{t^2 y}{x}) }+ \frac{t^{-B} y^{B}} { (1 - y^2) (1-\frac{t^2 x} {y})  (1-\frac{t^2}{y x}) } .\\
\label{rescon}
\end{eqnarray}
By setting $x=y=1$ and $t_1=b t,t_2=t/b$ we recover expansion (\ref{g1conit}).

\subsubsection{Conifold -- GKZ decomposition}

\begin{figure}[h!!!!!]
\begin{center}
\includegraphics[scale=0.5]{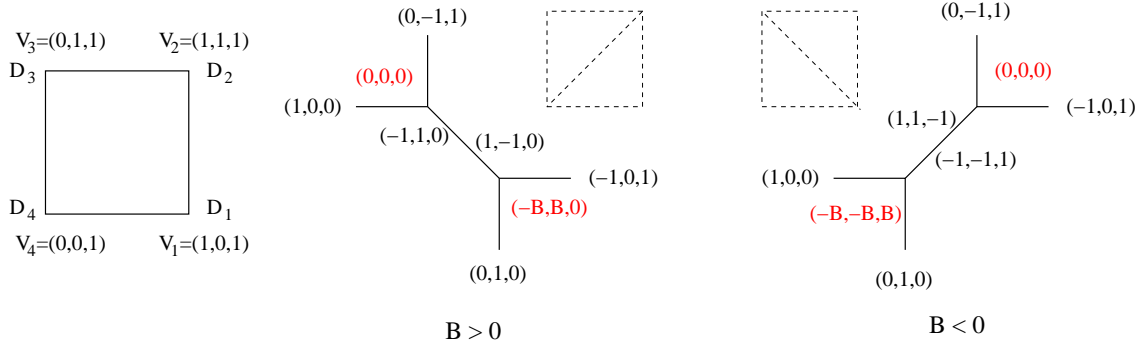}
\caption{Localization data for the $N=1$ baryonic generating functions. The vertices $V_i$ are in
correspondence with homogeneous coordinates $x_i$ and with a basis of  divisors $D_i$. Two
different resolutions, related by a flop, should be used for positive and negative $B$,
respectively. Each resolution has two fixed points, corresponding to the vertices of the
pq-webs; the weights $m^{(I)}_i,\, i=1,2,3$ and $m^{(I)}_B$ at the fixed points are indicated in
black and red, respectively.} \label{co2}
\end{center}
\end{figure}

We can similarly perform a GKZ decomposition of the $N=1$ generating function.
In Figure \ref{co2} the toric diagram and the two resolutions of the conifold
are reported. There are four divisors $D_i$ subject to three relations that
leave an independent divisor $D$,
$D_1=D_3=-D_2=-D_4\equiv D$.
Consider the cone of effective divisors $\sum a_i D_i, \, a_i\ge 0$
modulo linear equivalence in $\mathbb{R}$
$$\sum_{i=1}^4 a_i D_i \equiv (a_1+a_3-a_2-a_4) D \equiv B D$$
where we defined $B=a_1+a_3-a_2-a_4$. For each resolution, we solve Equation (\ref{weights}) for
the two triangles in the resolution, or, equivalently, the two vertices of the pq-web; the
resulting vectors $m_i^{(I)}$ and $m_B^{(I)}$ are reported in black and red respectively  in Figure \ref{co2}.

The convexity condition, Equation (\ref{convexity}), then tells us that the resolution on the
left corresponds to $B>0$ and the resolution on the right to $B<0$. Altogether we obtain two
half lines (cones) in $\mathbb{R}$ that form the GKZ fan as in Figure \ref{GKZconifold}. The
point of intersection $B=0$ of the two cones corresponds to the singular conifold and the two
cones in the fan are related by a flop transition.

We now compute the generating functions $g_{1,B}$ using localization. As mentioned in the
previous section, we must pay attention to the normalization of charges. The homogeneous
coordinates for the conifold are extremely simple:
$$({\bf A}_1\, ,\, {\bf B}_1\, ,\, {\bf A}_2\, ,\, {\bf B}_2\,) \quad\longrightarrow\quad (x_1\, ,\, x_2\, ,x_3\, ,\, x_4)$$
which can be easily translated in the notations of Section \ref{conif}.
The natural flavor $T^3$ basis is then given by Equation (\ref{rel})

\begin{eqnarray}
\nonumber
q_1&=&x_1 x_2 = t^2 x y,\\ \nonumber
q_2&=&x_2 x_3 = \frac{t^2 y}{x},\\
q_3&=&x_1 x_2 x_3 x_4 = t^4 .
\label{GLr}
\end{eqnarray}

We are ready to apply the localization formula. Each point in the GKZ fan is associated with a
resolution and a divisor: for $B>0$ we use the resolution on the left in Figure \ref{co2} and $B
D_1$, while for $B<0$ the resolution on the right and the divisor $|B| D_4$. The weights are
reported in Figure \ref{co2}. Equation (\ref{loc}) and Equation (\ref{GLr}) give

\begin{eqnarray}
g_{1,B\ge0}(t,x,y; {\cal C}) &=& \frac{t^B x^{B} } { (1 - \frac{1}{x^2}) (1-t^2 x y)  (1-\frac{t^2 x}{y}) }+ \frac{t^B x^{-B}} { (1 - x^2) (1-\frac{t^2 y} {x})  (1-\frac{t^2}{x y}) } , \nonumber \\ \nonumber
g_{1,B\le0}(t,x,y; {\cal C}) &=& \frac{t^{-B} y^{-B} } { (1-\frac{1}{y^2}) (1-t^2 x y)  (1-\frac{t^2 y}{x}) }+ \frac{t^{-B} y^{B}} { (1-y^2) (1-\frac{t^2 x} {y})  (1-\frac{t^2}{y x}) } .
\end{eqnarray}
which coincides with the result previously obtained in Equation (\ref{rescon}).

\subsubsection{Conifold -- multiplicities in the GKZ Decomposition}

The multiplicities in the GKZ decomposition of the $N=1$ generating function for the conifold are trivial since they are all equal to 1. Nevertheless it is instructive to follow the procedure which is outlined in Section \ref{auxiliary} in order to compute the multiplicities using the auxiliary GKZ partition function $Z_{\rm aux}(t)$ which counts independent sectors in the ring of invariants. As explained in Section \ref{auxiliary}, we assign a letter $a,b$ to the two types of arrows in the quiver ${\bf A}_i,{\bf B}_i$. There is only one relation $ab=0$ corresponding to the closed loop in the quiver. The polynomial ring for the GKZ decomposition of the conifold is therefore

\begin{equation}
\label{ZGKZconifold}
{\cal R}_{GKZ} ( {\cal C} ) = \mathbb{C}[a,b]/(ab),
\end{equation}

We thus compute the generating function for the polynomial ring (which can be easily computed by observing it is a complete intersection), with chemical potential $t_1$ to $a$ and $t_2$ to $b$ we find

\begin{equation}
\label{Zauxconifold}
Z_{\rm aux}(t_1, t_2; {\cal C}) = \frac{1-t_1 t_2}{(1-t_1) (1-t_2)} = 1+ \sum_{B = 1}^\infty t_1^B
+ \sum_{B = 1}^{\infty} t_2^{B}
\end{equation}

By expanding this auxiliary partition function we find multiplicity $1$ for the integer points $B>0$, multiplicity 1 for the integer points $B<0$, and multiplicity 1 for the point $B=0$, reproducing the lattice depicted in Figure \ref{GKZconifold}.
We finally have

\begin{equation}
g_1(t_1,t_2) = g_{1,0}(t_1,t_2) + \sum_{B=1}^\infty g_{1,B}(t_1,t_2) + \sum_{B=-\infty}^{-1} g_{1,B}(t_1,t_2) ,
\end{equation}
which appears to be a trivial observation as a Laurent series in the baryonic chemical potential
but in fact turns out to be nontrivial for more involved singularities.

\subsection{Generating functions for $\mathbb{C}^3/\mathbb{Z}_3$}
\label{N1C3Z3}

\begin{figure}[ht]
\begin{center}
  \epsfxsize = 10cm
  \centerline{\epsfbox{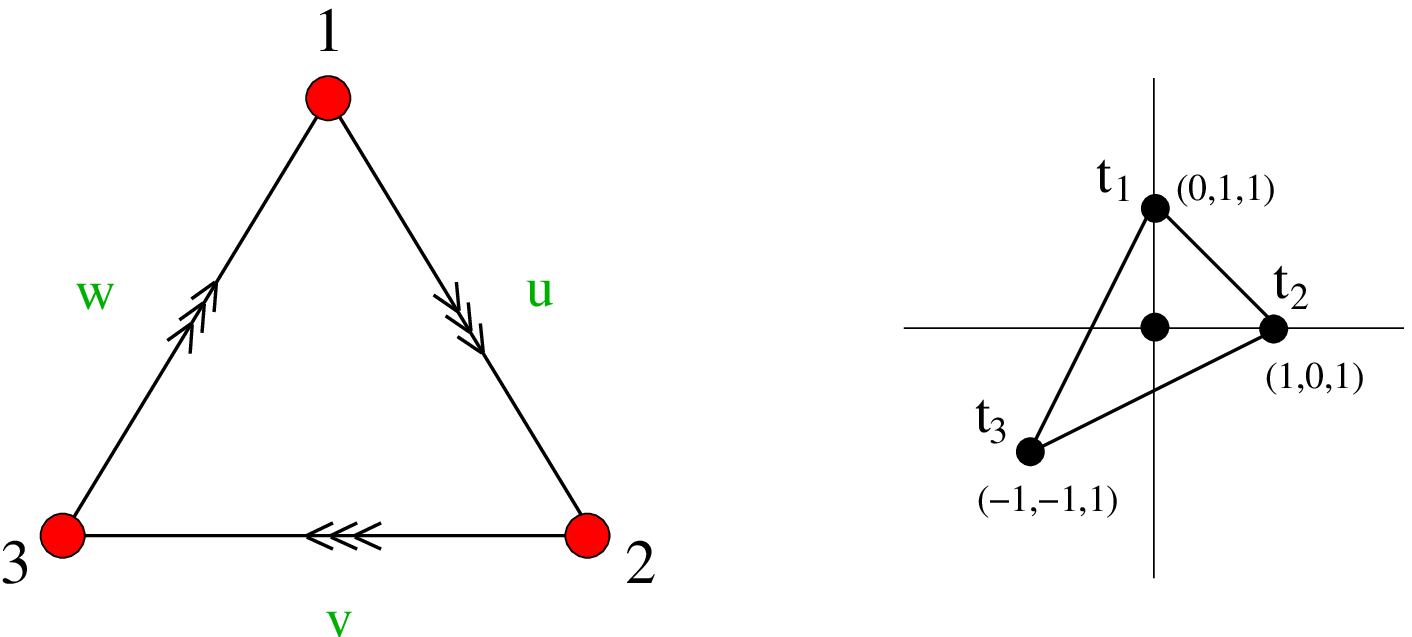}}
  \caption{Quiver and toric diagram for $\IC^3/\IZ_3$.}
  \label{dP0quiver}
\end{center}
\end{figure}

The quiver of the gauge theory for the $\mathbb{C}^3/\mathbb{Z}_3$ singularity is shown in \fref{dP0quiver}. The gauge theory has three sets of bifundamental fields ${\bf U}_i,{\bf V}_j,{\bf W}_k$ with $i,j,k=1,2,3$ and a superpotential $\epsilon_{ijk}{\bf U}_i{\bf V}_j{\bf W}_k$.

\paragraph{Symmetries and geometry}

The global flavor symmetry is $SU(3)\times U(1)_R$. All the fields have R-charge $2/3$ and each set of three fields transform in the fundamental representation of $SU(3)$. One can also define two anomalous baryonic $U(1)$ charges which can be chosen to be

\bean
&A_1:& ({\bf U,V,W}) \mapsto (a_1 {\bf U},\ a_1^{-1}{\bf V},\ {\bf W}) \\
&A_2:& ({\bf U,V,W}) \mapsto (a_2 {\bf U},\ a_2 {\bf V},\ a_2^{-2} {\bf W})
\eean

There are some non-anomalous discrete symmetries \cite{Gukov:1998kn} acting on the fields as follows,
\bean
&A:& ({\bf U,V,W}) \mapsto ({\bf W,U,V}) \\
&B:& ({\bf U,V,W}) \mapsto (b{\bf U},\ b^{-1}{\bf V},\ {\bf W}) \quad \mbox{where } b^3 = 1 \\
&C:& ({\bf U,V,W}) \mapsto (c{\bf U},\ c{\bf V},\ c^{-2}{\bf W}) = (c{\bf U},\ c{\bf V},\ c{\bf
W}) \quad \mbox{where } c^3 = 1
\eean
We see that $B$ is a subgroup of $A_1$ 
and $C$ is subgroup of $A_2$. $C$ is related to the torsion homology group for three cycles and, in a sense,
is a discrete baryonic charge.



We now have a look at the geometry of the CY using the
symplectic quotient construction outlined in Section \ref{D3sup}.
Since $d=3$ we introduce three homogeneous
coordinates $x_i$. The group $K$ of baryonic charges is defined by  Equation (\ref{Kgroup})
$$\prod_{j=1}^3 \mu_j^{\langle e_i,V_j\rangle }=1$$ which implies
$\mu_j=c$ with $c^3=1$.
The symplectic quotient description of the CY just reduces to the orbifold
description $\mathbb{C}[x_1,x_2,x_3]/\mathbb{Z}_3$, as expected.

As already discussed, the homogeneous coordinates can be used to give a full set of weights for
non-anomalous symmetries to the elementary fields. We can write  $x_i= c t_i$ in terms of the
discrete baryonic charge $c$ and the flavors $t_i$. The assignment of charges to the fields is
done using standard dimer techniques  and is reported in Table \ref{chargesC3Z3}. One can notice
that $t_i$ are the charges of the original ${\cal N}=4$ SYM. To keep track of the coordinates on
the two dimensional projection in Figure \ref{dP0quiver} we introduce three chemical potentials
$t, x, y$ which count the $R$-charge, and the $(x,y)$ integral positions, respectively and read
from the figure $t_1=t y, t_2 = t x, t_3 = \frac{t}{x y}$.

The full set of continuous charges, anomalous or not, is summarized in the Table \ref{chargesC3Z3}, as well as the chemical potentials and the assignment of homogeneous coordinates to the fields.

\begin{table}
\begin{center}
\begin{tabular}{|c|c|c|c|c||c||c|}
\hline
\ {\bf field} \ &   {\bf $SU(3)$}    & \ \  {\bf $R$} \ \  & \ \  {\bf $A_1$} \ \  & \ \  {\bf $A_2$}  \ \ &
\  chemical    \ & \  homogeneous \\
   & & & & & potentials & coordinates \\
\hline\hline $({\bf U}_1\, ,{\bf U}_2\, , {\bf U}_3) $  & ${\bf 3}$    & $\frac{2}{3}$  &   1 & $1$ & $(t a_1 a_2 y\, , t a_1 a_2 x\, ,t a_1 a_2 / x y)$  &
$(c t_1\, , c t_2\, , c t_3)$ \\
$({\bf V}_1\, ,{\bf V}_2\, , {\bf V}_3) $  & ${\bf 3}$    & $\frac{2}{3}$       & $-1$ & 1 & $(t a_2 y / a_1 \, , t a_2 x / a_1\, ,t a_2 / a_1 x y ) $  &
$(c t_1\, , c t_2\, , c t_3)$ \\
 $({\bf W}_1\, ,{\bf W}_2\, , {\bf W}_3) $ & ${\bf 3}$    & $\frac{2}{3}$ & 0 & $-2$  & $(t y / a_2^{2} \, , t x /a_2^{2} \, ,t / a_2^{2} x y)$  &
$(c t_1\, , c t_2\, , c t_3)$ \\
\hline
\end{tabular}
\end{center}
\caption{Global charges for the basic fields of the quiver gauge theory
living on the D-brane probing the orbifold $\IC^3 / \IZ_3$. The $x$ and $y$ chemical potentials count $SU(3)$ weights, while $A_1$ and $A_2$ count anomalous baryonic charges.}
\label{chargesC3Z3}
\end{table}

\subsubsection{The $N=1$ generating function}
The $N=1$ generating function is generated by the elementary fields ${\bf U}_i,{\bf V}_j,{\bf
W}_k$ modulo nine F-term relations which can be expressed through the ideal
\begin{eqnarray}
 {\cal I}= ({\bf V}_2 {\bf W}_3-{\bf V}_3 {\bf W}_2,\ {\bf V}_1 {\bf W}_3-{\bf V}_3 {\bf
W}_1, \ {\bf V}_1 {\bf W}_2-{\bf V}_2 {\bf W}_1, \nonumber \\
{\bf U}_2 {\bf W}_3-{\bf U}_3 {\bf W}_2, \ {\bf U}_1 {\bf W}_3-{\bf U}_3 {\bf W}_1,\ {\bf U}_1
{\bf W}_2-{\bf U}_2 {\bf W}_1,\nonumber \\ {\bf V}_2 {\bf U}_3-{\bf V}_3 {\bf U}_2,\ \ \ {\bf
V}_1 {\bf U}_3-{\bf V}_3 {\bf U}_1,\ \ \ {\bf V}_1 {\bf U}_2-{\bf V}_2 {\bf U}_1) . \ \nonumber
\end{eqnarray}

Each field carries R-charge $\frac{2}{3}$ and therefore we can give to all the same weight for the chemical potential, $t$.
Computing the Hilbert series of the polynomial ring

\begin{equation}
\label{ringC3Z3}
{\cal R}_{N=1} ( \IC^3/\IZ_3 ) = \mathbb{C}[\{ {\bf U}_i\},\{{\bf V}_j\},\{ {\bf W}_k\}]/{\cal I}
\end{equation}
with Macaulay2 we obtain

\begin{equation}
g_1(t; \IC^3/\IZ_3)=\frac{1+4 t+t^2}{(1-t)^5}\label{N1} .
\end{equation}

The dimension of an  irreducible algebraic variety $V$ can be computed from its Hilbert series
$g(t)$  by looking at the order of the pole for $t\rightarrow 1$
\begin{equation}
g_1(t) \sim   \frac{A}{(1-t)^{{\rm dim} V}},
\label{dim}
\end{equation}
with the residue $A$ a measure for the volume of this variety. The $N=1$ moduli space of vacua
for $\mathbb{C}^3/\mathbb{Z}_3$ has dimension five, as can be seen from
$$g_1(t)\sim \frac{6}{(1-t)^5}$$
This can be understood as three mesonic directions describing the CY plus two independent
baryonic parameters that correspond to the gauge theory FI terms. As usual, the mesonic moduli
space for $N=1$ is isomorphic to the CY geometry. The additional baryonic parameters come from
the fact that the gauge group is $SU(N)^G$ and not $U(N)^G/U(1)$. Since we do not have to impose
the $U(1)$ D-term conditions, this leaves $G-1$ additional free parameters that can be
identified with the FI terms in the gauge theory. Since in general the mesonic flat directions
are given by the symmetrized product of $N$ CY's for $N$ D-branes, giving $3N$ parameters, the
dimension of the moduli space for generic $N$ and $G$ is expected to be $3 N+G-1$.


\subsubsection{The GKZ decomposition}

In the singular CY there is only one independent divisor $D_i\equiv D$ with $3 D=0$. This reflects the
$\mathbb{Z}_3$ discrete charge. However, on the smooth resolution of the orbifold there is a new
divisor $D_4$ corresponding to the internal point. $D_1=D_2=D_3=D$ is still true but now $3 D$
is non-zero, but equal instead to $-D_4$. The cone of effective divisors in $\mathbb{R}$ is
given by
$$\sum_{i=1}^4 a_i D_i = (a_1+a_2+a_3-3 a_4) D \equiv \beta D\, , \qquad \qquad a_i\ge 0$$
and the convexity condition, Equation (\ref{convexity}), requires $\beta\ge 0$. The GKZ fan is
thus a half-line in $\mathbb{R}$. The integer parameter $\beta$ turns out to be the discrete K\"ahler modulus of the resolution of $\mathbb{C}^3/\mathbb{Z}_3$, measuring the discrete area of the two cycle.

The right basis for localization is given
by $$q_i=\prod_{j=1}^3 x_j^{\langle e_i,V_j \rangle}$$
and we compute (notice that the discrete baryonic charge $c$ correctly drops out from this formula)
 $q_1=\frac{t_2}{t_3}, q_2=\frac{t_1}{t_3}, q_3=t_1 t_2 t_3$. We thus obtain\footnote{
These partition functions reduce for $\beta=0,1,2$ to the three independent partition functions
for nontrivial divisors on the singular cone.}

\begin{eqnarray}
\label{g1bC3Z3}
g_{1,\beta} (t_1,t_2,t_3) &=& \frac{t_1^\beta}{(1-t_1^3)(1-\frac{t_2}{t_1})(1-t_3/t_1)} \\
&+&\frac{t_2^\beta}{(1-t_1/t_2)(1-t_2^3)(1-t_3/t_2)}+\frac{t_3^\beta}{(1-t_1/t_3)(1-t_2/t_3)(1-t_3^3)}.
\nonumber
\end{eqnarray}

\vskip 0.3cm $g_{1,0}$ is identified with the Molien invariant for the discrete group
$\mathbb{Z}_3$ and indeed computes the mesonic generating function as explained in detail in
\cite{Benvenuti:2006qr}. In the limit $t_i=t$ we find

\begin{eqnarray}
\label{g1geomZ3}
g_{1,\beta} (t,t,t) = t^\beta \left ( \frac{1+7t^3+t^6}{(1-t^3)^3} + \frac{3\beta(1+t^3)}{2(1-t^3)^2} + \frac{\beta^2}{2(1-t^3)} \right ) .
\end{eqnarray}

\subsubsection{Multiplicities}

The multiplicities in the GKZ decomposition of the $N=1$ generating function can be computed
using $Z_{\rm aux}(t)$, the auxiliary GKZ partition function counting independent sectors in the
ring of invariants. As explained in Section \ref{auxiliary}, we assign a letter $u,v,w$ to the
three types of arrows in the quiver ${\bf U}_i,{\bf V}_j,{\bf W}_k$. There is only one relation
$uvw=0$ corresponding to the closed loop in the quiver. We thus get the polynomial ring

\begin{equation}
\label{ZGKZC3Z3}
{\cal R}_{GKZ} ( \IC^3/\IZ_3 ) =\mathbb{C}[u,v,w]/(uvw),
\end{equation}
and compute the generating function (which can be easily computed by assuming it is a complete intersection), with charge $t$ to all letters obtaining

\begin{equation}
\label{ZauxC3Z3}
Z_{\rm aux}(t ; \IC^3/\IZ_3 ) = \frac{1-t^3}{(1-t)^3} = 1+ \sum_{\beta = 1}^\infty 3\beta t^\beta .
\end{equation}

By expanding this auxiliary partition function we find
multiplicity $3 \beta$ for the point $\beta>0$ and multiplicity 1 for the point $\beta=0$. This is easily understood:
the independent sectors contain determinants of the form
$(\det {\bf U})^n (\det {\bf V})^m$ with $n+m=\beta$ or similar with ${\bf U},{\bf V},{\bf W}$ permuted;
there are $3 \beta$ such sectors. This point will be further elaborated below.

We finally have

\begin{equation}
g_1(t_1,t_2,t_3) = g_{1,0}(t_1,t_2,t_3) + \sum_{\beta=1}^\infty 3 \beta g_{1,\beta}(t_1,t_2,t_3) .
\label{expa}
\end{equation}

This can be summed easily using Equations (\ref{g1bC3Z3}) and (\ref{ZauxC3Z3}) and gives

\begin{eqnarray}
\label{g1C3Z3}
g_{1} (t_1,t_2,t_3; \IC^3/\IZ_3) &=& \frac{1}{(1-t_1)^3 (1-\frac{t_2}{t_1})(1-t_3/t_1)} \\
&+&\frac{1}{(1-t_1/t_2)(1-t_2)^3(1-t_3/t_2)}+\frac{1}{(1-t_1/t_3)(1-t_2/t_3)(1-t_3)^3}.
\nonumber
\end{eqnarray}

For the special case $t_1=t_2=t_3=t$ we can take the limit or resum, using Equation (\ref{g1geomZ3}),

$$g_1(t,t,t; \IC^3/\IZ_3)=\frac{1+4 t+t^2}{(1-t)^5}$$

which is exactly Equation (\ref{N1}).

\subsubsection{Refining the GKZ decomposition}

Using Equation (\ref{ZauxC3Z3}) we summarize the multiplicities
\be
  m(\beta) =
  \left\{ \begin{array}{ll}
 1  & \textrm{for} \ \beta=0 \\
 3\beta & \textrm{for} \ \beta > 0.
  \end{array}
  \right.
\ee

For a dibaryon, the AdS/CFT dual object is a D3-brane that wraps a $\Sigma_3 = S^3 / \IZ_3$
cycle\footnote{Generically, $\Sigma_3$ is a Lens space.} in $S^5 / \IZ_3$. The homology tells us
that the wrapping number is characterized by an integer, modulo 3. By resolving the singular
Calabi-Yau, this gets promoted to a (non-negative) integer which is just the coordinate $\beta$ in the
GKZ cone. The GKZ fan does not take into account the possible topologically nontrivial flat
connections on the wrapped D3-brane. To avoid multiplicities, we include the U(1) extensions of
all the discrete charges. The R-charge is already a coordinate in the GKZ fan and its corresponding GKZ auxiliary generating function is given in Equation (\ref{ZauxC3Z3}). The
remaining charges are the anomalous charges $A_1$ and $A_2$, as given in Table \ref{chargesC3Z3}, which we now add to the lattice as
extra coordinates. The points in the resulting lattice form a ``hollow cone'' and have no
multiplicities.



The dressed auxiliary GKZ partition function which now also contains the anomalous charges can
be computed using the assumption that the polynomial ring, Equation (\ref{ZGKZC3Z3}), is a
complete intersection,
\begin{eqnarray}
\label{auxC3}
\nonumber
& &Z_{\rm aux}(t, a_1, a_2; \IC^3/\IZ_3 ) =
\frac{1-t^3}{(1-ta_1 a_2)(1-ta_2/a_1)(1-t/a_2^2)} = \\ &=& 1 +
(a_2^{-2} + a_1^{-1} a_2 + a_1 a_2)t + \\ \nonumber
&+& ( a_2^{-4} + a_1^{-1}a_2^{-1} + a_1 a_2^{-1} + a_2^2 + a_1^{-2}
a_2^2 + a_1^2 a_2^2 ) t^2+ \\
&+& (a_1^{-2}+a_1^2+a_2^{-6} + a_1^{-1} a_2^{-3} + a_1 a_2^{-3} +
a_1^{-3} a_2^3 + a_1^{-1}a_2^3
+ a_1 a_2^3 + a_1^3 a_2^3 ) t^3 + \ldots \nonumber
\end{eqnarray}

\begin{figure}[ht]
\begin{center}
  \epsfxsize = 8cm
  \centerline{\epsfbox{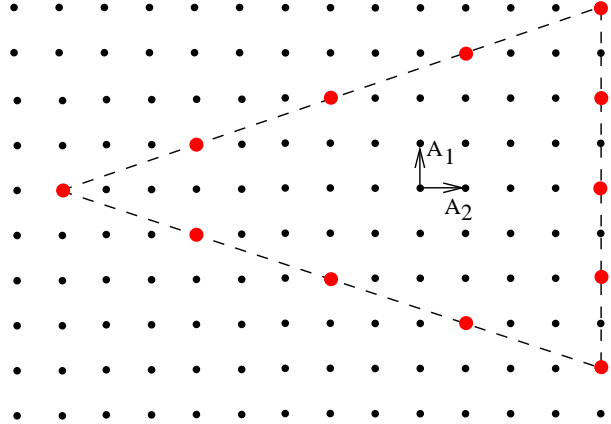}}
  \caption{The hollow triangle $C(4)$ above $R=4$, i.~e. the terms containing $t^4$. It gives the multiplicity $4\times 3 = 12$.}
  \label{emptyC3}
\end{center}
\end{figure}

\begin{figure}[ht]
\begin{center}
  \epsfxsize = 3cm
  \centerline{\epsfbox{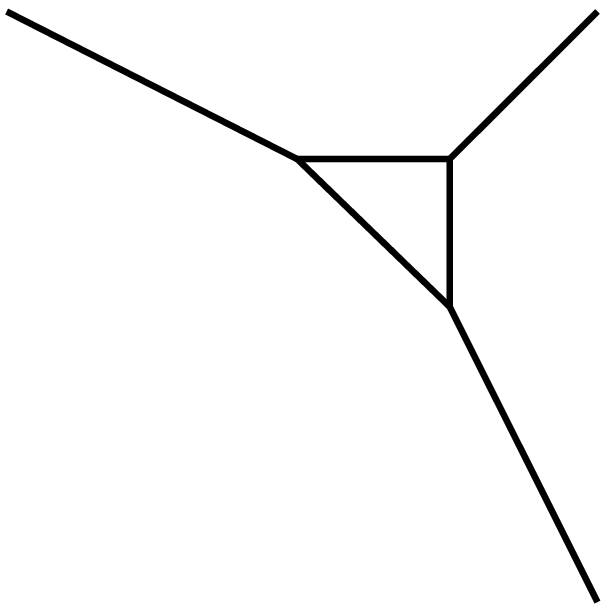}}
  \caption{The pq-web of $\IC^3 / \IZ_3$. The blown-up four-cycle is associated to the triangle in the middle.}
  \label{C3_Z3_pq}
\end{center}
\end{figure}

By drawing the lattice points in the $(A_1, A_2)$ lattice one can see that there is a ``hollow triangle" $C(\beta)$ above each point $\beta$
in the 1d GKZ cone (\fref{emptyC3}), with edge length measured by the R-charge. This gives the $1,3,6,9,\ldots$ multiplicities. The same triangle appears in the pq-web (\fref{C3_Z3_pq}) of the geometry. This is a general feature as we will see in other examples. The polygon in the fiber parameterized by the anomalous charges nicely matches the shape of the blown-up cycle in the pq-web.

We can now refine the decomposition (\ref{expa}) by introducing the anomalous
charges. We first write the expansion (\ref{auxC3}) in the form

$$Z_{\rm aux}(t, a_1, a_2; \IC^3/\IZ_3 ) = \sum_{\beta=0}^\infty  \left(\sum_{K\in C(\beta)} a_1^{K_1} a_2^{K_2} \right) t^\beta $$
where the two-dimensional index $K = (K_1, K_2)$ runs over the points of the hollow triangle
$C(\beta)$.

We can then refine the decomposition (\ref{expa}) by
replacing the multiplicity $m(\beta)$ with $\sum_{K\in C(\beta)} a_1^{K_1} a_2^{K_2} $,

\begin{equation}
g_1(t_1,t_2,t_3,a_1,a_2) = \sum_{\beta=0}^\infty \left(\sum_{K\in C(\beta)} a_1^{K_1} a_2^{K_2}\right) g_{1,\beta}(t_1,t_2,t_3) .
\label{expa2}
\end{equation}


By explicit computation we can resum the previous series and compare with
the expected field theory result, finding perfect agreement.
The left hand side of formula (\ref{expa2}) is indeed
the $N=1$ generating function
depending on all the five chemical potentials, which can be computed as
the Hilbert series for the polynomial ring, Equation (\ref{ringC3Z3})
using the grading in Table \ref{chargesC3Z3},
\begin{eqnarray}
& & g_1(t_1,t_2,t_3,a_1,a_2;\mathbb{C}^3/\mathbb{Z}_3) = \nonumber\\
& & \frac{P(t_1,t_2,t_3,a_1,a_2)}{(1 - \frac{t_1}{a_2^2})(1 - \frac{a_2 t_1}{a_1})(1 - a_1 a_2 t_1)(1 - \frac{t_2}{a_2^2})(1 - \frac{a_2 t_2}{a_1})(1 - a_1 a_2 t_2)(1 -\frac{t_3}{a_2^2})(1 - \frac{a_2 t_3}{a_1})(1 - a_1 a_2 t_3)}\nonumber\\
\label{N1C3}
\end{eqnarray}
where $P(t_1,t_2,t_3,a_1,a_2)$ is a polynomial in the gauge theory chemical potentials

\begin{eqnarray}
& & P(t_1,t_2,t_3,a_1,a_2)=1 - \frac{t_1 t_2}{a_1 a_2} - \frac{a_1 t_1 t_2}{a_2} - a_2^2 t_1 t_2 + t_1^2 t_2 + t_1 t_2^2 - \frac{t_1 t_3}{a_1 a_2} - \frac{a_1 t_1 t_3}{a_2} -
    a_2^2 t_1 t_3 + \nonumber\\
& & t_1^2 t_3 - \frac{t_2 t_3}{a_1 a_2} - \frac{a_1 t_2 t_3}{a_2} - a_2^2 t_2 t_3 + 4 t_1 t_2 t_3 + \frac{t_1 t_2 t_3}{a_1^2} + a_1^2 t_1 t_2 t_3 + \frac{t_1 t_2 t_3}{a_1 a_2^3} + \frac{a_1 t_1 t_2 t_3}{a_2^3} + \frac{a_2^3 t_1 t_2 t_3}{a_1} + \nonumber\\
& & a_1 a_2^3 t_1 t_2 t_3 - \frac{t_1^2 t_2 t_3}{a_2^2} - \frac{a_2 t_1^2 t_2 t_3}{a_1} -
     a_1 a_2 t_1^2 t_2 t_3 + t_2^2 t_3 - \frac{t_1 t_2^2 t_3}{a_2^2} - \frac{a_2 t_1 t_2^2 t_3}{a_1} - a_1 a_2 t_1 t_2^2 t_3 + t_1 t_3^2 + \nonumber\\
& &  t_2 t_3^2 - \frac{t_1 t_2 t_3^2}{a_2^2} - \frac{a_2 t_1 t_2 t_3^2}{a_1} - a_1 a_2 t_1 t_2 t_3^2 + t_1^2 t_2^2 t_3^2
\end{eqnarray}


We would like to stress that decomposition (\ref{expa2}) is highly nontrivial. The right hand
side has been computed from the geometrical localization formulae and the refined GKZ auxiliary
generating function. It is then remarkable that the sum on the right hand side coincides with
the field theory $N=1$ generating function.

Using Equations (\ref{g1bC3Z3}) and (\ref{auxC3}) we get the following simpler expression

\begin{eqnarray}
\label{g1a1a2C3Z3}
g_1(t_1,t_2,t_3,a_1,a_2; \mathbb{C}^3/\mathbb{Z}_3)
&=& \frac{1}{(1-t_1 a_1 a_2)(1- \frac{t_1 a_2}{a_1})(1- \frac{t_1}{a_2^2})(1-\frac{t_2}{t_1})(1-t_3/t_1)} \\ \nonumber
&+&\frac{1}{(1-t_1/t_2)(1-t_2 a_1 a_2)(1-t_2 a_2/a_1)(1-t_2 / a_2^2)(1-t_3/t_2)} \\ \nonumber
&+&\frac{1}{(1-t_1/t_3)(1-t_2/t_3)(1-t_3 a_1 a_2)(1-t_3 a_2/a_1)(1-t_3 / a_2^2)}.
\end{eqnarray}

The previous formula suggests the existence of a localization formula for the
holomorphic functions on the $N=1$ moduli space, which is a five-dimensional
variety with an action of five $U(1)$ symmetries, three flavor plus two
baryonic.

By projecting the refined GKZ expansion on the plane $(A_1,A_2)$ we would
get the expansion of the $N=1$ generating functions into sectors with definite
baryonic charge. The same result can be obtained by expanding $g_1$ in a
Laurent series by means of the residue theorem. It is easy to check that the
multiplicity of each sector is one.




\subsubsection{Generating functions for $N>1$}
The generating function $g_N$ is now obtained from the general formula (\ref{g1plet}) starting
from {\bf any} decomposition of the $N=1$ generating function, the GKZ decomposition
(\ref{expa}), the refined GKZ decomposition (\ref{expa2}) or the anomalous baryonic charge
decomposition. Since we are interested in writing generating functions depending on the
non-anomalous charges, at the end of the computation $a_i$ should be set to one.

The more economical way of obtaining $g_N$ is to start from decomposition (\ref{expa}). The
generating function for $N$ D-branes is now given by the plethystic exponentiation

\begin{equation}
\sum_{N=0}^\infty g_N(t_1,t_2,t_3) \nu^N = \hbox{PE}_\nu[g_{1,0}(t_1,t_2,t_3)] + \sum_{\beta=1}^\infty 3 \beta \hbox {PE}_\nu [g_{1,\beta}(t_1,t_2,t_3)]
\end{equation}

The cases $N=2$ and $N=3$ (with only one charge $t$) are given by

\begin{eqnarray}
& & g_2(t,t,t) =\nonumber \\
& & \frac{1+t+13 t^2+20 t^3+53 t^4+92 t^5+137 t^6+134 t^7+146 t^8+103 t^9+55 t^{10}+19 t^{11}+9 t^{12}}{(1-t)^8(1+t)^6(1+t+t^2)^3}
\nonumber
\end{eqnarray}

\begin{eqnarray}
& & g_3(t,t,t) =\nonumber \\
& &\frac{1+32t^3+394 t^6+2365t^9+7343t^{12}+12946t^{15}+13201t^{18}+7709 t^{21}+2314t^{24}+276t^{27}+3t^{30}}{(1-t^3)^{11}(1+t^3)^3}
\nonumber
\end{eqnarray}

Taking the Plethystic Logarithm for these expressions we find 9 generators for $N=1$, 18 baryonic and 10 mesonic for $N=2$, 30 baryonic and 10 mesonic for $N=3$, 45 baryonic, 10 mesonic of $R$ charge 2, 28 mesonic of $R$ charge 4 for $N=4$, etc. By taking the order of the pole at $t=1$ we find the dimension of the moduli space is $3N+2$. All of this agrees with the field theory expectations.

\subsection{Generating functions for $\mathbb{F}_0$}

\begin{figure}[ht]
\begin{center}
  \epsfxsize = 10cm
  \centerline{\epsfbox{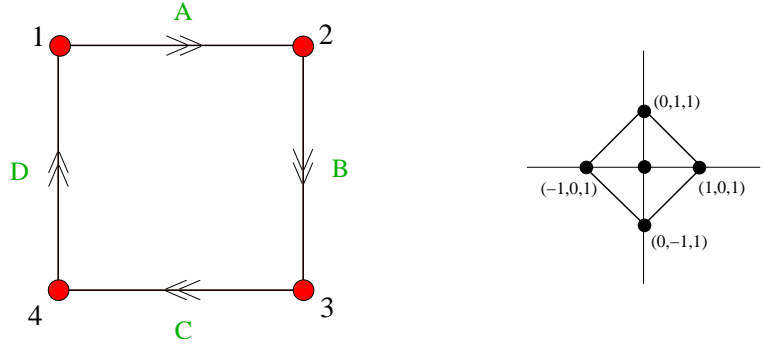}}
  \caption{Quiver and toric diagram for $\mathbb{F}_0$.}
  \label{F0quiver}
\end{center}
\end{figure}

$\mathbb{F}_0$, a $\mathbb{Z}_2$ freely acting orbifold of the conifold, has a quiver and the toric diagram given in Figure \ref{F0quiver}. The quiver gauge theory has four types of fields ${\bf A},{\bf B},{\bf C},{\bf D}$.
The superpotential is $\epsilon_{ij}\epsilon_{pq} {\bf A}_i{\bf B}_p{\bf C}_j{\bf D}_q$.

\paragraph{Symmetries and geometry}



Including flavor charges, we find a rank six global symmetry denoted by $SU(2)_1\times
SU(2)_2\times U(1)_R\times U(1)_B\times U(1)_{A_1}\times U(1)_{A_2}$. The basic fields have
transformation rules under the global symmetry which are summarized in Table \ref{globalF0}.

\begin{table}[htdp]
\begin{center}
\begin{tabular}{|c|c|c|c|c|c|c|c|c|}
\hline
\ {\bf field} \ & \ \  {\bf $F_1$}  \ \ & \ \  {\bf $F_2$}  \ \ & \ \  {\bf $R$} \ \ & \ \  {\bf $B$}  \ \ & \ \  {\bf $A_1$}  \ \ & \ \  {\bf $A_2$}  \ \ & \ \  chemical  \ \ & non-anomalous \\
 &  &  &  &  & &  & \ \  potentials \ \  & chemical potentials \\
\hline \hline
${\bf A}_1$  & $\frac{1}{2}$  & $0$      & $\frac{1}{2}$ & 1 & 1 & 0 & $t b x a_1$    & $t_1x =t b x$\\
${\bf A}_2$  & $-\frac{1}{2}$ & $0$       & $\frac{1}{2}$ & 1 & 1 & 0 & $\frac{t b a_1}{x}$ & $\frac{t_1}{x}=\frac {t b}{x}$ \\
${\bf B}_1$  & $0$          & $\frac{1}{2}$ & $\frac{1}{2}$ &$-1$ & 0 & 1 & $\frac{t y a_2}{b}$ & $t_2 y= \frac{ t y}{b}$   \\
${\bf B}_2$  & $0$          & $-\frac{1}{2}$ & $\frac{1}{2}$    &$-1$ & 0 & 1 & $\frac{t a_2}{b y}$ & $\frac{t_2}{y} =\frac{t}{b y}$  \\
${\bf C}_1$  & $\frac{1}{2}$    & $0$       & $\frac{1}{2}$ & 1 &$-1$ & 0 & $\frac{t b x}{a_1}$ & $t_1 x = t b x$  \\
${\bf C}_2$  & $-\frac{1}{2}$   & $0$       & $\frac{1}{2}$ & 1 &$-1$ & 0 & $\frac{t b}{x a_1}$  & $\frac{t_1}{x}=\frac{t b}{x}$ \\
${\bf D}_1$  & $0$ & $\frac{1}{2}$ & $\frac{1}{2}$ &$-1$ & 0 &$-1$ & $\frac{t y}{b a_2}$  & $t_2 y =\frac{t y}{b}$ \\
${\bf D}_2$  & $0$          & $-\frac{1}{2}$    & $\frac{1}{2}$ &$-1$ & 0 &$-1$ & $\frac{t}{b y a_2}$  & $\frac{t_2}{y}=\frac{t}{b y}$ \\
\hline
\end{tabular}
\end{center}
\caption{Global charges for the basic fields of the quiver gauge theory
living on the D-brane probing the CY with $\mathbb{F}_0$ base.}
\label{globalF0}
\end{table}

We can explicitly examine the geometry of $\mathbb{F}_0$.
To this purpose, since $d=4$, we  introduce four homogeneous coordinates $x_i$ in $\mathbb{C}^4$. As in Section \ref{D3sup} we define the group of rescalings
$\prod_{j=1}^4 \mu_j^{\langle e_i,V_j\rangle }=1$ which consists of a
continuous charge acting as $(b,1/b,b,1/b)$ on the $x_i$ and of a discrete one $(1,e,1,e)$ with
$e^2=1$. This implies that the manifold, as we know, is a $\mathbb{Z}_2$ quotient of the conifold

\begin{equation}
{\cal R} ( {\cal C} ) = \mathbb{C}[x_1, x_2, x_3, x_4]/(x_1x_2-x_3x_4); \qquad {\cal R} ( \mathbb{F}_0 ) = {\cal R} ( {\cal C} ) / \mathbb{Z}_2
\end{equation}

The homogeneous charges $x_i$ can be represented by chemical potentials as $$(x_1,x_2,x_3,x_4) \longrightarrow (t_1 x,e t_2 y,t_1/x,e t_2/y)$$ in terms of the discrete baryonic charge $e$ and the chemical potentials $t_i$ and $x,y$;
notations are inherited from original conifold theory.
For future reference, we notice that
the right basis for localization is given by $q_i=\prod_{j=1}^4 x_j^{\langle e_i,V_j\rangle }$ and we compute
 $q_1=x^2,q_2=y^2,q_3=t_1^2 t_2^2$.

\subsubsection{The $N=1$ generating function}

The F terms of the theory read:
\bean {\bf A}_1{\bf B}_i{\bf C}_2={\bf A}_2{\bf B}_i{\bf C}_1\,,\quad  {\bf
B}_1{\bf C}_i{\bf D}_2={\bf B}_2{\bf C}_i{\bf D}_1 \, ,\\
\quad {\bf C}_1{\bf D}_i{\bf A}_2={\bf C}_2{\bf D}_i{\bf A}_1\, ,\quad {\bf D}_1{\bf A}_i{\bf
B}_2={\bf D}_2{\bf A}_i{\bf B}_1 .
\eean

For $N=1$, the elementary fields are commuting numbers and in each equation a field factorizes.
For example, the first equation reduces to ${\bf B}_i({\bf A}_1{\bf C}_2-{\bf A}_2{\bf C}_1)=0$.
This implies that the $N=1$ moduli space is reducible to few different branches. At a generic point where all the
fields are different from zero, we can divide by the common factor and the F-term equations
reduce to
$$ {\bf A}_1 {\bf C}_2={\bf A}_2 {\bf C}_1\, , \qquad {\bf B}_1 {\bf D}_2={\bf B}_2 {\bf D}_1. $$
However, over the submanifold ${\bf B}_i={\bf D}_i=0$ the constraint $ {\bf A}_1 {\bf C}_2={\bf
A}_2 {\bf C}_1$ cannot be imposed and the dimension of the moduli space increases by one unit.
The same applies for the constraint ${\bf B}_1 {\bf D}_2={\bf B}_2 {\bf D}_1$ over the
submanifold ${\bf A}_i={\bf C}_i=0$.


This means that the moduli space is not irreducible and over particular submanifolds new
branches are opening up. This is similar to what happens with the Coulomb branch of ${\cal N}=2$ supersymmetric gauge theories.

We decide to study the irreducible components of the moduli space which contains the generic
point with all fields different from zero. Algebraically, this is obtained by taking the closure
of the open set ${\bf A},{\bf B},{\bf C},{\bf D}\ne 0$. We will see that the geometry of the CY
nicely captures this branch of the moduli space. The other branches can be added by performing
surgeries as in \cite{Hanany:2006uc}. The $N=1$ generating function of the generic branch is given
by the Hilbert series of the polynomial ring

\begin{equation}
{\cal R}_{N=1} ( \mathbb{F}_0 ) = \mathbb{C}[{\bf A}_i,{\bf B}_i,{\bf C}_i,{\bf D}_i]/({\bf
A}_1{\bf C}_2-{\bf A}_2{\bf C}_1, \ {\bf B}_1{\bf D}_2-{\bf B}_2{\bf D}_1)
\label{ringF0}
\end{equation}

In order to simplify expressions, we first set to 1 all chemical potentials in Table
\ref{globalF0} except for  $t_1$ and $t_2$. The Hilbert series is then easily computed by
observing that Equation (\ref{ringF0}) is a polynomial ring which is a complete intersection,

\begin{equation}
g_1(t_1, t_2; \mathbb{F}_0 ) = \frac{(1-t_1^2)(1-t_2^2)}{(1-t_1)^4(1-t_2)^4}
\label{N11}
\end{equation}

By taking order of the pole at $t_1=t_2=t=1$ we find the dimension of the moduli space to be 6; this can be easily understood by having three mesonic directions (parameterizing the CY) plus three baryonic directions given by the gauge theory FI terms, consistent with a dimension formula of $3N+G-1$.

\subsubsection{The GKZ decomposition.}

\begin{figure}[ht]
\begin{center}
  \epsfxsize = 6cm
  \centerline{\epsfbox{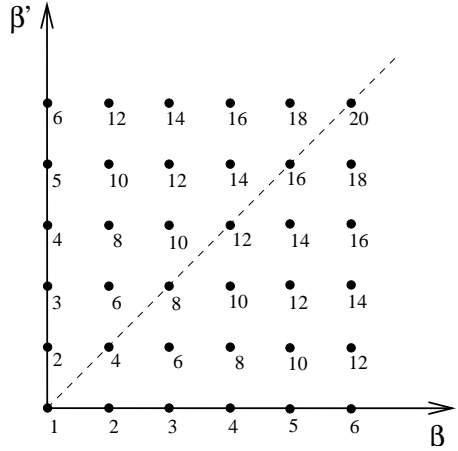}}
  \caption{GKZ decomposition for $\mathbb{F}_0$ with multiplicities.}
  \label{FF}
\end{center}
\end{figure}

On the singular cone, there is just one independent divisor $D_1=D_3=-D_2=-D_4$ as for the conifold. On the resolution, there is a new divisor $D_5$ corresponding to the internal point.
$D_1=D_3$ and $D_2=D_4$ are still true but now $D_1$ and $-D_2$ are different.
We can parametrize the GKZ fan in $\mathbb{R}^2$ with
$\beta D_1 +\beta^\prime D_2$. The integer parameters $\beta$ and $\beta^\prime$ have the interpretation as the discrete K\"ahler parameters of $\mathbb{F}_0$, namely the discrete areas of the two $P^1$'s.
In this case, the convexity condition requires $\beta,\beta^\prime\ge 0$. The GKZ cone is depicted in Figure \ref{FF} and the multiplicities are presented in Equation (\ref{dictfig}). Notice that the QFT baryonic charge is given by $B=\beta-\beta^\prime$, so the sector $\beta^\prime < \beta$ corresponds to the positive baryonic charges and  the sector $\beta^\prime > \beta$ corresponds to the negative ones.

It is interesting to note that the line $\beta^\prime = \beta$ contains the sectors with zero
baryonic charge. It is not however true that operators with zero baryonic charges are made with
traces; this is only true for $\beta=\beta^\prime=0$. The other sectors on the line
$\beta^\prime = \beta$ correspond to determinants of the form  $(\det {\bf A} \det {\bf B})^n$.

Localization now gives
(using the complete set of charges $(t_1x,t_2y,t_1/x,t_2/y)$)

\begin{eqnarray}
\label{ZBBF0}
g_{1,\beta,\beta^\prime} (t_1, t_2 , x, y; \mathbb{F}_0 ) &=& \frac{t_1^{\beta}t_2^{\beta^\prime} x^{-\beta} y^{-\beta^\prime}}{(1-x^2)(1- \frac{t_1^2 t_2^2}{x^2 y^2})(1-y^2)}+\frac{t_1^{\beta}t_2^{\beta^\prime} x^{\beta} y^{-\beta^\prime}}{(1-1/x^2)(1-t_1^2t_2^2x^2/y^2)(1-y^2)} \\ \nonumber
&+&\frac{t_1^{\beta}t_2^{\beta^\prime} x^{-\beta} y^{\beta^\prime}}{(1-x^2)(1-t_1^2t_2^2y^2/x^2)(1-1/y^2)}+\frac{t_1^{\beta}t_2^{\beta^\prime} x^{\beta} y^{\beta^\prime}}{(1-1/x^2)(1-t_1^2t_2^2x^2y^2)(1-1/y^2)}
\end{eqnarray}

The dependence on the baryonic charge can be obtained by replacing
$t_1\rightarrow tb$ and $t_2\rightarrow t/b$ and, as expected, is given
by a multiplicative factor
$$g_{1,\beta,\beta^\prime} (t_1, t_2 , x, y; \mathbb{F}_0 ) = b^{\beta -\beta^\prime} \hat g_{1,\beta,\beta^\prime} (t , x, y; \mathbb{F}_0 ) . $$
The generating function for $x,y=1$ can be nicely written as

\begin{equation}
g_{1,\beta,\beta^\prime} (t_1, t_2; \mathbb{F}_0 ) = \sum_{n=0}^\infty
(2 n+1+\beta)(2 n+1+\beta^\prime) t_1^{2 n+\beta} t_2^{2 n+\beta^\prime}
\label{seri}
\end{equation}

It is then obvious that, for example, the mesonic partition function $g_{1,0,0}$ can be obtained from the mesonic partition function for the conifold
\cite{Martelli:2006yb,Benvenuti:2006qr}
by projecting on the $\mathbb{Z}_2$ invariant part ($t_2\rightarrow -t_2$).

\subsubsection{Multiplicities}
To extract multiplicities we use the auxiliary partition function for the GKZ cone. We introduce
letters $a,b,c,d$ for the four possible classes of arrows. The only relation that they form is
related to the closed loop $abcd$. The generating function is the Hilbert series of the polynomial ring

\begin{equation}
{\cal R}_{\rm GKZ} ( \mathbb{F}_0 ) = \mathbb{C}[a,b,c,d]/(abcd)
\label{ringGKZF0}
\end{equation}

By assigning chemical potential $t_1$ to $a,c$ and $t_2$ to $b,d$ we obtain the
auxiliary GKZ partition function for multiplicities:

\begin{eqnarray}
\label{ZauxF0}
Z_{\rm aux}(t_1,t_2; \mathbb{F}_0 ) &=& \frac{1-t_1^2 t_2^2}{(1-t_1)^2 (1-t_2)^2} \\ \nonumber
&=& 1+ \sum_{\beta = 1}^\infty ( \beta + 1 ) t_1^\beta + \sum_{\beta' = 1}^\infty (\beta' + 1) t_2^{\beta'} + \sum_{\beta = 1}^\infty \sum_{\beta' = 1}^\infty 2 ( \beta + \beta' ) t_1^\beta t_2^{\beta'}
\end{eqnarray}

From which we can extract the following multiplicities,
$2(\beta+\beta^\prime)$ for $\beta,\beta^\prime\ge1$, $\beta+1$ for $\beta^\prime=0$, and
$\beta^\prime+1$ for $\beta=0$.

We thus have

\begin{equation}
 g_1(\{t_i\}; \mathbb{F}_0 ) = g_{1,0,0} + \sum_{\beta=1}^\infty (\beta+1)g_{1,\beta,0}
+ \sum_{\beta^\prime=1}^\infty (\beta^\prime+1)g_{1,0,\beta^\prime}
+ \sum_{\beta,\beta^\prime=1}^\infty 2 (\beta+\beta^\prime) g_{1,\beta,\beta^\prime}
\label{F0multiplicities}
\end{equation}
and one computes, using Equation (\ref{seri}),

$$ g_1(t_1, t_2; \mathbb{F}_0 )=\frac{(1-t_1^2)(1-t_2^2)}{(1-t_1)^4(1-t_2)^4} \label{g1F0} $$
which is exactly Equation (\ref{N11}).

\subsubsection{Refining the GKZ decomposition}

The auxiliary partition function in Equation (\ref{ZauxF0}) is derived by computing the Hilbert series of the GKZ ring in Equation (\ref{ringGKZF0}). By expanding this partition function we get multiplicities in the $(\beta, \beta^\prime)$ lattice given by the following infinite matrix,
\be
\left(
\begin{array}{ccccccc}
  1 & 2 & 3 & 4 & 5 & 6 & \\
  2 & 4 & 6 & 8 & 10 & 12 & \\
  3 & 6 & 8 & 10 & 12 & 14 & \ldots \\
  4 & 8 & 10 & 12 & 14 & 16 &   \\
  5 & 10 & 12 & 14 & 16 & 18 & \\
  6 & 12 & 14 & 16 & 18 & 20 & \\
  & & \vdots &  & & & \ddots
\end{array}
\right) .
\label{dictfig}
\ee

In order to avoid getting multiplicities, let us introduce the chemical potentials for the anomalous charges given in Table \ref{globalF0}. Using Macaulay2, or the fact that we are dealing with a complete
intersection, we can write the auxiliary partition function dressed with these new charges as

\be
Z_{\rm aux}(t_1,t_2, a_1, a_2; \mathbb{F}_0 ) =  \frac{1-t_1^2 t_2^{2}}{(1-t_1 a_1)(1-t_1/a_1)(1-t_2 a_2)(1-t_2/a_2)}
\label{f0gen}
\ee

\begin{figure}[ht]
\begin{center}
  \epsfxsize = 6cm
  \centerline{\epsfbox{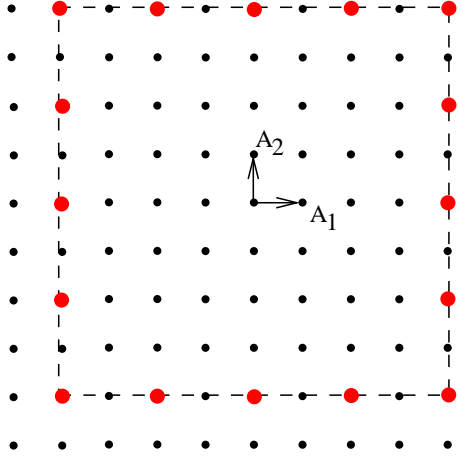}}
  \caption{The hollow rectangle $C(4,4)$ above $(\beta, \beta^\prime)=(4,4)$. It gives the multiplicity 16.}
  \label{emptyF0}
\end{center}
\end{figure}

\begin{figure}[ht]
\begin{center}
  \epsfxsize = 2.5cm
  \centerline{\epsfbox{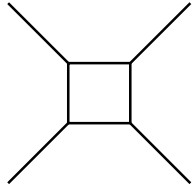}}
  \caption{The pq-web of $\mathbb{F}_0$. The blown-up four-cycle is associated to the square in the middle.}
  \label{F0_pq}
\end{center}
\end{figure}

By expanding this function, we see that above each point in the GKZ fan, parametrized by $(\beta, \beta^\prime)$, there is a rectangle $C(\beta,\beta^\prime)$ in the $(A_1, A_2)$ lattice as in \fref{emptyF0}. The related rectangle in the pq-web is shown in \fref{F0_pq}.

We can thus refine our decomposition of the $N=1$ partition function.
Equation (\ref{F0multiplicities}) can be replaced by the following formula
where all multiplicities are lifted:

\begin{equation}
g_1(t_1,t_2,a_1,a_2) = \sum_{\beta=0,\beta^\prime=0}^\infty \left(\sum_{K\in C(\beta,\beta^\prime)} a_1^{K_1} a_2^{K_2} \right) g_{1,\beta,\beta^\prime}(t_1,t_2) .
\label{expa2F0}
\end{equation}

One can explicitely resum the right hand side, computed from geometrical
data and the auxiliary GKZ generating functions, and compare it with
the $N=1$ generating function as computed by field theory.
By resumming the series in  Equation (\ref{expa2F0}) we indeed recover the
the generating function for $N=1$ with all anomalous and non-anomalous charges; this is given by
the Hilbert series for the polynomial ring of Equation (\ref{ringF0}) and is easily computed
using the fact that we deal with a complete intersection,

\begin{eqnarray}
& & g_1(x,y,t,b,a_1,a_2;\mathbb{F}_0)= \\
& & \frac{ (1 - \frac{t^2}{b^2}) (1 - b^2 t^2) }{(1 - \frac{b t}{a_1 x})(1 - \frac{a_1 b t}{x})(1 - \frac{b t x}{a_1})(1 - a_1 b t x)(1 - \frac{t}{a_2 b y})(1 - \frac{a_2 t}{b y})(1 - \frac{t y}{a_2 b})(1 - \frac{a_2 t y}{b})}\nonumber
\end{eqnarray}

For completeness we can rewrite this expression by summing using Equations (\ref{ZBBF0}) and (\ref{f0gen}),

\begin{eqnarray}
g_1(x,y,t,b,a_1,a_2;\mathbb{F}_0)
&=& \frac{1}{(1-x^2)(1-t_1 a_1/x)(1- \frac{t_1}{x a_1})(1-t_2 a_2/y)(1- \frac{t_2}{y a_2})(1-y^2)} \\ \nonumber
&+& \frac{1}{(1-1/x^2)(1-t_1 x a_1)(1-t_1 x/a_1)(1-t_2 a_2 / y)(1-t_2/ y a_2)(1-y^2)} \\ \nonumber
&+& \frac{1}{(1-x^2)(1-t_1 a_1 / x)(1-t_1/ x a_1)(1-t_2 y a_2)(1-t_2 y /a_2)(1-1/y^2)} \\ \nonumber
&+& \frac{1}{(1- \frac{1}{x^2})(1-t_1 x a_1)(1-t_1 x /a_1)(1-t_2 y a_2)(1-t_2 y /a_2)(1-1/y^2)} .
\end{eqnarray}

This formula suggests that some localization is at work in the
field theory $N=1$ moduli space, which is a six dimensional variety with the
action of six $U(1)$ flavor and baryonic symmetries.

\subsubsection{Expansion in baryonic charges}


Equation (\ref{N11}) can be refined by including the two anomalous chemical potentials $a_1$ and $a_2$,

\begin{equation}
g_1(t_1, t_2, a_1, a_2; \mathbb{F}_0)=\frac{(1-t_1^2)(1- t_2^2)}{(1-t_1 a_1)^2 (1-\frac{t_1}{a_1})^2 (1-{t_2 a_2})^2 (1-\frac{t_2}{a_2})^2 }
\label{g1t1t2F0}
\end{equation}

and by using residue formulae we can expand in terms of generating functions with fixed anomalous baryonic charges,

\begin{equation}
g_{1, A_1, A_2}(t_1, t_2; \mathbb{F}_0) = \frac{t_1^{|A_1|} t_2^{|A_2|} [1+t_1^2 + |A_1|(1-t_1^2)] [1+t_2^2 + |A_2|(1-t_2^2)]}{(1-t_1^2)^2 (1-t_2^2)^2}
\end{equation}

More generally, we can make explicit the dependence of the $N=1$ generating function on the full set of baryonic charges

\begin{equation}
g_1(t, b, a_1, a_2; \mathbb{F}_0)=\frac{(1-t^2 b^2)(1- \frac{t^2}{b^2})}{(1-t b a_1)^2 (1-\frac{t b}{a_1})^2 (1-\frac{t a_2}{b})^2 (1-\frac{t}{b a_2})^2 }
\label{g1tb}
\end{equation}
which can be expanded in sectors with definite baryonic charges. A contour integral argument
gives the following generating functions for 1 D-brane and for fixed baryonic charges







\begin{equation}
g_{1, B, A_1, A_2}(t; \mathbb{F}_0) = \frac{1+(-1)^{B+A_1+A_2}} {2(1-t^4)^3} \left \{
\begin{array}{l}

t^{-B-2A_1} [1+6t^4+t^8 - (B+2A_1)(1-t^8) + \\ + (A_1^2+B A_1)(1-t^4)^2] \qquad A_1 \le 0, \quad |A_2| \le -B-A_1 \\ \\

t^{-B+2A_1} [1+6t^4+t^8 - (B-2A_1)(1-t^8) + \\ + (A_1^2-B A_1)(1-t^4)^2] \qquad 
A_1 \ge 0, \quad |A_2| \le -B + A_1 \\ \\

t^{B-2A_2} [1+6t^4+t^8 + (B-2A_2)(1-t^8) + \\ + (A_2^2 - B A_2)(1-t^4)^2] \qquad 
A_2 \le 0, \quad |A_1| \le B - A_2 \\ \\

t^{B+2A_2} [1+6t^4+t^8 + (B+2A_2)(1-t^8) + \\ + (A_2^2+B A_2)(1-t^4)^2] \qquad 
A_2 \ge 0, \quad |A_1| \le B + A_2 \\

\end{array}
\right.
\label{barF0}
\end{equation}

The same result can be obtained by projecting the decomposition of the $N=1$
generating function on the refined GKZ lattice on the three dimensional space
containing the three baryonic charges. This is done by projecting the two
GKZ coordinates $(\beta,\beta^\prime)$ to the baryonic charge
$B=\beta-\beta^\prime$ and keeping fixed the two anomalous
charges $A_1$ and $A_2$, restricted to the conditions $-\beta\le A_1\le \beta$ and $-\beta^\prime\le A_2\le \beta^\prime$.



\subsubsection{Generating functions for $N>1$}

As before the higher $N$ generating function $g_N$ is given by the plethystic exponentiation using formula (\ref{g1plet}). We can start from {\bf any} decomposition of the $N=1$ generating function, the GKZ decomposition (\ref{F0multiplicities}), the refined GKZ decomposition (\ref{expa2F0}) or the expansion in baryonic charges (\ref{barF0}). Using the GKZ decomposition we have

\begin{eqnarray}
\sum_{N=0}^\infty g_N( \{t_i\}; \mathbb{F}_0 ) \nu^N &=& \hbox{PE}_\nu[g_{1,0,0}]
+ \sum_{\beta=1}^\infty(\beta + 1) \hbox {PE}_\nu [g_{1,\beta,0}] \\
&+& \sum_{\beta^\prime=1}^\infty(\beta^\prime + 1) \hbox {PE}_\nu [g_{1,0,\beta^\prime}]
+ \sum_{\beta=1}^\infty \sum_{\beta^\prime=1}^\infty 2(\beta + \beta^\prime ) \hbox {PE}_\nu [g_{1,\beta,\beta^\prime}] \nonumber
\end{eqnarray}

From which we can compute the generating function for $N=2$

\begin{eqnarray}
g_2(t_1,t_2) &=& \frac{F(t_1,t_2)}{(1-t_1^2)^4(1-t_1^2t_2^2)^3(1-t_2^2)^4}\nonumber\\
F(t_1,t_2) &=& 1+2t_2^2+t_1^8t_2^4(-6+8t_2^2+17t_2^4+2t_2^6)+t_1^2(2+14t_2^2+8t_2^4-3 t_2^6) \nonumber\\
& & -t_1^6t_2^2(3+20 t_2^2+10t_2^4-8t_2^6-t_2^8) +t_1^{10}(t_2^6+2 t_2^8)-t_1^4(-8 t_2^2+6 t_2^4+20 t_2^6+6 t_2^8)\nonumber
\end{eqnarray}

Taking the plethystic logarithm of the generating functions for $\mathbb F_0$,
it is  possible to find the generators of the moduli space. For $N=1$ the
plethystic logarithm of (\ref{N11}) is very simple:
\[
\left(4 t_1 + 4 t_2 \right)  - \left( t_1^2 + t_2^2 \right)
\]
and it correctly reproduces the 8 chiral field generators and
the two relations among them.
For $N=1$ the moduli space is a complete intersection, and this is reflected
by the fact that the plethystic logarithm has a finite number of terms.
In the $N=2$ case there are 12 baryonic generators and 13 mesonic ones and the
moduli space is no more a complete intersection.
By looking at the order of the pole for the computed generating functions
near $t_1=t_2=t=1$, we checked that the dimension of the moduli space is $3 N
+ G - 1$ as expected.

\subsection{Counting in $\frac{1}{2} \mathbb{F}_0$ and in $\frac{3}{4} \mathbb{F}_0$}

We can also consider, as was done for the conifold in \cite{Forcella:2007wk}, ficticious
theories counting subsets of the BPS chiral operators.

The half $\mathbb{F}_0$ corresponds to considering only the BPS operators
containing ${\bf A}_1,{\bf B}_1,{\bf C}_1,{\bf D}_1$ and no occurences of
the other four elementary fields. This is done by truncating
Equation (\ref{seri}) for $g_{1,\beta,\beta^\prime}$

\begin{equation}
g_{1,\beta,\beta^\prime} = \sum_{n=0}^\infty
 t_1^{2 n+\beta} t_2^{2 n+\beta^\prime}
\end{equation}

We easily obtain the generating functions for $N=1,2,$

\begin{eqnarray}
g_1(t_1,t_2; \frac{1}{2} \mathbb{F}_0) &=& \frac{1}{(1-t_1)^2(1-t_2)^2}\nonumber\\
g_2(t_1,t_2; \frac{1}{2} \mathbb{F}_0) &=&\frac{1}{(1-t_1^2)^2(1-t_1^2t_2^2)(1-t_2^2)^2}\end{eqnarray}

Moreover, by generalizing arguments given in \cite{Forcella:2007wk} for the $1/2$ conifold, it
is easy to write a formula for generic $N$

\begin{equation}
g_N(t_1,t_2; \frac{1}{2} \mathbb{F}_0) = \frac{1}{(1-t_1^N)^2(1-t_2^N)^2}\prod_{i=1}^{N-1}\frac{1}{1-t_1^{2 i} t_2^{2 i}}
\end{equation}
which is interpreted as the fact that the ring of invariants is freely generated by four
determinants $\det {\bf A}_1,\det {\bf B}_1,\det {\bf C}_1,\det {\bf D}_1$ and $N-1$ mesons
${\rm Tr} ({\bf A}_1{\bf B}_1{\bf C}_1{\bf D}_1)^i$ with $i=1,...,N-1$. This can be understood
easily from the absence of nontrivial F term conditions and the fact that all baryons factorize
into an alementary determinant times mesons.

The $\frac{3}{4} \mathbb{F}_0$ corresponds to considering only the BPS operators
containing ${\bf A}_1,{\bf B}_1,{\bf C}_1,{\bf D}_1$ and ${\bf A}_2,{\bf C}_2$
and no occurences of the other two elementary fields. This is done by taking

\begin{equation}
g_{1,\beta,\beta^\prime} = \sum_{n=0}^\infty (2n+1+\beta) t_1^{2 n+\beta} t_2^{2 n+\beta^\prime} .
\end{equation}

Using the same multiplicities as in Equation (\ref{F0multiplicities}) we obtain

\begin{eqnarray}
g_1(t_1,t_2; \frac{3}{4} \mathbb{F}_0) &=& \frac{1+t_1}{(1-t_1)^3(1-t_2)^2}\nonumber\\
g_2(t_1,t_2; \frac{3}{4} \mathbb{F}_0) &=&\frac{(1+t_1^2t_2^2)(1+2 t_1^2+2 t_1^4t_2^2+t_1^6t_2^2)}{(1-t_1^2)^4(1-t_1^2t_2^2)^2(1-t_2^2)^2}
\label{threequarter}
\end{eqnarray}

\subsection{Generating functions for del Pezzo 1}

\begin{figure}[ht]
\begin{center}
  \epsfxsize = 10cm
  \centerline{\epsfbox{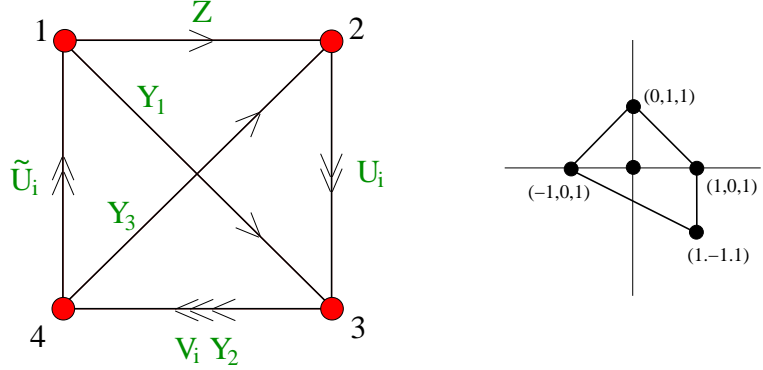}}
  \caption{Quiver and toric diagram for $dP_1$.}
  \label{dP1quiver}
\end{center}
\end{figure}

\begin{figure}[h!!!]
\begin{center}
\includegraphics[scale=0.4]{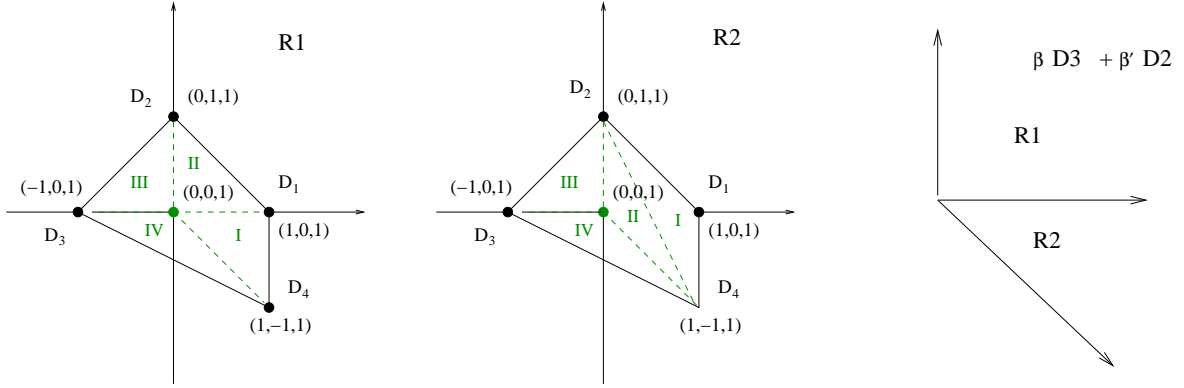}
\caption{The two resolutions, the quiver and GKZ decomposition for $dP_1$.}
\label{Y21a}
\end{center}
\end{figure}

The quiver and the toric diagram are depicted in Figures \ref{dP1quiver} and \ref{Y21a}.

Including flavor charges, we find a rank six global symmetry denoted by $SU(2) \times U(1)_{F}
\times U(1)_{R_0} \times U(1)_B\times U(1)_{A_1}\times U(1)_{A_2}$. $R_0$ is not the
$R$-symmetry of the superconformal theory but rather a more convenient parametrization of the
fields for the purpose of the counting problem. It counts the fields in way such that if the
$dP_1$ theory is Higgsed down to the $\IC^3/\IZ_3$ theory by giving a vev to the ${\bf Z}$
field, then $R_0$ measures the $R$-symmetry of this resulting theory. The basic fields have
transformation rules under the global symmetry which are summarized in Table \ref{globaldP1}.

\begin{table}[htdp]
\begin{center}
\begin{tabular}{|c|c|c|c|c|c|c|c|c|c|}
\hline
\ {\bf field} \ & \   {\bf $SU(2)$}   \ & \ \  $F$  \ \ & \ \  $R_0$ \ \ & \ \  {\bf $B$}  \ \ & \ \  {\bf $A_1$}  \ \ & \ \  {\bf $A_2$}  \ \ &  chemical   & non-anomalous  & GKZ\\
 &  &  &  &  & &  &   potentials  & ch. potentials  & letters\\
\hline \hline
${\bf U}_1$             & $\frac{1}{2}$  & $0$       & $\frac{2}{3}$ & $-2$ & 1 & 0 & $\frac{t x} {b^2} a_1$ & $x_2 =\frac{t x} {b^2}$  &  d\\
${\bf U}_2$             & $-\frac{1}{2}$     & $0$   & $\frac{2}{3}$ & $-2$ & 1 & 0 & $\frac{t} {x b^2} a_1$ & $x_4=\frac{t} {x b^2}$ &  d\\
${\bf \tilde U}_1$      & $\frac{1}{2}$ & $0$  & $\frac{2}{3}$ &$-2$ & $-1$ & 0 & $\frac{t x} {b^2 a_1}$ & $x_2 =\frac{t x} {b^2}$ & b\\
${\bf \tilde U}_2$      & $-\frac{1}{2}$ & $0$ & $\frac{2}{3}$ &$-2$ & $-1$ & 0 & $\frac{t}{x b^2 a_1}$ & $x_4=\frac{t} {x b^2}$  & b\\
${\bf V}_1$             & $\frac{1}{2}$    & $-1$     & $\frac{2}{3}$ & 1 &$0$ & $-1$ & $\frac{t b x}{y a_2}$ & $x_1 x_2 = \frac{t b x}{y}$ & a\\
${\bf V}_2$  & $-\frac{1}{2}$   & $-1$       & $\frac{2}{3}$ & 1 &$0$ & $-1$ & $\frac{t b}{x y a_2}$  & $x_1 x_4 = \frac{t b}{x y}$  & a\\
${\bf Y}_1$ & $0$ & $1$ & $\frac{2}{3}$ &$1$ & $1$&$1$ & $t b y a_1 a_2$  & $x_3= t b y$ & f\\
${\bf Y}_2$ & $0$ & $1$ & $\frac{2}{3}$ &$1$ & 0 &$-1$ & $\frac{t b y}{a_2}$ & $x_3=t b y$ & a\\
${\bf Y}_3$ & $0$ & $1$ & $\frac{2}{3}$ &$1$ & $-1$ & $1$ & $\frac{t b y a_2}{a_1}$  & $x_3=t b y$ & e\\
${\bf Z}$  & $0$     & $-1$ & 0 &$3$ & $0$ & $1$ & $\frac{b^3 a_2}{y}$  & $x_1=\frac{b^3}{y}$ & c\\
\hline
\end{tabular}
\end{center}
\caption{Global charges for the basic fields of the quiver gauge theory
living on the D-brane probing the CY with $dP_1$ base.}
\label{globaldP1}
\end{table}

\subsubsection{The $N=1$ generating function}

The polynomial ring for the $N=1$ moduli space is

\begin{equation}
{\cal R}_{N=1} ( dP_1 ) = \mathbb{C}[{\bf Y}_1,{\bf Y}_2,{\bf Y}_3,{\bf Z},{\bf U}_1,{\bf U}_2,{\bf \tilde U}_1,{\bf \tilde U}_2,{\bf V}_1,{\bf V}_2]/\{\partial W =0\}
\label{ringdP1}
\end{equation}
where $W=\epsilon_{ab}{\bf Y}_1 {\bf V}_a {\bf \tilde U}_b +\epsilon_{ab}{\bf Y}_3  {\bf U}_a {\bf
V}_b +\epsilon_{ab}{\bf Y}_2 {\bf \tilde U}_a {\bf Z} {\bf U}_b$.

There are 10 different F-term equations. Consider in particular the equations for ${\bf Z}$ and ${\bf Y}_2$
\be \epsilon_{ab}{\bf \tilde U}_a {\bf Z U}_b = 0 \, \qquad\qquad \epsilon_{ab}{\bf U}_b
{\bf Y}_2 {\bf \tilde U}_a = 0 \label{fterm}
\ee
The fact that we can factorize a field in each of these equations implies that
the moduli space of vacua is not irreducible. Over the submanifold ${\bf Z}={\bf Y}_2=0$ the dimension of the moduli space increases by one unit.


Instead of the two conditions (\ref{fterm}) we will impose the simpler
condition
\be \epsilon_{ab}{\bf \tilde U}_a {\bf U}_b=0. \label{fterm1} \ee
As in the
$\mathbb{F}_0$ example, this means that we are considering one irreducible component of the moduli space, which is the closure of the open set ${\bf Z},{\bf Y}_2\ne 0$.
This is the branch that is nicely described by the CY geometry.

We give chemical potentials to the fields using four homogeneous coordinates $x_i$ as in Table \ref{globaldP1} as described in
\cite{Butti:2005vn,Butti:2005ps}.
With these weights we can compute the Hilbert series for the graded ring, Equation (\ref{ringdP1}), by using Macaulay2 and obtain the $N=1$ generating function $g_1(x_i)$,

\begin{equation}
g_1(\{x_i\}; dP_1)= \frac{Q(x_i)}{(1-x_1)(1-x_2)^2(1-x_1 x_2)(1-x_3)^3(1-x_4)^2(1-x_1 x_4)}
\label{dP1gen}
\end{equation}
where
\begin{eqnarray} Q(x_i) &=& 1- x_2 x_3-2 x_1 x_2 x_3 +x_1 x_2^2 x_3+x_1 x_2 x_3^3-x_2 x_4-2 x_1 x_2 x_4+x_1 x_2^2 x_4-x_3 x_4 \nonumber\\
& -& 2 x_1 x_3 x_4 + 2 x_2 x_3 x_4+6 x_1 x_2 x_3 x_4+2 x_1^2 x_2 x_3 x_4-2 x_1 x_2^2 x_3 x_4-x_1^2x_2^2x_3 x_4 +x_1 x_3^2 x_4 \nonumber\\ &-&
2 x_1 x_2 x_3^2 x_4 -x_1^2 x_2 x_3^2 x_4 +
 x_1 x_2 x_4^2 +x_1 x_3 x_4^2 -2 x_1 x_2 x_3 x_4^2 -x_1^2 x_2 x_3 x_4^2 +x_1^2 x_2^2 x_3^2 x_4^2 \nonumber
\end{eqnarray}



Looking at the order of the pole at $x_1=x_2=x_3=x_4\rightarrow 1$ we find that the dimension of the moduli space is $6$, in agreement with our general formula $3 N+G-1$.

\subsubsection{The GKZ decomposition.} The GKZ fan for
$dP_1$ is depicted in Figure \ref{Y21a}.
Including the internal points we have five divisors $D_i$ subject to the equivalence
relations $D_2=D_4,\, D_5=-D_2-2D_3,\, D_1=D_3-D_2$. This leaves a $I+d-3=2$-dimensional space
of K\"ahler parameters, which we can identify with the plane $D_2$, $D_3$. We have
\begin{equation}
\sum_{i=1}^5 a_i D_i = (a_3+a_1-2 a_5)D_3+(a_2-a_1+a_4-a_5)D_2 \equiv \beta D_3 +\beta^\prime D_2
\end{equation}

Even if $a_i\ge 0$, $\beta$ and $\beta^\prime$ can be negative.
There are two smooth resolutions $R1$ and $R2$.
 Assign the number $\beta$ to the vertex
$D_3$ and $\beta^\prime$ to the vertex $D_2$ of the toric diagram, and zero to all other vertices.
The convexity conditions for R1 give $\beta,\beta^\prime\ge 0$ and the conditions for R2 give
$\beta\ge 0,\, \beta^\prime \le 0,\beta+\beta^\prime \ge 0$. These two sets of conditions determine the two adjacent cones
in the GKZ decomposition. Notice that the three boundary lines can be associated naturally with
the directions $D_2$, $D_3$ and $D_1=D_3-D_2$.
We go from R1 to R2 with a flop.

Collecting this together and using Equation (\ref{loc})  we obtain

\begin{eqnarray}
g^{R1}_{1,\beta,\beta^\prime}(x_i) &=& \frac{x_2^{\beta^\prime} x_3^{\beta}}{(1-x_1 x_2/ x_3)(1-x_2 x_3^2)(1-x_4/x_2)} + \frac{x_3^{\beta}x_4^{\beta^\prime}}{(1-x_2/x_4)(1-x_1 x_4/x_3)(1-x_3^2 x_4)}
\nonumber \\
& +&  \frac{x_4^{\beta+\beta^\prime} x_1^{\beta}}{(1-x_2/ x_4)(1-x_3/x_1x_4)(1-x_4^3x_1^2)} +
 \frac{x_1^{\beta} x_2^{\beta+\beta^\prime}}{(1-x_1^2 x_2^3)(1-x_3/x_1x_2)(1-x_4/x_2)}
\nonumber \\
g_{1,\beta,\beta^\prime}^{R2}(x_i) & =&
\frac{x_3^{\beta-2\beta^\prime}}{(1-x_1/x_3^3)(1-x_2 x_3^2)(1-x_3^2 x_4)} +
\frac{x_1^{-\beta^\prime} x_3^{\beta+\beta^\prime}}{(1-x_1 x_2/ x_3)(1-x_3^3/x_1)(1-x_1x_4/x_3)}
\nonumber \\
& +&
\frac{x_1^{\beta} x_4^{\beta+\beta^\prime}}{(1-x_2/ x_4)(1-x_3/x_1x_4)(1-x_4^3x_1^2)} +
\frac{x_1^{\beta} x_2^{\beta+\beta^\prime}}{(1-x_1^2 x_2^3)(1-x_3/x_1x_2)(1-x_4/x_2)}\nonumber\\
\nonumber\\
\label{ZbbdP1}
\end{eqnarray}

\subsubsection{Multiplicities}
We determine multiplicities using the auxiliary partition functions for the GKZ
cone. We have six different equivalence classes of fields ${a,b,c,d,e,f}$ corresponding to the fields as in Table \ref{globaldP1}.
The relations are
$${\cal I}= \{ abcd, \  afd,  \ abe,  \ cd,  \ bc,  \ eaf,  \
fd-be \} $$
The first three relations correspond to closed loops in the quiver, the composition of arrows $cd$, $bc$ and $eaf$ are equivalent to the $e$,$f$ and $c$ respectively and should be set to zero in order to avoid overcounting and, finally,
$fd$ and $be$ are identified since they have the same starting and ending
point. Not all the relations are independent.

We can assign charges to the letters $a,b,c\ldots$ by considering their
representative in terms of elementary fields and using the homogeneous charge
assignment given in Table \ref{globaldP1}.
In the GKZ plane we want
to use the charges $x_2$ and $x_3$ associated with $D_2$ and $D_3$. The restriction to the GKZ
plane requires using the equivalence relations $D_4=D_2$ and $D_1=D_3-D_2$ that
translate to the restriction to $x_4=x_2$ and $x_1=x_3/x_2$.
We thus obtain the assignment

\begin{equation}
\{a, b, c, d, e, f\} \longrightarrow \{x_3, x_2, x_3/x_2, x_2, x_3, x_3\}.
\end{equation}
Using Macaulay2 to compute the Hilbert series for the polynomial ring

\begin{equation}
{\cal R}_{\rm GKZ} (dP_1) = \mathbb{C}[a,b,c,d,e,f]/{\cal I}
\label{GKZringdP1}
\end{equation}
we obtain the auxiliary GKZ partition function \footnote{The auxiliary polynomial ring of
Equation (\ref{GKZringdP1}) has in fact the same Hilbert series as for a simpler polynomial ring
$ \IC[ a,b,c,d] / \{ abcd \} $.}
\begin{equation}
Z_{\rm aux}(x_2,x_3; dP_1) = \frac{1-x_3^2 x_2}{(1-x_3)(1-x_3/x_2)(1-x_2)^2} = \sum_{\beta, \beta^\prime}
m(\beta, \beta^\prime) x_3^\beta x_2^{\beta^\prime}
\end{equation}
which is expanded in power series for $x_3<1,x_2<1,x_3/x_2<1$.
It exactly fills the regions R1 and R2 of the GKZ fan. We can also extract
the multiplicities: in the internal points of region R1 $m(\beta,\beta^\prime)=3 \beta+2 \beta^\prime$, in the internal points of region R2 $m(\beta,\beta^\prime)=3(\beta+\beta^\prime)$; at the origin $m(0,0) = 1$, on the vertical axis $m(0, \beta^\prime) = \beta^\prime +1$, on the horizontal axis $m(\beta, 0) = 3\beta$ and finally on the diagonal $m(\beta, -\beta) = 1$.

The $N=1$ generating function is a sum over the two GKZ regions:

\begin{equation}
g_1(\{ x_i\})= \sum_{\beta,\beta^\prime\in R1} m(\beta,\beta^\prime) g_{1,\beta,\beta^\prime}^{R1}(x_i) + \sum_{\beta,\beta^\prime\in R2} m(\beta,\beta^\prime) g_{1,\beta,\beta^\prime}^{R2}(x_i)
\label{GKZdP1}
\end{equation}

By resumming this formula, we obtain precisely Equation (\ref{dP1gen}).

\subsubsection{Refinement of the GKZ Lattice}



Now we add the anomalous charges according to Table \ref{globaldP1}. Macaulay2 computes the GKZ partition function to be

\be
Z_{\rm aux}(x_2, x_3, a_1, a_2; dP_1) = \frac{1- x_2 x_3^2}{(1- \frac{x_3 a_2}{x_2})(1- {x_2 a_1}{})(1- \frac{x_2}{a_1})(1-\frac{x_3}{a_2})}
\label{ZauxdP1}
\ee

\begin{figure}[ht]
\begin{center}
  \epsfxsize = 7cm
  \centerline{\epsfbox{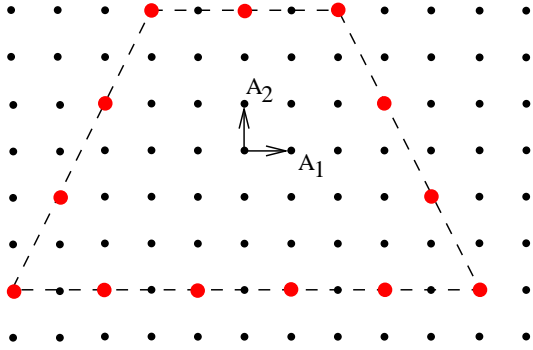}}
  \caption{The hollow trapezoid above $(\beta, \beta ^\prime)=(4,3)$ which is in region R1. It gives the multiplicity 13.}
  \label{emptydP1_1}
\end{center}
\end{figure}

\begin{figure}[ht]
\begin{center}
  \epsfxsize = 7cm
  \centerline{\epsfbox{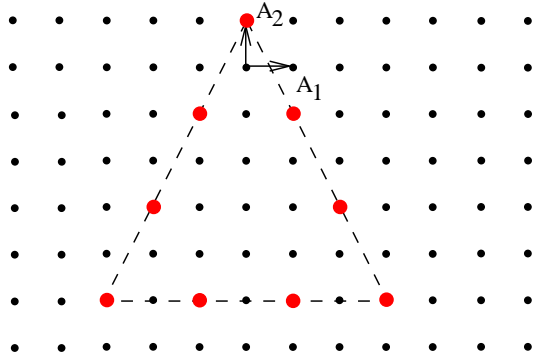}}
  \caption{The hollow triangle above $(\beta,\beta^\prime)=(6,-1)$ which is in region R2. It gives the multiplicity 9.}
  \label{emptydP1_2}
\end{center}
\end{figure}

The expansion gives a hollow cone in the four dimensional $(B,B^\prime, A_1, A_2)$ lattice. In the
R1 region, we have a trapezoid $C_{R1}(\beta,\beta^\prime)$ in the $(A_1, A_2)$ lattice above a point in the GKZ cone
parameterized by $(\beta,\beta^\prime)$ (see \fref{emptydP1_1}). This degenerates to a triangle $C_{R2}(\beta,\beta^\prime)$ as we
move to the R2 region (\fref{emptydP1_2}). This ``explains'' the $m(\beta,\beta^\prime)=2\beta^\prime+3\beta$ multiplicities.


\begin{figure}[ht]
\begin{center}
  \epsfxsize = 3cm
  \centerline{\epsfbox{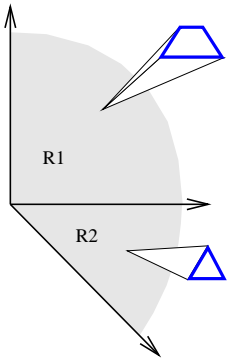}}
  \caption{The two cones (grey areas) of the $dP_1$ GKZ fan. Above the points in $R1$ a trapezoid sits in the fiber. The edge lengths are controlled by the position in the base. The trapezoid degenerates to a triangle
  in the $R2$ region.}
  \label{dP1_r1r2}
\end{center}
\end{figure}

\begin{figure}[ht]
\begin{center}
  \epsfxsize = 3cm
  \centerline{\epsfbox{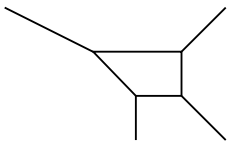}}
  \caption{The pq-web of $dP_1$. The blown-up four-cycle is associated to the trapezoid in the middle.}
  \label{dP1_pq}
\end{center}
\end{figure}


The $N=1$ decomposition in Equation (\ref{GKZdP1}) can be refined to

\begin{equation}
g_1(x_i,a_1,a_2; dP_1)= \sum_{\beta,\beta^\prime\in R1} \left(\sum_{K\in C_{R1}(\beta,\beta^\prime)} a_1^{K_1} a_2^{K_2} \right) g_{1,\beta,\beta^\prime}^{R1}(x_i) + \sum_{\beta,\beta^\prime\in R2} \left(\sum_{K\in C_{R2}(\beta,\beta^\prime)} a_1^{K_1} a_2^{K_2} \right) g_{1,\beta,\beta^\prime}^{R2}(x_i)
\end{equation}


where the generating function for $N=1$ depending on all charges is:
\begin{eqnarray}
& & g_1(x_1,x_2,x_3,x_4,a_1,a_2;dP_1)= \nonumber\\
& &  \hspace{-0.9cm} \frac{Q(x_1,x_2,x_3,x_4,a_1,a_2)}{(1 - a_2 x_1)(1 - \frac{x_2}{a_1})(1 - a_1 x_2)(1 - \frac{x_1 x_2}{a_2})(1 - \frac{x_3}{a_2})(1 - \frac{a_2 x_3}{a_1})(1 -  a_1 a_2 x_3)(1 - \frac{x_4}{a_1})(1 - a_1 x_4)(1-\frac{x_1 x_4}{a_2})}\nonumber\\
\end{eqnarray}
where $Q(x_1,x_2,x_3,x_4,a_1,a_2)$ is the polynomial:
\begin{eqnarray}
& & Q(x_1,x_2,x_3,x_4,a_1,a_2)= 1 -
    a_2 x_2 x_3 - \frac{x_1 x_2 x_3}{a_1} - a_1 x_1 x_2 x_3 + x_1 x_2^2 x_3 + a_2 x_1 x_2 x_3^2 - x_2 x_4 - \nonumber\\
& & \frac{x_1 x_2 x_4}{a_1 a_2} - \frac{a_1 x_1 x_2 x_4}{a_2} + \frac{x_1 x_2^2 x_4}{a_2} - a_2 x_3 x_4 - \frac{x_1 x_3 x_4}{a_1} - a_1 x_1 x_3 x_4 + \frac{a_2 x_2 x_3 x_4}{a_1} + a_1 a_2 x_2 x_3 x_4 + \nonumber\\
& &  4 x_1 x_2 x_3 x_4 + \frac{x_1 x_2 x_3 x_4}{a_1^2} + a_1^2 x_1 x_2 x_3 x_4 + \frac{x_1^2 x_2 x_3 x_4}{a_1 a_2} + \frac{a_1 x_1^2 x_2 x_3 x_4}{a_2} - \frac{x_1 x_2^2 x_3 x_4}{a_1} - a_1 x_1 x_2^2 x_3 x_4 - \nonumber\\
& & \frac{x_1^2 x_2^2 x_3 x_4}{a_2} + a_2 x_1 x_3^2 x_4 - \frac{a_2 x_1 x_2 x_3^2 x_4}{a_1} - a_1 a_2 x_1 x_2 x_3^2 x_4 - x_1^2 x_2 x_3^2 x_4 + \frac{x_1 x_2 x_4^2}{a_2} + x_1 x_3 x_4^2 -\nonumber\\
& & \frac{x_1 x_2 x_3 x_4^2}{a_1} - a_1 x_1 x_2 x_3 x_4^2 - \frac{x_1^2 x_2 x_3 x_4^2}{a_2} + x_1^2 x_2^2 x_3^2 x_4^2
\end{eqnarray}

This expression can be rewritten in a more symmetric form by using Equations (\ref{ZbbdP1}) and (\ref{ZauxdP1}) and some algebraic manipulation,

\begin{eqnarray}
g_1(x_1, x_2, x_3, x_4, a_1, a_2; dP_1)
&=& \frac{1}{(1- x_1 a_2)(1- x_2 a_1)(1- \frac{x_2}{a_1})(1-\frac{x_1 x_2}{a_2})(1- \frac{x_3}{x_1x_2})(1-\frac{x_4}{x_2})}  \nonumber \\ \nonumber
&+& \frac{1}{(1- \frac{x_1 x_2}{x_3})(1- \frac{x_3 a_2}{x_2})(1- {x_2 a_1}{})(1- \frac{x_2}{a_1})(1-\frac{x_3}{a_2})(1-\frac{x_4}{x_2})} \\ \nonumber
&+& \frac{1}{(1-\frac{x_2}{x_4})(1-\frac{x_1 x_4}{x_3})(1- \frac{x_3 a_2}{x_4})(1- x_4 a_1)(1- \frac{x_4}{a_1})(1-\frac{x_3}{a_2})} \\ \nonumber
&+& \frac{1}{(1-\frac{x_2}{ x_4})(1-\frac{x_3}{x_1 x_4})(1- x_1 a_2)(1- x_4 a_1)(1- \frac{x_4}{a_1})(1-\frac{x_1 x_4}{a_2})} \\ \nonumber
&+& \frac{1}{(1 - a_2 x_1)(1 - \frac{x_2}{a_2 x_3})(1 - \frac{x_3}{a_2})(1 - \frac{a_2 x_3}{a_1})(1 - a_1 a_2 x_3)(1
- \frac{x_4}{a_2 x_3})} \, .
\nonumber
\end{eqnarray}

As for $\mathbb{C}^3/Z_3$ and $\mathbb{F}_0$, also the $N=1$ generating function
for $dP_1$ can be written in a form that recalls a localization formula for the
$N=1$ field theory moduli space, which is a six dimensional variety acted by
a total of six flavor and baryonic symmetries. It would be interesting to
investigate the general properties of the $N=1$ moduli space varieties and
to see in particular whether these varieties are toric and the previous
formulae can be interpreted as a localization.

\subsubsection{Generating functions for $N>1$}

The generating functions for $N>1$ can be obtained as usual by plethystic
exponentiation and resummation over the points of the considered decomposition,
according to the general formula (\ref{g1plet}). Using the GKZ decomposition, with
the multiplicities already obtained for the $N=1$ case, the generating functions
for general $N$ are given by:

\begin{equation}
\sum_{N=0}^\infty g_N(\{ x_i\}) \nu^N = \sum_{\beta,\beta^\prime\in R1}
m_1(\beta,\beta^\prime) {\rm PE}_\nu [g_{1,\beta,\beta^\prime}^{R1}(x_i)] +
\sum_{\beta,\beta^\prime\in R2} m_2(\beta,\beta^\prime) {\rm PE}_\nu
[g_{1,\beta,\beta^\prime}^{R2}(x_i)] \label{dp1n}
\end{equation}

For the case $N=2$ we compute:

\begin{equation}
g_2(t,b) = \frac{F(t,b)}{\left(b^6-1\right) \left(b^4-t^2\right)^4
 \left(b^2 t^2-1\right)^5 \left(t^3-1\right)^3 \left(b^6
   t^3-1\right)}
\label{g2dp1}
\end{equation}
\[
\begin{array}{l}
F(t,b) = b^8 \left(t^{14} \left(-t^6+6 t^3+3\right) b^{22}+
   t^{12} \left(4 t^6-21 t^3-15\right) b^{20}
  +t^{10}  \left(-2 t^{12}-6 t^9+5 t^6 \right. \right. \\ \left. \left.
 + 3 t^3+32\right) b^{18}
 -t^5 \left(3 t^{15}-61 t^{12}+39 t^9-24 t^6+16
   t^3+3\right) b^{16}-t^6 \left(9 t^{15}-30 t^{12}+85 t^9
 \right. \right. \\ \left. \left.
 +7 t^6-30 t^3+7\right) b^{14}+
 \left(-7 t^{22}-6 t^{19}+25 t^{16}-32 t^{13}+67 t^{10}-48 t^7+t^4\right)
 b^{12}+t^2 \left(-5 t^{18}\right. \right. \\ \left. \left.
 +64 t^{15}-91 t^{12}+48 t^9-29 t^6+6 t^3+7\right) b^{10}+\left(-t^{21}+14
   t^{18}-53 t^{15}+44
 t^{12}+56 t^9 -19 t^6 \right. \right. \\ \left. \left.
 +6 t^3+1\right) b^8+t^4 \left(3 t^{15}+15 t^{12}-17 t^9+24 t^6-48 t^3-1\right)
   b^6 +t^2 \left(4 t^{15}-44 t^{12}+9
   t^9-9 t^6 \right. \right. \\ \left. \left.
 +6 t^3+2\right) b^4+t^9 \left(-4 t^6+27 t^3+9\right) b^2+t^7
 \left(t^6-6 t^3-3\right)\right)
\end{array}
\]

\vspace{1em}

We explicitly checked for the considered cases that the dimension of the moduli space, equal to
the order of the pole for the generating function when $t$ and $b$ approach 1, is $3N+G-1$ as
expected.

The generators of the moduli space and their relations can be studied by
computing the plethystic logarithm of the generating functions. In the $N=1$
case we checked that the plethystic logarithm correctly reproduces
the 10 chiral fields generators together with their 9 F-term relations (recall
that we get only one relation (\ref{fterm1}) from the two (\ref{fterm})).

The case $N=2$ for $dP_1$ is more interesting. The first terms in the
expansion of the plethystic logarithm of $g_2(t,b)$ in (\ref{g2dp1}) are:
\begin{equation}
b^6 + \left( \frac{6}{b^4} + 12 b^2 \right) t^2 + \left( 9+b^6 \right) t^3
- \ldots
\end{equation}
The first positive terms can be matched with 29 generators of the moduli space
for $N=2$: in fact at level $t^3$ we find the 9 known mesons of $dP_1$ (this
matches with the number of generators over the integers for the dual fan),
and 20 baryonic generators. The baryonic generators can be identified as
follows:
\[
\begin{array}{l}
b^6 \rightarrow  ({\bf Z},{\bf Z}) \qquad \quad
 t^3 b^6 \rightarrow ({\bf Y_1 Y_2 Y_3},{\bf Z})\\[1em]
\displaystyle \frac{6t^2}{b^4}
\rightarrow  ({\bf U}_1, {\bf U}_1),({\bf U}_1, {\bf
  U}_2),({\bf U}_2, {\bf U}_2), \, \,
 ({\bf \tilde U}_1, {\bf \tilde U}_1),({\bf
  \tilde U}_1, {\bf \tilde U}_2),({\bf \tilde U}_2, {\bf \tilde U}_2) \\[1.5em]
12 b^2 t^2 \rightarrow
\begin{array}{l}
 ({\bf Y_1},{\bf Y_1}), ({\bf Y_2},{\bf Y_2}), ({\bf
    Y_3},{\bf Y_3}), \,\,  ({\bf V_1},{\bf V_1}),  ({\bf V_1},{\bf V_2}),
    ({\bf V_2},{\bf V_2}) \\[0.3em]
  ({\bf Y_2},{\bf V_1}), ({\bf Y_2},{\bf V_2}),
\,\, ({\bf Z U_1},{\bf Y_1}), ({\bf Z U_2},{\bf Y_1}), \, \,
({\bf \tilde U_1 Z},{\bf Y_3}),({\bf \tilde U_2 Z},{\bf Y_3})
\end{array}
\end{array}
\]
where $({\bf X},{\bf Y})$ stands for the $N=2$ color indices contraction:
$\epsilon_{i,j} \epsilon^{a,b} {\bf X}_a^i {\bf Y}_b^j$.

\section {The Molien formula: checks for $N>1$ }

The baryonic generating functions found in the previous sections can be checked against an
explicit field theory computation, at least for small values of $N$. The problem of finding
invariants under the action of a continuous group is the hearth of invariant theory and goes
back to the nineteenth  century, as most of the concepts necessary for its solution, like
syzygies and free resolutions, all amenable to Hilbert. Modern advances, such as the discovery
of Groebner basis, gave an algorithmic way of solving such problems and the advent of computer
algebra programs made some computations really doable.

The generating functions for a fixed number of colors $N$ can be reduced to a problem for polynomial rings as follows. Consider an $\mathcal{N}=1$ supersymmetric gauge theory with $F$ elementary fields $X$ and a gauge group ${\cal G}$. Since we are discussing the chiral ring we can replace ${\cal G}$ with its complexification ${\cal G}_c$.
For quiver theories, the elementary fields consist of $N\times N$ matrices.
Consider a polynomial ring in $F N^2$ variables $\mathbb{C}[X_{ij}]$
made with the entries of these matrices. The F-terms give matrix relations
whose entries are polynomial equations. We can collect all the polynomial
F-term equations in an ideal $\cal I$ and define the quotient ring
\begin{equation}
{\cal R}[X_{ij}]=\mathbb{C}[X_{ij}]/\cal I
\end{equation}
The gauge group ${\cal G}_c$ and the global symmetry group act naturally on the ring and we can grade
the elements of ${\cal R}$ with gauge and global charges. Denoting with $t_i$ the global Abelian
charges and with $z_i$ the charges under the Cartan subgroup of the gauge group ${\cal G}_c$, we can
write the generating function, or Hilbert series, of the graded ring ${\cal R}$,
\begin{equation}
H_{\cal R}(t;z)=\sum_{nm} a_{nm} z^n t^m \label{ser}
\end{equation}
which can be arranged to be a power series in the global charges $t$ and a
Laurent expansion in the gauge charges $z$. The full gauge group ${\cal G}_c$
acts on the quotient ring ${\cal R}$ and, since the gauge symmetry commutes with
the global symmetry, all the elements of ${\cal R}$ with given charge $t^m$ form
a (not necessarily irreducible) representation of ${\cal G}_c$. Therefore,
the coefficient of $t^n$ in Equation (\ref{ser}) is the
character of a ${\cal G}_c$ representation,

\begin{equation}
H_{\cal R}(t;z)=\sum_{m=0}^\infty \chi^m(z) t^m =\sum_{m=0}^\infty \left(\sum_i a^m_i \chi^{(i)}(z)\right) t^m
\end{equation}

Here we have denoted with $\chi^{(i)}$ the irreducible representations of
${\cal G}_c$ and decomposed the representation on the elements of charge $t^m$ into
irreducible ones. The generating function for invariants is given by the
projection onto the trivial representation with character $\chi^{(0)}=1$,

\begin{equation}
H_{\cal R}^{inv}(t)=\sum_{m=0}^\infty  a^m_0  t^m
\end{equation}

The projection can be easily done by averaging $H(t;z)$ on the gauge group
with the Weyl measure. The latter has indeed the property to keep only
the contribution of the trivial representation

\begin{equation}
\int d\mu(z) \chi^{(i)}(z) =\delta_{i,0}
\end{equation}

For a given group $G$ with rank $r$ we can explicitly write the measure as
a multi-contour integral

\begin{equation}
\frac{1}{|W|}\prod_{j=1}^r \int_{|z_j|=1} \frac{dz_j}{2\pi i z_j} (1-z^{h(\alpha)})
\end{equation}
where $h(\alpha)$ are the weights of the adjoint representation and $|W|$ is
the order of the Weyl group.
We finally get the Molien formula:
\begin{equation}
H_{\cal R}^{inv}(t)= \frac{1}{|W|}\prod_{j=1}^r \int_{|z_j|=1} \frac{dz_j}{2\pi i z_j} (1-z^{h(\alpha)}) H_{\cal R}(t;z)
\label{Molien}
\end{equation}

Since the multi-contour integrals can be evaluated with the residue theorem, the real problem in
using Equation (\ref{Molien}) is the determination of the integrand, that is the Hilbert series
of the quotient ring ${\cal R}$. Fortunately, this is the kind of problems that modern commutative
algebra made algorithmic and that can be easily solved with computer algebra programs. For
example Macaulay2 naturally deals with polynomial rings and it has a build-in command {\it
hilbertSeries}. For moderate values of $F N^2$, the computation takes fractions of second, but
it can become too hard with a standard computer already at $N=3$ and a number of fields $F$ of order 10. In these cases, one can still truncate the computation at a maximum degree in t and get a sensible
result.

It is worth mentioning that special care has to be taken when the moduli space is not irreducible and at certain points in moduli space new branches are opening up. This is the case for example for ${\cal N}=2$ theories that have additional Coulomb branches and was treated in \cite{Hanany:2006uc}.
The general case that we want to address is the following. Suppose that we have an $N=1$ F-term equation where one of the elementary fields can be factorized: $X_0 F(X)=0$ where $F(X)$ is a polynomial not containing $X_0$. Considering the $N=1$ moduli space as a fibration over the line parametrized by $X_0$, we see that the dimension of the fiber increases by one unit over $X_0=0$: indeed, for $X_0\ne 0$ we can impose the further constraint $F(X)=0$ which reduces by one the dimension of the fiber. This means that a new branch opens up at $X_0$ and the full moduli
space is reducible. This is the case for $\mathbb{F}_0$ and $dP_1$ as
discussed above in detail. We may want to determine the generating function
for a given irreducible component of the moduli space, in particular
the closure of the open set $X_0\ne 0$, or, for $N>1$, of $\det(X_0)\ne 0$.
This is done with a standard trick. Add a new element $q$ to the ring and
a new equation, $q \det(X_0)-1$ to the ideal $\cal I$. Clearly, the new equation
prevents $\det(X_0)$ from being zero. The irreducible component of the
moduli space we are interested in is obtained by projecting the variety
defined by the new ideal $\tilde{\cal I}=({\cal I},q\det(X_0)-1)$ on the space parameterized
by the $X,X_0$ and taking the closure. This can be done by eliminating $q$
from the ideal $\tilde{\cal I}$. This defines the elimination ideal $\cal J$ that
can be computed with the Macaulay2 command {\it eliminate}.
If we define

\begin{equation}
{\cal R}[X_{ij}]=\mathbb{C}[X_{ij}]/\cal J
\end{equation}
we can now proceed as before, compute the Hilbert series of this ring
and project it onto gauge invariants with the Molien formula. This would
give us the generating function for the particular irreducible component
of the moduli space.

We now present some explicit examples based on the conifold and $\mathbb{F}_0$. The other cases
presented in this paper can be checked similarly, at least for small values of number of fields,
$F$, and number of D-branes, $N$. When $N$ increases it is necessary to truncate the series to a
maximum degree.

\subsection{Example: $N=2$ for the conifold}

The generating function of the conifold for $N=2$ was explicitly computed in
\cite{Forcella:2007wk} and is given in Equation (\ref{conifoldN2}). For $N=2$ we have four
fields ${\bf A}_i,{\bf B}_i$ that are two-by-two matrices, whose entries we denote by
$a_i^{pq},b_i^{pq}$. The four matrix F-term equations
$${\bf A}_1 {\bf B}_i {\bf A}_2={\bf A}_2 {\bf B}_i {\bf A}_1\, ,\qquad\qquad\qquad {\bf B}_1 {\bf A}_i {\bf B}_2={\bf B}_2 {\bf A}_i {\bf B}_1$$
give rises to sixteen polynomial equations for the $a_i^{p,q},b_i^{p,q}$
which generates an ideal $\cal I$ in the polynomial ring $\mathbb{C}[a_i^{p,q},b_i^{p,q}]$.
The element $(g,\bar g)$ of the
complexified gauge group $SL(2)\times SL(2)$ acts on the matrices
as ${\bf A}_i\rightarrow g {\bf A}_i \bar g^{-1}$ and ${\bf B}_i\rightarrow \bar g {\bf B}_i g^{-1}$.
All the entries $a_i^{p,q}$ and $b_i^{p,q}$ transform with a definite
charge under the Cartan subgroup

\begin{eqnarray}
g &=& \left(\begin{array}{ll} z & 0 \\ 0 & 1/z \end{array}\right ) \nonumber\\
\bar g &= & \left(\begin{array}{ll} w & 0 \\ 0 & 1/w \end{array}\right )
\end{eqnarray}

We further assign chemical potential $t_1$ to the eight fields  $a_i^{p,q}$
and chemical potential $t_2$ to the eight fields  $b_i^{p,q}$. The sixteen F-term
constraints generating the ideal $\cal I$
transform homogeneously under the gauge and global charges.
We can thus grade the quotient ring

\begin{equation}
{\cal R}_{N=2} = \mathbb{C}[a_i^{p,q},b_i^{p,q}]/\cal I
\end{equation}
with four charges, corresponding to chemical potentials, two gauge $z,w$ and two global
$t_1,t_2$. The Hilbert series of ${\cal R}_{N=2}$ can be computed using Macaulay2

\begin{equation}
H_{\cal R}(t_1, t_2; z, w)=\frac{P(t_1,t_2;z,w)}{(1-t_1 zw)(1-t_1 \frac{z}{w})(1-t_1 \frac{w}{z})(1-\frac{t_1}{zw})(1-t_2 zw)(1-t_2 \frac{z}{w})(1-t_2 \frac{w}{z})(1-\frac{t_2}{zw})}
\end{equation}

\begin{eqnarray}
P(t_1,t_2;z,w)&=&1+4 t_1^3 t_2+4 t_1 t_2^3+ 6 t_1^2 t_2^2 + t_1^4 t_2^4-2(t_2 t_1^2+t_1 t_2^2+t_2^2 t_1^3+t_1^2 t_2^3) \left(w + \frac{1}{w}\right) \left(z + \frac{1}{z}\right) 
\nonumber\\
&+&t_1^2 t_2^2 \left(w + \frac{1}{w}\right)^2 \left(z + \frac{1}{z}\right)^2
\end{eqnarray}

The molien formula now reads

\begin{equation}
g_2(t_1,t_2; {\cal C})=\int_{|w|=1}\frac{dw (1-w^2)}{2\pi i w} \int_{|z|=1}\frac{dz (1-z^2)}{2\pi i z}  H_{\cal R}(t_1,t_2;z,w)
\end{equation}

Some attention should be paid in performing the contour integrals. Recall that $H_{\cal R}$ gives the
generating function for the ring ${\cal R}$ when expanded in power series in $t_1$ and in $t_2$ which are supposed to be complex numbers of modulus less than one. This should be taken into account
when performing the contour integrals on the unit circles $|z|=|w|=1$. For example, the first
contour integration in $z$ takes contribution only from the residues in the points $t_1 w, t_1/w,
t_2 w, t_2/w$ lying inside the unit circle $|z|=1$ (we take $|t_i|<1,|w|=1$). Similar arguments
apply to the second integration. After performing the two integrals we obtain
\begin{equation}
g_2(t_1,t_2)= \frac{1 + t_1 t_2 + t_1^2 t_2^2 - 3 t_1^4 t_2^2 - 3 t_1^2 t_2^4 + t_1^5 t_2^3 + t_1^3 t_2^5  - 3 t_1^3 t_2^3 + 4 t_1^4 t_2^4}{(1 - t_1^2)^3(1 - t_1 t_2)^3 (1 - t_2^2)^3} ,
\end{equation}
which perfectly coincides with Equation (\ref{conifoldN2}).

\subsection{Example: $N=1$ and $N=2$ for $\frac{3}{4}\mathbb{F}_0$ - reducibility of the moduli space}

We now consider an example where the moduli space is not irreducible.
We consider the $\frac{3}{4} \mathbb{F}_0$ case in order to limit the number
of equations involved. The following discussion applies to $\mathbb{F}_0$ and
$dP_1$ as well.
For $N=1$ we consider the polynomial ring

\begin{equation}
{\cal R}
 = \mathbb{C}[{\bf A}_i,{\bf B}_1,{\bf C}_i,{\bf D}_1] / \cal I
\end{equation}
with six variables. There are two F-term equations
$$ {\cal I}=( \, {\bf A}_1 {\bf B}_1 {\bf C}_2 = {\bf A}_2 {\bf B}_1 {\bf C}_1 \,\,\,  ,  {\bf C}_1 {\bf D}_1 {\bf A}_2 = {\bf C}_2 {\bf D}_1 {\bf A}_1 \, ) $$
The Hilbert series for this polynomial ring is

\begin{equation}
H_{\cal R}(t_1,t_2)=\frac{1-2 t_1^2 t_2+t_1^2 t_2^2}{(1-t_1)^4(1-t_2)^2} .
\end{equation}

As already discussed, the variety defined by $\cal I$ is not irreducible:
we are interested in the closure of the open set ${\bf B}_1,{\bf D}_1\ne 0$.
We then define a new ideal by adding two new variables $q_1,q_2$ to ${\cal R}$,

\begin{equation}
\tilde {\cal R}= \mathbb{C}[{\bf A}_i,{\bf B}_1,{\bf C}_i,{\bf D}_1, q_1,q_2] / \tilde{\cal I}
\end{equation}
and two new generators to the ideal $\cal I$

$$ \tilde{\cal I} =({\cal I}, q_1 {\bf B}_1-1, q_2 {\bf D}_1-1)$$
The closure of the open set ${\bf B}_1,{\bf D}_1\ne 0$ is obtained
by eliminating $q_1$ and $q_2$. This can be done in a polynomial
way by using the Groebner basis and the algorithm
 is implemented in Macaulay2 in the command {\it eliminate}. In our
case the elimination ideal is just
$${\cal J}=(\, {\bf A}_1 {\bf C}_2 - {\bf A}_2 {\bf C}_1 \, ) $$
and the Hilbert series of

\begin{equation}
{\cal R}^\prime = \mathbb{C}[{\bf A}_i,{\bf B}_1,{\bf C}_i,{\bf D}_1] / \cal J
\end{equation}
is

$$H_{\cal R^\prime}(t_1,t_2)= \frac{1+t_1}{(1-t_1)^3(1-t_2)^2}$$
which indeed coincides with the $g_1(t_i, \frac{3}{4} \mathbb{F}_0)$ generating
function given in Equation (\ref{threequarter}). The $N=2$ generating function
should be computed in a similar way. The fields ${\bf A}_i,{\bf B}_1,{\bf C}_i,{\bf D}_1$ are now two-by-two matrices, for a total of $24$ independent entries.
The ideal $\cal I$ now contains $8$ polynomial equations, given by
$$ \tilde{\cal I} =({\cal I}, q_1 \det {\bf B}_1 -1, q_2 \det {\bf D}_1-1) ,$$
and the elimination ideal $\cal J$ is obtained by {\it eliminating} $q_1$ and $q_2$. The Hilbert
series of $\cal J$ graded with the gauge charges is a rational function $H_{\cal
R}(t_1,t_2;z_1,z_2,z_3,z_4)$ whose expression is too long to be reported here. Four integrations
using the residue theorem finally give the $N=2$ generating function given in
(\ref{threequarter}).




\section{Conclusions}
In this paper we have performed a further step in the understanding and the computation of the
generating functions for the chiral ring of the superconformal gauge theories living on branes
at CY singularities. We have reinforced the conjecture that the generating functions for $N$
colors can be computed simply in terms of the $N=1$ generating functions through the plethystic
program. We tested this conjecture, suggested by a computation with D3-branes in the dual
background, in field theory for small values of $N$. It would be interesting to perform checks
for large $N$ as well as to investigate the statistical properties of the resulting generating
functions.

In particular we have made an explicit investigation of the properties of the complete $N=1$
generating function and we have compared the result with a geometrical computation. The emerging
structure reveals once more the deep interplay between the quiver gauge theory and the algebraic
geometry of the CY. In particular we found an intriguing relation between the decomposition of
the $N=1$ generating function in sector of given baryonic charge and the discretized K\"ahler
moduli space of the CY.

The geometrical structure of the complete moduli space for $N$ colors, which is
obtained by the $N$-fold symmetrized product of the CY by adding the baryonic
directons is still poorly understood. We have seen that already for $N=1$
the moduli space is rich and interesting. We leave for future work the
understanding of the geometric structure of these moduli spaces.

\paragraph{Acknowledgments}

We would like to thank Freddy Cachazo, Bo Feng, Yang-Hui He and Sergio Benvenuti for useful discussions. A.Z. thanks the Laboratoire de Physique Theorique
in Jussieu and the
Galileo Galilei Institute for Theoretical Physics for the hospitality during the completion of
this work. D.F and A.Z. are supported in part by INFN and MURST under contract 2005-024045-004
and 2005-023102 and by the European Community's Human Potential Program MRTN-CT-2004-005104 and
D.F also by the European Superstring Network MRTN-CT-2004-512194. D.V. is supported by the U.S.
Department of Energy under cooperative research agreement \#DE-FC02-94ER40818. D.F. would like
to thank Constantin Bachas, Raphael Benichou, Umut Gursoy, Bernard Julia, Bruno Machet, Ruben
Minasian, Michela Petrini, Boris Pioline, Giuseppe Policastro and Jan Troost for useful
discussions and for their wonderful hospitality in Paris. Even more D.F. heartily acknowledge
Gianna, Franco, Stefano, Roberto and Nathalie for make him feeling at home in Paris.

\appendix
\section{Appendix: Singular ${\cal N}=2$ horizons}

In this appendix we discuss the ${\cal N}=2$ supersymmetric theories obtained at orbifold
singularities of the form $\mathbb{C}^2/\mathbb{Z}_n\times \mathbb{C}$ and we make some general
observations about the corresponding generating functions. Since the factor $\mathbb{C}$ in the
geometry factorizes, we can immediately write the following expression,
\begin{equation}
g_1 \hspace{-0.1cm} \left(\{ t_i \}, t; \ \frac{\mathbb{C}^2}{\mathbb{Z}_n} \times \mathbb{C}
\right)= \frac{1}{1-t} \cdot  g_1(\{ t_i \};  \mathbb{C}^2/\mathbb{Z}_n   ) \label{tota}
\end{equation}
where $t$ is a chemical potential for the $\mathbb{C}$ factor and $t_i$ a set of chemical
potential for the four dimensional singularity $\mathbb{C}^2/\mathbb{Z}_n$. We count BPS
operators on the Higgs branch of the theory without including mixed Higgs-Coulomb branches.

\subsection{$\mathbb{C}^2/\mathbb{Z}_2$}
The case of the CY singularity $\mathbb{C}^2/\mathbb{Z}_2\times \mathbb{C}$ was considered in
detail in \cite{Forcella:2007wk}. The theory has ${\cal N}=2$ supersymmetry with two vector
multiplets in the adjoint representation of $SU(N)$ and two bi-fundamental hypermultiplets. In
${\cal N}=1$ notation we have six chiral multiplets denoted as ${\bf \phi}_1, {\bf \phi}_2, {\bf
A}_1, {\bf A}_2, {\bf B}_1, {\bf B}_2$, with a superpotential
\begin{equation}
W = {\bf \phi}_1 ({\bf A}_1 {\bf B}_1 - {\bf A}_2 {\bf B}_2) + {\bf \phi}_2 ({\bf B}_2 {\bf A}_2
- {\bf B}_1 {\bf A}_1)
\end{equation}

\begin{figure}[ht]
\begin{center}
  \epsfxsize = 11cm
  \centerline{\epsfbox{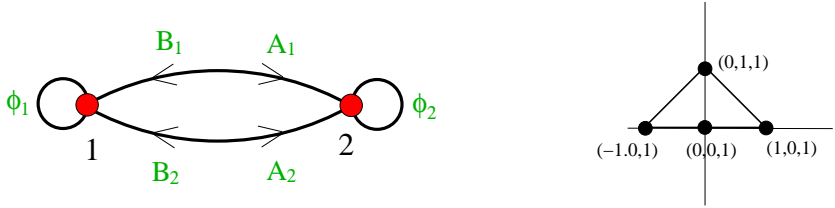}}
  \caption{Quiver and toric diagram for $\mathbb{C}^2/\mathbb{Z}_2\times \mathbb{C}$.}
  \label{c2z2quiver}
\end{center}
\end{figure}

Let us consider the $N=1$ generating function, which is given by
\begin{equation}
g_1(t_1,t_2,x;  \frac{\mathbb{C}^2}{\mathbb{Z}_2} ) = \frac{1-t_1 t_2}{(1- t_1 x)(1-\frac{t_1}{x})(1-t_2 x)(1-\frac{t_2}{x})}
\label{g1C2Z2}
\end{equation}
This function describes a three dimensional moduli space which is a  complete intersection,
generated by four fields satisfying one F-term relation ${\bf A}_1 {\bf B}_1 = {\bf A}_2 {\bf
B}_2$. We assign chemical potential $t_1=t b$ to ${\bf A}_i$ and $t_2=t/b$ to ${\bf B}_i$
where $b$ measures the baryonic charge. The parameter $x$ counts the difference between indices
1 and 2. Let us introduce two more chemical potentials, $q_1=t x$ and $q_2 = t/x$. These
potentials are natural conjugate variables for the two coordinates of $\mathbb{C}^2$. They count
the number of fields which descend by the orbifold action from the first (second) adjoint field,
respectively.

Similarly to the conifold case, we can expand the $N=1$ generating function in baryonic charges
\begin{eqnarray}
g_1(t_1,t_2; \ \frac{\mathbb{C}^2}{\mathbb{Z}_2} ) &=& \sum_{B=-\infty}^\infty b^B g_{1,B}(t;\  \frac{\mathbb{C}^2}{\mathbb{Z}_2} ) \nonumber \\
g_{1,B}(t,x; \ \frac{\mathbb{C}^2}{\mathbb{Z}_2} ) &=& \frac{t^{|B|} x^{|B|} } { (1 -
\frac{1}{x^2}) (1-t^2 x^2)} + \frac{t^{|B|} x^{-|B|}} { (1 - x^2)(1-\frac{t^2}{x^2}) }
\label{poi}
\end{eqnarray}
The same result can be obtained by localization and by using the auxiliary GKK
partition function which reproduces two one-dimensional cones corresponding
to the two Weyl chambers of $SU(2)$.



\subsubsection{$\mathbb{C}^2/\mathbb{Z}_2$  as sum over Young tableaux}
The baryonic generating functions for $\mathbb{C}^2/\mathbb{Z}_2$ can be given an interpretation
as sum over Young tableaux.

There is indeed an intriguing relation between the BPS partition functions for CY of the form $X\times
\mathbb{C}$ and Nekrasov's partition function for ${\cal N}=2$ $U(1)$ gauge theories defined on
the surface $X$. The relation is defined by the following identity
\cite{Nakajima:2003pg,Fujii:2005dk,Noma:2006pe},
\begin{equation}
\sum_{N=1}^\infty \nu^N {\rm Ch} H^0(S^N(X),{\cal O}) = \exp\left ( \sum_{k=1}^\infty \frac{\nu^k g_{1,0}(t_i^k; X)}{k} \right ) = \prod_I Z(q_1^I,q_2^I,\nu;\mathbb{C}^2)
\label{mes}
\end{equation}
In this identity, $g_{1,0}(t_i)$ is the mesonic $N=1$ generating function which is also the
partition function of holomorphic functions on $X$. The first equality in Equation (\ref{mes})
is precisely the statement that the mesonic BPS operators count holomorphic functions on the
$N$-fold symmetric product of $X$. The last equality follows from the computation of $g_{1,0}$
in terms of localization
\begin{equation}
g_{1,0}=\sum_I \frac{1}{(1-q_1^I)(1-q_2^I)}
\end{equation}
Here $I$ labels the fixed points and $(q_1^I,q_2^I)$ are the weights for the
$T^2$ action in a smooth resolution of $X$.
Since the $g_{1,0}$ partition function decomposes as the sum over elementary
partition functions for copies of $\mathbb{C}^2$, the last equality follows.

Equation (\ref{mes}) can be reinterpreted as the K-theory version of Nekrasov's $U(1)$ partition
function for the case of a surface $X$ \cite{Noma:2006pe} and
written in terms of Young tableaux. Indeed the partition function for $\mathbb{C}^2$ $Z(q_1,q_2,\nu ; \ \mathbb{C}^2)$ can be written as a sum over Young
tableaux
\begin{equation}
Z(q_1,q_2,\nu ;\mathbb{C}^2)= \sum_{Y} \frac{\nu^{|Y|}}{\prod_{s\in Y}(1-q_1^{-l(s)}q_2^{a(s)+1})(1-q_1^{1+l(s)})(1-q_2^{-a(s)})}
\end{equation}

There is a similar result for baryonic partition functions. Since we can write
\begin{equation}
g_{1,B}=\sum_I \frac{q_0^I}{(1-q_1^I)(1-q_2^I)}
\end{equation}
we have the following expression,
\begin{equation}
\exp\left ( \sum_{r=1}^\infty \frac{\nu^r g_{1,B}(t_i^r;X)}{r} \right )= \sum_{N=1}^\infty \nu^N
{\rm Ch} H^0(S^N(X),{\cal O}(B)) = \prod_I Z(q_1^I,q_2^I,\nu q_0^I;\mathbb{C}^2).
\end{equation}
This expresses the generating function as an expansion in products of series over Young
tableaux. For the case of orbifolds, we expect that this simplifies to a sum over a single set
of tableaux as in \cite{Noma:2006pe}.

The identity seems to be related to the blow-up formula of Nakajima \cite{Nakajima:2003pg} thus
reenforcing the relation of the baryonic charge $B$ with the K\"ahler modulus.

\subsubsection{An expression for $g_N$ for $\mathbb{C}^2/\mathbb{Z}_2$}

Here we give a computation inspired by the previous discussion that may help
in simplifying the higher $N$ generating functions.



The generating function for 1 D-brane and baryonic charge $B$ was computed
to be

\begin{equation}
g_{1,B} ( q_1, q_2) = \frac{q_1^{|B|}}{(1-q_1^2)(1-\frac{q_2}{q_1})} + \frac{q_2^{|B|}}{(1-\frac{q_1}{q_2})(1-q_2^2)}
\end{equation}

We can now take the Plethystic Exponential for this expression, keeping track of the baryon
number. The simplest way of doing this is to take $\hbox{PE}[b^B g_{1,B}]$:
\begin{equation}
\hbox{PE}[b^B g_{1,B}] =
\exp\biggl(\sum_{k=1}^\infty \frac{\nu^k b^{k B} q_1^{k |B|}}{ k (1-q_1^{2k})(1-\frac{q_2^k}{q_1^k})}\biggr)
\exp\biggl(\sum_{k=1}^\infty \frac{\nu^k b^{k B} q_2^{k |B|}}{ k (1-\frac{q_1^k}{q_2^k})(1-q_2^{2k})}\biggr)
\end{equation}

This form of the equation can be compared with the generating function for the two dimensional
complex plane, $\mathbb{C}^2$,
\begin{equation}
g (\nu; t_1, t_2; \mathbb{C}^2 ) =
\exp\biggl(\sum_{k=1}^\infty \frac{\nu^k }{ k (1-t_1^{k})(1 - t_2^k)}\biggr) =
\sum_{N=0}^\infty \nu^N g_N(t_1, t_2; \mathbb{C}^2) ,
\label{grandC2}
\end{equation}
to write an expression

\begin{equation}
\hbox{PE}[b^B g_{1,B}] = g (\nu b^B q_1^{|B|} ; q_1^2, \frac{q_2}{q_1} ; \mathbb{C}^2 ) \cdot g
(\nu b^B q_2^{|B|} ; \frac{q_1}{q_2}, q_2^2 ; \mathbb{C}^2 ) ,
\end{equation}
precisely as explained above. We can now use the expansion in the number of branes to
demonstrate that the baryon number dependence is simple,
\begin{equation}
\hbox{PE}[b^B g_{1,B}] =
\sum_{N_1=0}^\infty \nu^{N_1} b^{B N_1} q_1^{|B| N_1} g_{N_1} (q_1^2, \frac{q_2}{q_1}; \mathbb{C}^2)
\sum_{N_2=0}^\infty \nu^{N_2} b^{B N_2} q_2^{|B| N_2} g_{N_2} (\frac{q_1}{q_2}, q_2^2; \mathbb{C}^2)
,
\end{equation}

In order to perform the sum over all baryon numbers, we use the identity

\begin{equation}
\sum_{B=-\infty}^\infty b^B t^{|B|} = \frac {1-t^2} {(1-t b) (1-\frac{t}{b})}
\end{equation}
and get
\bea
\sum_{B=-\infty}^\infty \hbox{PE}[b^B g_{1,B}] & = & \sum_{N_1=0}^\infty \sum_{N_2=0}^\infty
\nu^{N_1 + N_2} \frac {1-q_1^{2 N_1} q_2^{2 N_2}} {[1- (q_1 b)^{N_1} (q_2 b)^{N_2}] [1- (
\frac{q_1}{b} )^{N_1} ( \frac{q_2}{b} )^{N_2}]} \times \nonumber \\
& & \hspace{2cm} \times g_{N_1} (q_1^2, \frac{q_2}{q_1}; \mathbb{C}^2) \ g_{N_2}
(\frac{q_1}{q_2}, q_2^2; \mathbb{C}^2) .
\eea

From this expression we can easily extract the generating function for a fixed $N$ number of
D-branes,
\bea
g_N (q_1, q_2, b; \frac{\mathbb{C}^2}{\mathbb{Z}_2} ) & = &  \sum_{N_1 + N_2 = N} \frac
{1-q_1^{2 N_1} q_2^{2 N_2}} {[1- (q_1 b)^{N_1} (q_2 b)^{N_2}] [1- ( \frac{q_1}{b} )^{N_1} (
\frac{q_2}{b})^{N_2}]} \times \nonumber \\
& & \hspace{2cm} \times g_{N_1} (q_1^2, \frac{q_2}{q_1}; \mathbb{C}^2) \ g_{N_2}
(\frac{q_1}{q_2}, q_2^2; \mathbb{C}^2) .
\eea
By determining the first few terms for $\mathbb{C}^2$ in the expansion (\ref{grandC2}),
\begin{eqnarray}
g_1 (t_1, t_2; \mathbb{C}^2 ) &=& \frac{1}{(1-t_1)(1-t_2)} \nonumber\\
g_2 (t_1, t_2; \mathbb{C}^2 ) &=& \frac{1+t_1 t_2}{(1-t_1)(1-t_2)(1-t_1^2)(1-t_2^2)},
\end{eqnarray}
we are able to compute the generating functions for $\mathbb{C}^2/\mathbb{Z}_2$ with the
following results. For $N=1$ we obtain,
\begin{equation}
g_1 (q_1, q_2, b; \frac{\mathbb{C}^2}{\mathbb{Z}_2} ) = \frac {1} {(1- q_1 b) (1- \frac{q_1}{b})
(1- \frac{q_2}{q_1} )} + \frac {1}{ ( 1- q_2 b ) ( 1- \frac{q_2}{b} ) (1 - \frac{q_1}{q_2} ) } .
\end{equation}
For $N=2$,
\begin{eqnarray}
\nonumber
g_2 (q_1, q_2, b; \frac{\mathbb{C}^2}{\mathbb{Z}_2} ) &=&
\frac {1 + q_1 q_2}{(1- q_1^2 b^2) (1- \frac{q_1^2}{b^2}) (1-q_1^2) (1- \frac{q_2}{q_1} )(1- \frac{q_2^2}{q_1^2} )} \\ \nonumber
&+& \frac {1-q_1^2 q_2^2}{(1- q_1 q_2 b^2 ) (1- \frac{q_1 q_2}{b^2}) (1-q_1^2)(1- \frac{q_2}{q_1} )(1- \frac{q_1}{q_2} ) (1-q_2^2)} \\
&+& \frac {1 + q_1 q_2}{(1- q_2^2 b^2) (1- \frac{q_2^2}{b^2}) (1- \frac{q_1}{q_2} )(1- \frac{q_1^2}{q_2^2} ) (1-q_2^2) } .
\end{eqnarray}

\subsection{$\mathbb{C}^2/\mathbb{Z}_3$}

Let us consider the orbifold geometry $\mathbb{C}^2/\mathbb{Z}_3\times \mathbb{C}$. The theory
has ${\cal N}=2$ supersymmetry with three vector multiplets in the adjoint representation and
three bi-fundamental hypermultiplets. In ${\cal N}=1$ notation, this transforms to nine chiral
multiplets denoted by ${\bf \phi}_i,\ {\bf X}_i, {\bf Y}_i$ with $i=1,2,3$. (The charges are
written in Table \ref{globalC2Z3}.) The superpotential is
\begin{equation}
W = {\bf \phi}_1 ({\bf X}_1 {\bf Y}_1 - {\bf Y}_3 {\bf X}_3) + {\bf \phi}_2 ({\bf X}_2 {\bf Y}_2
- {\bf Y}_1 {\bf X}_1)+ {\bf \phi}_3 ({\bf X}_3 {\bf Y}_3 - {\bf Y}_2 {\bf X}_2)
\end{equation}

\begin{figure}[ht]
\begin{center}
  \epsfxsize = 11cm
  \centerline{\epsfbox{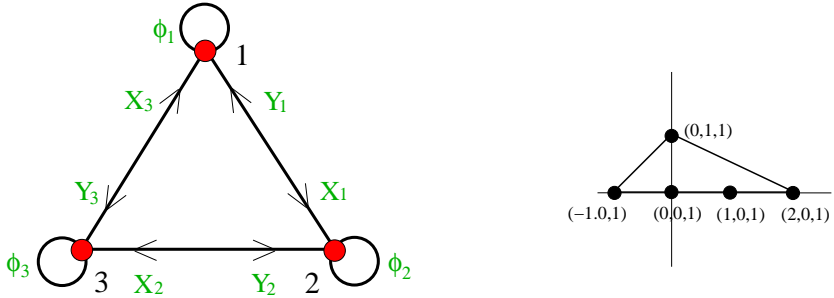}}
  \caption{Quiver and toric diagram for $\mathbb{C}^2/\mathbb{Z}_3\times \mathbb{C}$.}
  \label{c2z3quiver}
\end{center}
\end{figure}

The generating function for one D-brane is relatively easy to obtain once one makes the
observation that the moduli space is a complete intersection. The argument goes as follows.
Excluding the adjoint fields, there are six fields in the quiver. Besides the two flavor
symmetries that are dual to isometries in the $\mathbb{C}^2$ directions, there are also two
baryonic symmetries coming from the two Fayet-Iliopoulos terms which can be introduced for three
gauge groups. All together we count four symmetries leading to a four-dimensional moduli space
generated by six fields. Since there are only two F-term relations, ${\bf X}_1{\bf Y}_1={\bf
X}_2{\bf Y}_2={\bf X}_3{\bf Y}_3$, this manifold is a complete intersection.

\begin{table}[htdp]
\begin{center}
\begin{tabular}{|c|c|c|c|c|c|}
\hline
\ {\bf field} \ &  {\bf $U(1)_X$}   &   {\bf $U(1)_Y$}  & \ \  {\bf $B_1$} \ \ & \ \  {\bf $B_2$}  \ \ & \ \  charges \ \ \\ 
\hline \hline
${\bf X}_1$  & $1$ & $0$ & $1$ & $0$ & $t_1 b_1$ \\
${\bf Y}_1$  & $0$ & $1$ & $-1$ & $0$ & $t_2/b_1$ \\
${\bf X}_2$  & $1$ & $0$ & $-1$ & $1$ & $t_1 b_2/b_1$ \\
${\bf Y}_2$  & $0$ & $1$ & $1$ & $-1$ & $t_2 b_1/b_2$ \\
${\bf X}_3$  & $1$ & $0$ & $0$ & $-1$ & $t_1/b_2$ \\
${\bf Y}_3$  & $0$ & $1$ & $0$ & $1$ & $t_2 b_2$ \\
\hline
\end{tabular}
\end{center}
\caption{Global charges for the bi-fundamental fields of the quiver gauge theory
living on the D-brane probing the $\mathbb{C}^2/\mathbb{Z}_3\times\mathbb{C}$ singularity.}
\label{globalC2Z3}
\end{table}

This enables us to immediately write down the generating function,
\begin{equation}
\label{g1C2Z3} g_1(t_1, t_2; b_1, b_2; \frac{\mathbb{C}^2}{\mathbb{Z}_3} ) = \frac{(1-t_1
t_2)^2}{ (1-t_1 b_1) (1-\frac{t_2}{b_1}) (1-\frac{t_1 b_2}{b_1}) (1-\frac{t_2 b_1}{b_2})
(1-\frac{t_1}{b_2}) (1-t_2 b_2) }
\end{equation}
where we have set the generator charges in the denominator according to Table \ref{globalC2Z3}.
The numerator is obtained from the charges of the F-terms.

\begin{figure}[ht]
\begin{center}
  \epsfxsize = 5cm
  \centerline{\epsfbox{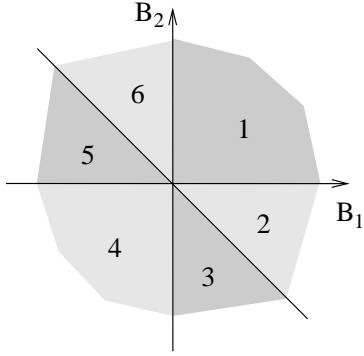}}
  \caption{The six cones for $\mathbb{C}^2/\mathbb{Z}_3\times \mathbb{C}$.}
  \label{c2z3gkz}
\end{center}
\end{figure}

Alternatively, one arrives at the exact same result using localization and
the auxiliary GKZ polynomial ring. The latter leads to six
two dimensional cones parametrized by $B_1$ and $B_2$. These cones correspond to the
Weyl chambers of SU(3) depicted in \fref{c2z3gkz}. For fixed baryonic charges, localization
gives

\be
\label{g1z3B}
 g_{1,B_1,B_2}(t_1,t_2; \ \IC^2 / \IZ_3) =
\frac{t_1^{\tilde B_1+2 \tilde B_2}}{(1-t_1^3)(1-\frac{t_2}{t_1^2})}+\frac{t_1^{\tilde B_1}
t_2^{\tilde B_2}}{(1-\frac{t_1^2}{t_2})(1-\frac{t_2^2}{t_1})}+\frac{t_2^{2\tilde B_1+ \tilde
B_2}}{(1-\frac{t_1}{t_2^2})(1-t_2^3)}  \nonumber
\ee
where $(\tilde B_1, \tilde B_2)$ is a lattice point in the fundamental chamber, and can be
obtained by Weyl reflection of $(B_1, B_2)$. This is similar to the $\IC^3 / \IZ_2$ case, where
the Weyl reflection boils down to taking the absolute value of the baryonic charge, cf.
(\ref{poi}).

In detail, we have

\begin{equation}
g_{1,B_1,B_2}(t_1, t_2; \frac{\mathbb{C}^2}{\mathbb{Z}_3}) = \left \{
\begin{array}{l}
\frac{t_1^{B_1+2B_2}} {(1-t_1^3)(1-\frac{t_2}{t_1^2})} + \frac{ t_1^{B_1} t_2^{B_2} }
{(1-\frac{t_1^2}{t_2} ) (1- \frac{t_2^2}{t_1})} +
\frac{t_2^{2B_1+B_2}} {(1-\frac{t_1}{t_2^2})(1-t_2^3)} \quad B_1\ge0, B_2\ge0 \\ \\

\frac{t_1^{B_1-B_2}} {(1-t_1^3)(1-\frac{t_2}{t_1^2})} + \frac{ t_1^{B_1+B_2} t_2^{-B_2} }
{(1-\frac{t_1^2}{t_2} ) (1- \frac{t_2^2}{t_1})} +
\frac{t_2^{2B_1+B_2}} {(1-\frac{t_1}{t_2^2})(1-t_2^3)} \quad B_1 \ge 0 \ge B_2\ge -B_1 \\ \\

\frac{t_1^{B_1-B_2}} {(1-t_1^3)(1-\frac{t_2}{t_1^2})} + \frac{ t_1^{-B_1-B_2} t_2^{B_1} }
{(1-\frac{t_1^2}{t_2} ) (1- \frac{t_2^2}{t_1})} +
\frac{t_2^{-B_1-2B_2}} {(1-\frac{t_1}{t_2^2})(1-t_2^3)} \quad 0 \le B_1\le -B_2 \\ \\

\frac{t_1^{-2B_1-B_2}} {(1-t_1^3)(1-\frac{t_2}{t_1^2})} + \frac{ t_1^{-B_2} t_2^{-B_1} }
{(1-\frac{t_1^2}{t_2} ) (1- \frac{t_2^2}{t_1})} +
\frac{t_2^{-B_1-2B_2}} {(1-\frac{t_1}{t_2^2})(1-t_2^3)} \quad B_1\le0, B_2\le0 \\ \\

\frac{t_1^{-2B_1-B_2}} {(1-t_1^3)(1-\frac{t_2}{t_1^2})} + \frac{ t_1^{B_2} t_2^{-B_1-B_2} }
{(1-\frac{t_1^2}{t_2} ) (1- \frac{t_2^2}{t_1})} +
\frac{t_2^{-B_1+B_2}} {(1-\frac{t_1}{t_2^2})(1-t_2^3)} \quad B_1 \le 0 \le B_2 \le -B_1 \\ \\

\frac{t_1^{2B_1+B_2}} {(1-t_1^3)(1-\frac{t_2}{t_1^2})} + \frac{ t_1^{-B_1} t_2^{B_1+B_2} }
{(1-\frac{t_1^2}{t_2} ) (1- \frac{t_2^2}{t_1})} + \frac{t_2^{-B_1+B_2}}
{(1-\frac{t_1}{t_2^2})(1-t_2^3)} \quad B_2\ge -B_1 \ge 0 \\ \nonumber
\end{array}
\right.
\end{equation}

Now $g_{1,B_1,B_2}$ can be summed up for $B_1$ and $B_2$, giving precisely the result of
(\ref{g1C2Z3}).

\subsection{$\mathbb{C}^2/\mathbb{Z}_n$}

For sake of completeness, we derive here the general structure of the $N=1$ generating function
for $\mathbb{C}^2/\mathbb{Z}_n$ by making use of brane tilings \cite{Hanany:2005ve,
Franco:2005rj}. It is easy to see that the polynomial ring is a complete intersection and
therefore the $N=1$ generating function is easy to compute. This could also be computed on a
case-by-case basis with Hilbert series in Macaulay2.

\begin{figure}[ht]
\begin{center}
  \epsfxsize = 6.5cm
  \centerline{\epsfbox{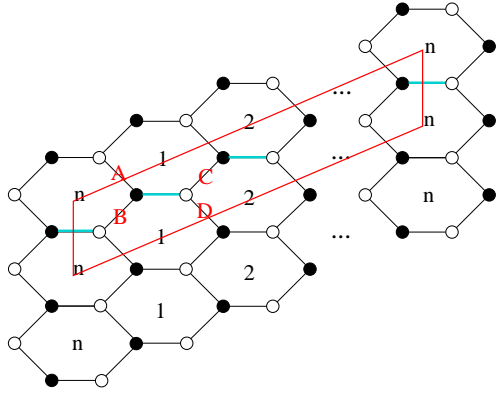}}
  \caption{Tiling for $\IC^2 / \IZ_n$. The blue adjoint fields give nontrivial constraints.}
  \label{c2zn_tiling}
\end{center}
\end{figure}

The brane tiling for the geometry is shown in \fref{c2zn_tiling}. The edges are in one-to-one
correspondence with the $2n$ bifundamental and $n$ adjoint fields in the theory. Black and white
vertices give terms in the superpotential. Gauge groups are labeled by the faces in the tiling.

F-terms for the bifundamental fields set the blue adjoint fields equal to each other. On the other hand, F-terms for the adjoint fields give $n-1$ constraints on the bifundamental fields. For instance,
\be
{\bf A B} = {\bf C D}
\ee

where ${\bf A}, {\bf B} ,{\bf C}$ and $\bf D$ are indicated in \fref{c2zn_tiling}. The baryonic
charges of these equations are zero. Flavor charges are assign $t_1 (t_2) $ for a field that
goes southwest (southeast) in \fref{c2zn_tiling}, respectively. For example, $\bf A$ and $\bf D$
are assigned a chemical potential $t_1$ and $\bf B$ and $\bf C$ are assigned a chemical
potential $t_2$. Hence, each F-term carries weight $t_1 t_2$, leading to a $(1-t_1 t_2)$ factor
in the numerator of the generating function. There are $n-1$ such relations. The general formula
for $\IC^2 /\IZ_n$,

\[ g_1(t_1, t_2, b_1, \ldots, b_{n-1} ; \ \IC^2 / \IZ_n)= \frac{(1-t_1 t_2)^{n-1}}{\prod_{i=1}^n (1-t_1 b^{\vec B_i})(1-t_2 b^{-\vec B_i})}
\]
where $b^{\vec B} \equiv b_1^{B_1} b_2^{B_2} \cdots b_{n-1}^{B_{n-1}}$. Here the assignment of
baryonic charges in $n$ vectors $\vec B^i$ which live in $n-1$ dimensions can be chosen with any
convenient basis which is isomorphic to the simple roots of the $A_{n-1}$ Lie algebra. A
possible choice for these $n$ vectors can be a straightforward generalization of Table
\ref{globalC2Z3}, $\vec B_1 = (1,0,0\ldots0), \vec B_2 = (-1,1,0\ldots0), \vec B_3 =
(0,-1,1,0\ldots0), \ldots,  \vec B_{n-1} = (0\ldots0, -1,1), \vec B_{n} = (0\ldots0,-1).$

\section{Appendix: A look at the shiver}

In order to understand anomalous baryonic charges, we consider the mirror Calabi-Yau. The
geometric description of the mirror \cite{Hori:2000kt, Hori:2000ck, Hanany:2001py} consists of a
double fibration over $W \in \IC$,
\bean
  W &=& P(w,z) \equiv \sum c_{p,q} w^p z^q \\
  W &=& uv
\eean
where $w,z \in \IC^*$ and $u,v \in \IC$. $P(w,z)$ is the Newton polynomial of the toric diagram
and describes a punctured Riemann surface fiber over the $W$ plane. The genus of this surface
equals $I$, i.~e. the number of internal points in the toric diagram.

According to the mirror conjecture, the gauge theory arises from D6-branes wrapping
three-cycles. These three-cycles intersect over $W=0$, and open strings at such intersection
points give chiral bifundamental matter fields \cite{Berkooz:1996km}. The three-cycles wrap
one-cycles in the Riemann-surface fiber at $W=0$. They determine a mirror tiling which we will
call here the ``shiver''. This graph is related to the brane tiling (or dimer graph) by the
so-called ``antimap''. The detailed description can be found in \cite{Feng:2005gw}.

The brane tiling lives on a torus and the two nontrivial cycles are related to the flavor
charges. The shiver lives on the Riemann surface where the $2I$ nontrivial cycles are related to
the anomalous charges. We have the analogy

\vspace{0.3cm} \qquad {\bf flavor charges : brane tiling :: anomalous charges : shiver}
\vspace{0.3cm}

In the following, we study this analogy through examples.

\subsection{Anomalous charges of $\IC^3/\IZ_3$}

\begin{figure}[ht]
\begin{center}
  \epsfxsize = 6cm
  \centerline{\epsfbox{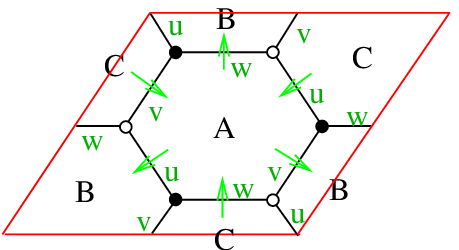}}
  \caption{Shiver for $\IC^3/\IZ_3$.}
  \label{dP0shiver}
\end{center}
\end{figure}

Let us consider the shiver of the $\IC^3/\IZ_3$ theory which is given in \fref{dP0shiver}. The
edges are in one-to-one correspondence with the arrows in the quiver. The $u,v$ and $w$
represent the three types of fields which differ in their (torsion) D1-charges. It follows that
their anomalous U(1) charges are also different,
\be
  u=a_1 a_2 \qquad
  v=\frac{a_2}{a_1} \qquad
  w=\frac{1}{a_2^2}.
\ee

We see that trivial loops along the green arrows have vanishing charges. The two nontrivial
loops have charges $\frac{1}{a_1^2}$ and $\frac{1}{a_1 a_2^3}$, respectively. Therefore,
anomalous charges distinguish between the two cycles and in this sense they are similar to the
flavor charges in the tiling.

\subsection{Anomalous charges of $\mathbb{F}_0$}

\begin{figure}[ht]
\begin{center}
  \epsfxsize = 11cm
  \centerline{\epsfbox{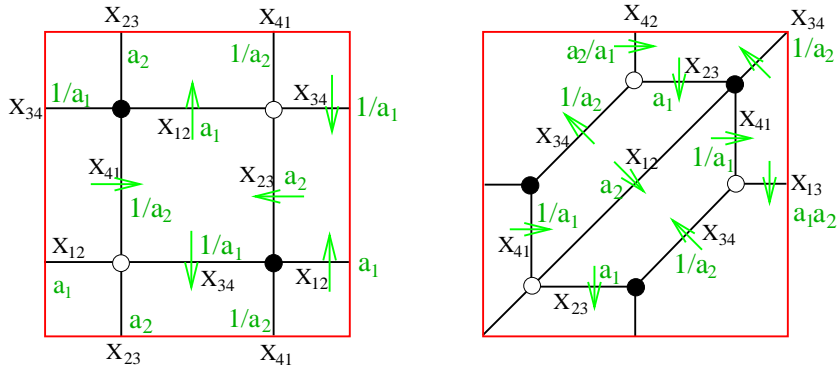}}
  \caption{(i) Shiver for $\mathbb{F}_0$. \  (ii) Shiver for  $dP_1$.}
  \label{F0dP1shiver}
\end{center}
\end{figure}

The assignment of anomalous charges is shown in \fref{F0quiver}. In the shiver
(\fref{F0dP1shiver}), trivial loops along the green arrows again have vanishing charges. The two
nontrivial loops have charges $a_1^2$ and $a_2^2$, respectively.

\subsection{Anomalous charges of $dP_1$}

The assignment of anomalous charges is shown in \fref{dP1quiver}. In the shiver
(\fref{F0dP1shiver}), the two nontrivial loops have charges $\frac{a_2^2}{a_1}$ and
$\frac{1}{a_1 a_2^2}$, respectively.

\clearpage

\section{Appendix: List of notations}

\vskip 0.5cm

\begin{tabular}{lll}
&  $N$ & number of D-branes \\
&  $G$ & number of $SU(N)$ gauge group factors in the theory \\
&  $F$ & number of fields (chiral bi-fundamental multiplets) in the theory \vspace{0.3cm} \\
&  $t_i$ & chemical potentials (weights) \\
&  $x,y, q_i$ & flavor charge weights \\
&  $b_i$ & non-anomalous baryonic weights \\
&  $a_i$ & anomalous baryonic weights \vspace{0.3cm} \\
&  $R$ & R-charge \\
&  $B_i, B^\prime$ & baryonic charges from the gauge theory \\
&  $\beta, \beta^\prime$ & baryonic charges from the geometry\\
&  $A_i$ & anomalous baryonic charges \vspace{0.3cm} \\
&  ${\bf A,B,C,\ldots }$ & fields in the gauge theory \\
&  ${\cal R}$ & polynomial ring \\
&  ${\cal I, J}$ & ideals for a polynomial ring \\
&  $a,b,c,\ldots$ & generators of the auxiliary GKZ ring \\
&  $m(B_1, B_2, \ldots)$ & multiplicities in the GKZ fan \vspace{0.3cm} \\
&  $I$ & number of internal (integral) points in toric diagram \\
&  $d$ & number of external points in toric diagram \\
&  $D_i$ & toric divisors assigned to the integral points in the toric diagram \\
&  $x_i$ & homogeneous coordinates \vspace{0.3cm} \\
&  $g_N(\{t_i\};CY)$ & baryonic generating function for $N$ D-branes probing CY \\
&  $\mbox{PE}_\nu[\ \cdot \ ]$ & plethystic exponential with weight $\nu$ for the number of
D-branes \\
&  $Z_{\beta, \beta^\prime} (\{t_i\};CY)$ & partition function from the geometry \\
&  $C(\beta)$ & hollow polygon in the fiber over the GKZ lattice \\
\end{tabular}


\newpage
\bibliography{Molien}

\providecommand{\href}[2]{#2}\begingroup\raggedright\begin{thebibliography}{10}

\bibitem{Romelsberger:2005eg}
C.~Romelsberger, {\it Counting chiral primaries in {N=1, d=4} superconformal
  field theories},  {\em Nucl. Phys.} {\bf B747} (2006) 329--353,
  [\href{http://xxx.lanl.gov/abs/hep-th/0510060}{{\tt hep-th/0510060}}].

\bibitem{Kinney:2005ej}
J.~Kinney, J.~M. Maldacena, S.~Minwalla, and S.~Raju, {\it An index for 4
  dimensional super conformal theories},
  \href{http://xxx.lanl.gov/abs/hep-th/0510251}{{\tt hep-th/0510251}}.

\bibitem{Nakayama:2005mf}
Y.~Nakayama, {\it Index for orbifold quiver gauge theories},  {\em Phys. Lett.}
  {\bf B636} (2006) 132--136,
  [\href{http://xxx.lanl.gov/abs/hep-th/0512280}{{\tt hep-th/0512280}}].

\bibitem{Biswas:2006tj}
I.~Biswas, D.~Gaiotto, S.~Lahiri, and S.~Minwalla, {\it Supersymmetric states
  of {N = 4 Yang-Mills} from giant gravitons},
  \href{http://xxx.lanl.gov/abs/hep-th/0606087}{{\tt hep-th/0606087}}.

\bibitem{Mandal:2006tk}
G.~Mandal and N.~V. Suryanarayana, {\it Counting {1/8-BPS} dual-giants},  {\em
  JHEP} {\bf 03} (2007) 031,
  [\href{http://xxx.lanl.gov/abs/hep-th/0606088}{{\tt hep-th/0606088}}].

\bibitem{Benvenuti:2006qr}
S.~Benvenuti, B.~Feng, A.~Hanany, and Y.-H. He, {\it Counting {BPS} operators
  in gauge theories: Quivers, syzygies and plethystics},
  \href{http://xxx.lanl.gov/abs/hep-th/0608050}{{\tt hep-th/0608050}}.

\bibitem{Martelli:2006vh}
D.~Martelli and J.~Sparks, {\it Dual giant gravitons in {Sasaki-Einstein}
  backgrounds},  {\em Nucl. Phys.} {\bf B759} (2006) 292--319,
  [\href{http://xxx.lanl.gov/abs/hep-th/0608060}{{\tt hep-th/0608060}}].

\bibitem{Basu:2006id}
A.~Basu and G.~Mandal, {\it Dual giant gravitons in {AdS(m) x Y**n
  (Sasaki-Einstein)}},  \href{http://xxx.lanl.gov/abs/hep-th/0608093}{{\tt
  hep-th/0608093}}.

\bibitem{Butti:2006au}
A.~Butti, D.~Forcella, and A.~Zaffaroni, {\it Counting {BPS} baryonic operators
  in {CFT}s with {Sasaki}-{Einstein} duals},
  \href{http://xxx.lanl.gov/abs/hep-th/0611229}{{\tt hep-th/0611229}}.

\bibitem{Hanany:2006uc}
A.~Hanany and C.~Romelsberger, {\it Counting {BPS} operators in the chiral ring
  of {N=2} supersymmetric gauge theories or {N = 2} braine surgery},
  \href{http://xxx.lanl.gov/abs/hep-th/0611346}{{\tt hep-th/0611346}}.

\bibitem{Feng:2007ur}
B.~Feng, A.~Hanany, and Y.-H. He, {\it Counting gauge invariants: {The}
  plethystic program},  {\em JHEP} {\bf 03} (2007) 090,
  [\href{http://xxx.lanl.gov/abs/hep-th/0701063}{{\tt hep-th/0701063}}].

\bibitem{Forcella:2007wk}
D.~Forcella, A.~Hanany, and A.~Zaffaroni, {\it Baryonic generating functions},
  \href{http://xxx.lanl.gov/abs/hep-th/0701236}{{\tt hep-th/0701236}}.

\bibitem{Grant:2007ze}
L.~Grant and K.~Narayan, {\it Mesonic chiral rings in {Calabi-Yau} cones from
  field theory},  \href{http://xxx.lanl.gov/abs/hep-th/0701189}{{\tt
  hep-th/0701189}}.

\bibitem{Nakayama:2007jy}
Y.~Nakayama, {\it Finite {N} index and angular momentum bound from gravity},
  \href{http://xxx.lanl.gov/abs/hep-th/0701208}{{\tt hep-th/0701208}}.

\bibitem{Lee:2007kv}
S.~Lee, S.~Lee, and J.~Park, {\it Toric {AdS}(4)/{CFT}(3) duals and {M}-theory
  crystals},  \href{http://xxx.lanl.gov/abs/hep-th/0702120}{{\tt
  hep-th/0702120}}.

\bibitem{Dolan:2007rq}
F.~A. Dolan, {\it Counting {BPS} operators in {N=4 SYM}},
  \href{http://xxx.lanl.gov/abs/arXiv:0704.1038 [hep-th]}{{\tt arXiv:0704.1038
  [hep-th]}}.

\bibitem{Dav}
D.~Forcella, {\it {BPS Partition Functions for Quiver Gauge Theories: Counting
  the Fermionic d.o.f.}}, . work in progress.

\bibitem{Douglas:1996sw}
M.~R. Douglas and G.~W. Moore, {\it D-branes, {Q}uivers, and {ALE}
  {I}nstantons},  \href{http://xxx.lanl.gov/abs/hep-th/9603167}{{\tt
  hep-th/9603167}}.

\bibitem{Johnson:1996py}
C.~V. Johnson and R.~C. Myers, {\it Aspects of type {IIB} theory on {ALE}
  spaces},  {\em Phys. Rev.} {\bf D55} (1997) 6382--6393,
  [\href{http://xxx.lanl.gov/abs/hep-th/9610140}{{\tt hep-th/9610140}}].

\bibitem{Kachru:1998ys}
S.~Kachru and E.~Silverstein, {\it 4d conformal theories and strings on
  orbifolds},  {\em Phys. Rev. Lett.} {\bf 80} (1998) 4855--4858,
  [\href{http://xxx.lanl.gov/abs/hep-th/9802183}{{\tt hep-th/9802183}}].

\bibitem{Lawrence:1998ja}
A.~E. Lawrence, N.~Nekrasov, and C.~Vafa, {\it On conformal field theories in
  four dimensions},  {\em Nucl. Phys.} {\bf B533} (1998) 199--209,
  [\href{http://xxx.lanl.gov/abs/hep-th/9803015}{{\tt hep-th/9803015}}].

\bibitem{Douglas:1997de}
M.~R. Douglas, B.~R. Greene, and D.~R. Morrison, {\it Orbifold resolution by
  {D-branes}},  {\em Nucl. Phys.} {\bf B506} (1997) 84--106,
  [\href{http://xxx.lanl.gov/abs/hep-th/9704151}{{\tt hep-th/9704151}}].

\bibitem{Morrison:1998cs}
D.~R. Morrison and M.~R. Plesser, {\it Non-spherical horizons. {I}},  {\em Adv.
  Theor. Math. Phys.} {\bf 3} (1999) 1--81,
  [\href{http://xxx.lanl.gov/abs/hep-th/9810201}{{\tt hep-th/9810201}}].

\bibitem{Beasley:1999uz}
C.~Beasley, B.~R. Greene, C.~I. Lazaroiu, and M.~R. Plesser, {\it {D3-branes}
  on partial resolutions of {Abelian} quotient singularities of {Calabi-Yau}
  threefolds},  {\em Nucl. Phys.} {\bf B566} (2000) 599--640,
  [\href{http://xxx.lanl.gov/abs/hep-th/9907186}{{\tt hep-th/9907186}}].

\bibitem{Feng:2000mi}
B.~Feng, A.~Hanany, and Y.-H. He, {\it {D-brane} gauge theories from toric
  singularities and toric duality},  {\em Nucl. Phys.} {\bf B595} (2001)
  165--200, [\href{http://xxx.lanl.gov/abs/hep-th/0003085}{{\tt
  hep-th/0003085}}].

\bibitem{Cachazo:2001sg}
F.~Cachazo, B.~Fiol, K.~A. Intriligator, S.~Katz, and C.~Vafa, {\it A geometric
  unification of dualities},  {\em Nucl. Phys.} {\bf B628} (2002) 3--78,
  [\href{http://xxx.lanl.gov/abs/hep-th/0110028}{{\tt hep-th/0110028}}].

\bibitem{Feng:2001xr}
B.~Feng, A.~Hanany, and Y.-H. He, {\it Phase structure of {D-brane} gauge
  theories and toric duality},  {\em JHEP} {\bf 08} (2001) 040,
  [\href{http://xxx.lanl.gov/abs/hep-th/0104259}{{\tt hep-th/0104259}}].

\bibitem{Feng:2002zw}
B.~Feng, S.~Franco, A.~Hanany, and Y.-H. He, {\it Symmetries of toric duality},
   {\em JHEP} {\bf 12} (2002) 076,
  [\href{http://xxx.lanl.gov/abs/hep-th/0205144}{{\tt hep-th/0205144}}].

\bibitem{Hanany:1997tb}
A.~Hanany and A.~Zaffaroni, {\it On the realization of chiral four-dimensional
  gauge theories using branes},  {\em JHEP} {\bf 05} (1998) 001,
  [\href{http://xxx.lanl.gov/abs/hep-th/9801134}{{\tt hep-th/9801134}}].

\bibitem{Hanany:1998ru}
A.~Hanany, M.~J. Strassler, and A.~M. Uranga, {\it Finite theories and marginal
  operators on the brane},  {\em JHEP} {\bf 06} (1998) 011,
  [\href{http://xxx.lanl.gov/abs/hep-th/9803086}{{\tt hep-th/9803086}}].

\bibitem{Hanany:1998it}
A.~Hanany and A.~M. Uranga, {\it Brane boxes and branes on singularities},
  {\em JHEP} {\bf 05} (1998) 013,
  [\href{http://xxx.lanl.gov/abs/hep-th/9805139}{{\tt hep-th/9805139}}].

\bibitem{Hanany:2005ve}
A.~Hanany and K.~D. Kennaway, {\it Dimer models and toric diagrams},
  \href{http://xxx.lanl.gov/abs/hep-th/0503149}{{\tt hep-th/0503149}}.

\bibitem{Franco:2005rj}
S.~Franco, A.~Hanany, K.~D. Kennaway, D.~Vegh, and B.~Wecht, {\it Brane dimers
  and quiver gauge theories},  {\em JHEP} {\bf 01} (2006) 096,
  [\href{http://xxx.lanl.gov/abs/hep-th/0504110}{{\tt hep-th/0504110}}].

\bibitem{Butti:2005vn}
A.~Butti and A.~Zaffaroni, {\it R-charges from toric diagrams and the
  equivalence of a- maximization and {Z}-minimization},  {\em JHEP} {\bf 11}
  (2005) 019, [\href{http://xxx.lanl.gov/abs/hep-th/0506232}{{\tt
  hep-th/0506232}}].

\bibitem{Hanany:2005ss}
A.~Hanany and D.~Vegh, {\it Quivers, tilings, branes and rhombi},
  \href{http://xxx.lanl.gov/abs/hep-th/0511063}{{\tt hep-th/0511063}}.

\bibitem{Feng:2005gw}
B.~Feng, Y.-H. He, K.~D. Kennaway, and C.~Vafa, {\it Dimer models from mirror
  symmetry and quivering amoebae},
  \href{http://xxx.lanl.gov/abs/hep-th/0511287}{{\tt hep-th/0511287}}.

\bibitem{Hanany:2006nm}
A.~Hanany, C.~P. Herzog, and D.~Vegh, {\it Brane tilings and exceptional
  collections},  {\em JHEP} {\bf 07} (2006) 001,
  [\href{http://xxx.lanl.gov/abs/hep-th/0602041}{{\tt hep-th/0602041}}].

\bibitem{Garcia-Etxebarria:2006aq}
I.~Garcia-Etxebarria, F.~Saad, and A.~M. Uranga, {\it Quiver gauge theories at
  resolved and deformed singularities using dimers},  {\em JHEP} {\bf 06}
  (2006) 055, [\href{http://xxx.lanl.gov/abs/hep-th/0603108}{{\tt
  hep-th/0603108}}].

\bibitem{Brini:2006ej}
A.~Brini and D.~Forcella, {\it Comments on the non-conformal gauge theories
  dual to {Y(p,q)} manifolds},  {\em JHEP} {\bf 06} (2006) 050,
  [\href{http://xxx.lanl.gov/abs/hep-th/0603245}{{\tt hep-th/0603245}}].

\bibitem{Butti:2006hc}
A.~Butti, {\it Deformations of toric singularities and fractional branes},
  {\em JHEP} {\bf 10} (2006) 080,
  [\href{http://xxx.lanl.gov/abs/hep-th/0603253}{{\tt hep-th/0603253}}].

\bibitem{Gubser:1998fp}
S.~S. Gubser and I.~R. Klebanov, {\it Baryons and domain walls in an {N = 1}
  superconformal gauge theory},  {\em Phys. Rev.} {\bf D58} (1998) 125025,
  [\href{http://xxx.lanl.gov/abs/hep-th/9808075}{{\tt hep-th/9808075}}].

\bibitem{McGreevy:2000cw}
J.~McGreevy, L.~Susskind, and N.~Toumbas, {\it Invasion of the giant gravitons
  from anti-de {Sitter} space},  {\em JHEP} {\bf 06} (2000) 008,
  [\href{http://xxx.lanl.gov/abs/hep-th/0003075}{{\tt hep-th/0003075}}].

\bibitem{Hashimoto:2000zp}
A.~Hashimoto, S.~Hirano, and N.~Itzhaki, {\it Large branes in {AdS} and their
  field theory dual},  {\em JHEP} {\bf 08} (2000) 051,
  [\href{http://xxx.lanl.gov/abs/hep-th/0008016}{{\tt hep-th/0008016}}].

\bibitem{Balasubramanian:2001nh}
V.~Balasubramanian, M.~Berkooz, A.~Naqvi, and M.~J. Strassler, {\it Giant
  gravitons in conformal field theory},  {\em JHEP} {\bf 04} (2002) 034,
  [\href{http://xxx.lanl.gov/abs/hep-th/0107119}{{\tt hep-th/0107119}}].

\bibitem{Berenstein:2002ke}
D.~Berenstein, C.~P. Herzog, and I.~R. Klebanov, {\it Baryon spectra and
  {AdS/CFT} correspondence},  {\em JHEP} {\bf 06} (2002) 047,
  [\href{http://xxx.lanl.gov/abs/hep-th/0202150}{{\tt hep-th/0202150}}].

\bibitem{Herzog:2003wt}
C.~P. Herzog and J.~McKernan, {\it Dibaryon spectroscopy},  {\em JHEP} {\bf 08}
  (2003) 054, [\href{http://xxx.lanl.gov/abs/hep-th/0305048}{{\tt
  hep-th/0305048}}].

\bibitem{Mikhailov:2000ya}
A.~Mikhailov, {\it Giant gravitons from holomorphic surfaces},  {\em JHEP} {\bf
  11} (2000) 027, [\href{http://xxx.lanl.gov/abs/hep-th/0010206}{{\tt
  hep-th/0010206}}].

\bibitem{Beasley:2002xv}
C.~E. Beasley, {\it {BPS} branes from baryons},  {\em JHEP} {\bf 11} (2002)
  015, [\href{http://xxx.lanl.gov/abs/hep-th/0207125}{{\tt hep-th/0207125}}].

\bibitem{Martelli:2006yb}
D.~Martelli, J.~Sparks, and S.-T. Yau, {\it {Sasaki-Einstein} manifolds and
  volume minimisation},  \href{http://xxx.lanl.gov/abs/hep-th/0603021}{{\tt
  hep-th/0603021}}.

\bibitem{Benvenuti:2005ja}
S.~Benvenuti and M.~Kruczenski, {\it From {Sasaki}-{Einstein} spaces to quivers
  via {BPS} geodesics: {L(p,q|r)}},  {\em JHEP} {\bf 04} (2006) 033,
  [\href{http://xxx.lanl.gov/abs/hep-th/0505206}{{\tt hep-th/0505206}}].

\bibitem{Benvenuti:2005cz}
S.~Benvenuti and M.~Kruczenski, {\it Semiclassical strings in
  {Sasaki}-{Einstein} manifolds and long operators in {N = 1} gauge theories},
  {\em JHEP} {\bf 10} (2006) 051,
  [\href{http://xxx.lanl.gov/abs/hep-th/0505046}{{\tt hep-th/0505046}}].

\bibitem{Franco:2005sm}
S.~Franco {\em et.~al.}, {\it Gauge theories from toric geometry and brane
  tilings},  {\em JHEP} {\bf 01} (2006) 128,
  [\href{http://xxx.lanl.gov/abs/hep-th/0505211}{{\tt hep-th/0505211}}].

\bibitem{Butti:2005sw}
A.~Butti, D.~Forcella, and A.~Zaffaroni, {\it The dual superconformal theory
  for {L(p,q,r)} manifolds},  {\em JHEP} {\bf 09} (2005) 018,
  [\href{http://xxx.lanl.gov/abs/hep-th/0505220}{{\tt hep-th/0505220}}].

\bibitem{Butti:2005ps}
A.~Butti and A.~Zaffaroni, {\it From toric geometry to quiver gauge theory: The
  equivalence of a-maximization and {Z}-minimization},  {\em Fortsch. Phys.}
  {\bf 54} (2006) 309--316, [\href{http://xxx.lanl.gov/abs/hep-th/0512240}{{\tt
  hep-th/0512240}}].

\bibitem{Franco:2006gc}
S.~Franco and D.~Vegh, {\it Moduli spaces of gauge theories from dimer models:
  {Proof} of the correspondence},  {\em JHEP} {\bf 11} (2006) 054,
  [\href{http://xxx.lanl.gov/abs/hep-th/0601063}{{\tt hep-th/0601063}}].

\bibitem{Benvenuti:2006xg}
S.~Benvenuti, L.~A. Pando~Zayas, and Y.~Tachikawa, {\it Triangle anomalies from
  {Einstein} manifolds},  {\em Adv. Theor. Math. Phys.} {\bf 10} (2006)
  395--432, [\href{http://xxx.lanl.gov/abs/hep-th/0601054}{{\tt
  hep-th/0601054}}].

\bibitem{Lee:2006ru}
S.~Lee and S.-J. Rey, {\it Comments on anomalies and charges of toric-quiver
  duals},  {\em JHEP} {\bf 03} (2006) 068,
  [\href{http://xxx.lanl.gov/abs/hep-th/0601223}{{\tt hep-th/0601223}}].

\bibitem{Imamura:2006ub}
Y.~Imamura, {\it Anomaly cancellations in brane tilings},
  \href{http://xxx.lanl.gov/abs/hep-th/0605097}{{\tt hep-th/0605097}}.

\bibitem{Ueda:2006wy}
K.~Ueda and M.~Yamazaki, {\it Brane tilings for parallelograms with application
  to homological mirror symmetry},
  \href{http://xxx.lanl.gov/abs/math.ag/0606548}{{\tt math.ag/0606548}}.

\bibitem{Butti:2006nk}
A.~Butti, D.~Forcella, and A.~Zaffaroni, {\it Deformations of conformal
  theories and non-toric quiver gauge theories},  {\em JHEP} {\bf 02} (2007)
  081, [\href{http://xxx.lanl.gov/abs/hep-th/0607147}{{\tt hep-th/0607147}}].

\bibitem{Imamura:2006ie}
Y.~Imamura, {\it Global symmetries and 't {Hooft} anomalies in brane tilings},
  {\em JHEP} {\bf 12} (2006) 041,
  [\href{http://xxx.lanl.gov/abs/hep-th/0609163}{{\tt hep-th/0609163}}].

\bibitem{Oota:2006eg}
T.~Oota and Y.~Yasui, {\it New example of infinite family of quiver gauge
  theories},  {\em Nucl. Phys.} {\bf B762} (2007) 377--391,
  [\href{http://xxx.lanl.gov/abs/hep-th/0610092}{{\tt hep-th/0610092}}].

\bibitem{Kato:2006vx}
A.~Kato, {\it Zonotopes and four-dimensional superconformal field theories},
  \href{http://xxx.lanl.gov/abs/hep-th/0610266}{{\tt hep-th/0610266}}.

\bibitem{Imamura:2007dc}
Y.~Imamura, H.~Isono, K.~Kimura, and M.~Yamazaki, {\it Exactly marginal
  deformations of quiver gauge theories as seen from brane tilings},
  \href{http://xxx.lanl.gov/abs/hep-th/0702049}{{\tt hep-th/0702049}}.

\bibitem{M2}
D.~R. Grayson and M.~E. Stillman, ``Macaulay 2, a software system for research
  in algebraic geometry.'' Available at http://www.math.uiuc.edu/Macaulay2/.

\bibitem{Cox:2000vi}
D.~A. Cox and S.~Katz, {\it Mirror symmetry and algebraic geometry}, .
  Providence, USA: AMS (2000) 469 p.

\bibitem{OdaTadao:19910900}
T.~Oda and H.~S. Park, {\it Linear {Gale} transforms and
  {Gelfand-Kapranov-Zelevinskij} decompositions},  {\em Tohoku mathematical
  journal. Second series} {\bf 43} (19910900), no.~3.

\bibitem{Gelf}
M.~Gelfand, I.~Kapranov and A.~Zelevinsky, {\it Discriminants, {R}esultants and
  {M}ultidimensional {D}eterminants}, . Birkhauser, Boston, 1994.

\bibitem{Fulton}
W.~Fulton, {\it Introduction to {T}oric {V}arieties}, . Princeton University
  press, 1993.

\bibitem{Gukov:1998kn}
S.~Gukov, M.~Rangamani, and E.~Witten, {\it Dibaryons, strings, and branes in
  {AdS} orbifold models},  {\em JHEP} {\bf 12} (1998) 025,
  [\href{http://xxx.lanl.gov/abs/hep-th/9811048}{{\tt hep-th/9811048}}].

\bibitem{Nakajima:2003pg}
H.~Nakajima and K.~Yoshioka, {\it Instanton counting on blowup. {I}},
  \href{http://xxx.lanl.gov/abs/math.ag/0306198}{{\tt math.ag/0306198}}.

\bibitem{Fujii:2005dk}
S.~Fujii and S.~Minabe, {\it A combinatorial study on quiver varieties},
  \href{http://xxx.lanl.gov/abs/math.ag/0510455}{{\tt math.ag/0510455}}.

\bibitem{Noma:2006pe}
Y.~Noma, T.~Nakatsu, and T.~Tamakoshi, {\it Plethystics and instantons on {ALE}
  spaces},  \href{http://xxx.lanl.gov/abs/hep-th/0611324}{{\tt
  hep-th/0611324}}.

\bibitem{Hori:2000kt}
K.~Hori and C.~Vafa, {\it Mirror symmetry},
  \href{http://xxx.lanl.gov/abs/hep-th/0002222}{{\tt hep-th/0002222}}.

\bibitem{Hori:2000ck}
K.~Hori, A.~Iqbal, and C.~Vafa, {\it D-branes and mirror symmetry},
  \href{http://xxx.lanl.gov/abs/hep-th/0005247}{{\tt hep-th/0005247}}.

\bibitem{Hanany:2001py}
A.~Hanany and A.~Iqbal, {\it Quiver theories from {D6}-branes via mirror
  symmetry},  {\em JHEP} {\bf 04} (2002) 009,
  [\href{http://xxx.lanl.gov/abs/hep-th/0108137}{{\tt hep-th/0108137}}].

\bibitem{Berkooz:1996km}
M.~Berkooz, M.~R. Douglas, and R.~G. Leigh, {\it Branes intersecting at
  angles},  {\em Nucl. Phys.} {\bf B480} (1996) 265--278,
  [\href{http://xxx.lanl.gov/abs/hep-th/9606139}{{\tt hep-th/9606139}}].

\end{thebibliography}\endgroup
\bibliographystyle{JHEP}

\end{document}